\documentclass[journal,twocolumn,10pt]{IEEEtran}
\usepackage[T1]{fontenc}
\usepackage{float}
\usepackage{amsmath}
\usepackage{amsthm}
\usepackage{amssymb}
\usepackage{stmaryrd}
\usepackage{graphicx}
\usepackage{multirow}
\usepackage{xcolor}
\usepackage{tabularx}
\usepackage{booktabs}
\usepackage{verbatim}
\providecommand{\tabularnewline}{\\}
\usepackage{cite}
\usepackage{balance}

\usepackage[switch]{lineno}
\usepackage[named]{algo}
\usepackage{algpseudocode}
\usepackage{algorithm}

\floatstyle{ruled}
\newtheorem{thm}{Theorem}

\newtheorem{lem}[thm]{Lemma}

\usepackage[inline]{enumitem}

\ifCLASSOPTIONcompsoc
  \usepackage[caption=false,font=normalsize,labelfont=sf,textfont=sf]{subfig}
\else
  \usepackage[caption=false,font=footnotesize]{subfig}
\fi

\begin{document}

\title{Resource Allocation in Heterogeneously-Distributed \\Joint Radar-Communications under \\Asynchronous Bayesian Tracking Framework}
%\title{Resource Allocation in Joint Heterogeneously-Distributed Radars and HetNets under Bayesian Tracking Framework}
\author{Linlong~Wu,~\IEEEmembership{Member~IEEE,} Kumar~Vijay~Mishra,~\IEEEmembership{Senior Member~IEEE,}
Bhavani~Shankar~M.~R.,~\IEEEmembership{Senior Member~IEEE,} and~Bj\"{o}rn~Ottersten,~\IEEEmembership{Fellow~IEEE}
\thanks{
The authors are with the Interdisciplinary Centre for Security, Reliability and Trust (SnT), University of Luxembourg, Luxembourg City L-1855, Luxembourg. E-mail: \{linlong.wu@, kumar.mishra@ext., bhavani.shankar@, bjorn.ottersten@\}uni.lu. \emph{Corresponding author: Linlong Wu.}
}
\thanks{Their work is supported in part by ERC AGNOSTIC under grant EC/H2020/ERC2016ADG/742648 and in part by FNR CORE SPRINGER under grant C18/IS/12734677.}
%\thanks{L. Wu acknowledges support via  Luxembourg National Research Fund (FNR) through the  ``\textit{S}ignal \textit{Pr}ocess\textit{i}ng for \textit{N}ext \textit{Ge}neration \textit{R}adar'' (SPRINGER) project and ERC Advanced Grant AGNOSTIC \textcolor{red}{write full form as I have written for SPRINGER}.}
\thanks{The conference precursor of this work was presented in the 2021 IEEE International Workshop on Signal Processing Advances in Wireless Communications (SPAWC).}
}
\maketitle
\begin{abstract}
Optimal allocation of shared resources is key to deliver the promise of jointly operating radar and communications systems. In this paper, unlike prior works which examine synergistic access to resources in colocated joint radar-communications or among identical systems, we investigate this problem for a distributed system comprising heterogeneous radars and multi-tier communications. In particular, we focus on resource allocation in the context of multi-target tracking (MTT) while maintaining stable communications connections. By simultaneously allocating the available power, dwell time and shared bandwidth, we improve the MTT performance under a Bayesian tracking framework and guarantee the communications throughput. Our \textit{a}lter\textit{n}ating allo\textit{c}ation of \textit{h}eterogene\textit{o}us \textit{r}esources (ANCHOR) approach solves the resulting non-convex problem based on the alternating optimization method that monotonically improves the Bayesian Cram\'{e}r-Rao bound. Numerical experiments demonstrate that ANCHOR significantly improves the tracking error over two baseline allocations and stability under different target scenarios and radar-communications network distributions.
\end{abstract}


\IEEEpeerreviewmaketitle{}

\section{Introduction}

%\item Background and justification of JRC. Two aspects of JRC. The resource allocation is also important for JRC.
Spectrum sharing between radar and communications systems has lately garnered significant research interest in order to efficiently exploit the electromagnetic spectrum, a scarce natural resource \cite{mishra2019toward,petropulu2020dfrc}. While radars require wider bandwidths to deliver a high-resolution sensing performance \cite{skolnik2008radar}, mobile communications are witnessing an ever-increasing demand for broader bandwidths in order to provide a designated quality-of-service (QoS) \cite{Digicomm}. Spectrum-sharing solutions address these competing requirements either by \textit{opportunistic spectral access} of broadcast systems by passive radars \cite{khan2016opportunistic}; \textit{spectrally coexisting} legacy systems that operate by mitigating the mutual interference with minimum design modifications at the cost of performance \cite{interferencealignment,ayyar2019robust,wu2019sequence}; or \textit{spectrally co-designed} joint radar-communications (JRC) systems \cite{Dokhanchi2019AmmWave,liu2018toward,liu2021integrated} that utilize a common transmit/receive hardware and waveform. %{\color{red} Although resource allocation strategies for all the three types of radar-communications We consider optimized resource allocation for the three most commonly used types of JRC schemes}. 

%\item Existing work on RA of JRC.
A successful operation of these spectrum sharing systems relies on an efficient radio resource optimization \cite{shajaiah2014resource,tan2016optimizing,zatman2016joint}. In coexistence paradigms with single antenna or with a fixed beamforming, the resources may include total transmit power, transmit signal bandwidth, and transmission time slots \cite{Dokhanchi2019AmmWave}. Additionally, in phased array or multiple-input multiple-output (MIMO) systems, antennas need to be reserved for optimal sharing \cite{Athina1}. In single antenna JRC systems, the resource allocation objective is to optimize the transmit energy of the dual-purpose waveform based on the propagation channels of radar and communications users. In JRC systems employing a transmit antenna array, highly directional beamforming toward the radar surveillance area while also ensuring sidelobes on the users offers an interesting solution \cite{DFRC}. In JRC systems involving hybrid analog-digital transceiver architectures \cite{elbir2021terahertz}, antenna selection may also be viewed as a component of resource allocation objective.

% distributed systems
While colocated JRC solutions continue to be comprehensively investigated, distributed JRC systems involving widely separated radars and communications transmitters remain relatively unexamined. Among prior works, the spectrum-sharing model in \cite{he2019performance} comprises a widely distributed MIMO radar \cite{sun2019target} but studies coexistence with a simplistic point-to-point MIMO communications. The distributed radar proposed in \cite{sedighi2021localization} exploits a communications waveform but operates in passive (receive-only) mode. The co-design framework in \cite{liu2020co} develops new waveforms and precoders for a distributed but identical units of radar and communications. The displaced sensor imaging in \cite{wang2020displaced} considers identical units of radar sensors on the same vehicular platform that coordinate timing via communications protocols.

There are very few resource allocation studies for  identical or \textit{homogeneous} distributed JRC systems \cite{yimin}. In such a system, the objective is to optimize the transmission strategies (e.g., transmit power, duration, bandwidth) for identical transmit systems in order to minimize the target localization error in a dynamic scenario and maximize the communications capacity. In the context of distributed radar resource allocation for target localization and tracking, seminal works have appeared recently \cite{yan2016joint,yan2017cooperative,yan2017joint,yan2020optimal,yan2021target}. In \cite{yan2016joint}, a joint beam and power allocation scheme was developed for a colocated MIMO radar network. In \cite{yan2017cooperative}, target assignment and dwell time allocation were considered simultaneously in a phased array radar network. Subsequently, in \cite{yan2017joint}, an adaptive power allocation was proposed with dynamic tracking threshold adjustment based on the asynchronous working manner of the radar networks. This was extended in \cite{yan2020optimal} by considering \textit{heterogeneous} radar (HetRad) network, i.e. comprising different types of radar units, for multi-target tracking (MTT). 
%It proposed corresponding resource allocation algorithms based on the Bayesian Cram\'{e}r-Rao bound (CRB) on the estimates of target parameters. 
%Further, to increase the target capacity, the power and dwell time allocation was investigated in \cite{yan2021target}, and the formulated problem was solved by a two-step procedure.  

However, aforementioned prior works do not consider spectrum sharing paradigm with wireless communications. %, whose emerging importance has been already stressed. 
Additionally, the distributed nature of radar deployment, strong likelihood of encountering heterogeneous radar units in practice, and an ever-increasing deployment of cellular communications potentially makes optimal resource allocation difficult in a dynamic scenario. In each of these cases, the resulting spectrum reuse leads to interference between the two systems and impacts the resource allocation strategy. In \cite{yimin}, resource allocation is considered for target localization excluding additional complexities of a heterogeneous radar network and multi-tier communications. %only a point-to-point communications system was examined. 
Some recent state-of-the-art works \cite{zafar1,karimi2021energy} on combining sensing functionalities in heterogeneous wireless networks (HetNets) explore the effect of varied communications coverage areas on the joint system performance. However, these studies do not explore resource allocation and do not, \textit{per se}, qualify as JRC systems. 

In this paper, we consider a heterogeneous mix of radar units and multi-tier communications to propose optimal allocation of resources such as power, dwell time, and bandwidth. Our approach optimizes the radar tracking performance while maintaining a reliable QoS in communications systems. Preliminary results of this work appeared in our conference publication \cite{wu2021heterogeneously} where multi-tier communications and various systems configurations were ignored. %Comparing to our preliminary work \cite{wu2021heterogeneously}, 
In this work, we further generalize the study in \cite{wu2021heterogeneously} to the following aspects:
First, from system model perspective, we include target Doppler in the radar measurement model and estimate all key parameters, $-$ range, angle-of-arrival (AOA), and Doppler velocity $-$ of a target. Next, we model the communications system as a multi-tier cellular network. We also consider the allocation of the shared frequency bands besides power and dwell time in the joint HetRad-HetNet scenario. Finally, a novel algorithm based on the alternating optimization framework is developed to solve the resulting non-convex problem. For each subproblem, we further propose an iterative method to solve with low computational complexity and proven convergence analyses. Our proposed \textit{a}lter\textit{n}ating allo\textit{c}ation of \textit{h}eterogene\textit{o}us \textit{r}esources (ANCHOR) algorithm for distributed resource allocation shows roughly $10$ times improvement of tracking performance in the sense of root  mean  square  error over the baseline allocations.
%\cite{van1987simulated,kirkpatrick1983optimization} approach.

%\item Structure (write as follows)
The rest of this paper is organized as follows. In the next section, we introduce the configurations and system model of the heterogeneous JRC network. We formulate the resource allocation problem in Section~\ref{sec:problem} and describe the proposed ANCHOR algorithm in Section~\ref{sec:algo}. We validate our models and methods via numerical experiments in Section~\ref{sec:simu}. We conclude in Section~\ref{sec:summ}.

% \item Notations (write as shown in the following example)
%\emph{Notation:}
Throughout the paper, we reserve boldface lowercase and boldface uppercase for vectors and matrices, respectively. 
%See Table \ref{tab:notation} for other notations used throughout the paper.
We denote the transpose, conjugate, and Hermitian by $(\cdot)^T$, $(\cdot)^*$, and $(\cdot)^H$, respectively. The notation $\mathbf{I}_N$ is the $N\times N$ identity matrix, and $\boldsymbol{1}_{N}$ is a vector with all $N$ elements being 1. The Kronecker product is denoted by $\otimes$; $||\cdot||_p$ is the $\ell_p$ norm; and $\left\Vert \cdot\right\Vert _{0}$ is the number of non-zero elements of the vector. The notation $\text{Tr}\left\lbrace \cdot \right\rbrace $ is the trace of the matrix, $|\cdot|$ is the cardinality of a set,  $\mathbb{E}\left[ \cdot \right]$ is the statistical expectation function, $\nabla$ is the gradient operator, and $\mathcal{N}\left(\cdot,\cdot\right)$ represents the probability density function of the normal distribution. 
%\begin{table}[h]
%	\centering{}
%	\caption{Notations}
%	\label{tab:notation}
%	\begin{tabular}{p{0.25\columnwidth}p{0.5\columnwidth}}
%		\toprule
%		$(\cdot)^T$ & Transpose \\
%		$(\cdot)^*$ & Conjugate \\
%		$(\cdot)^H$ & Conjugate transpose \\
%		$\mathbf{I}_N$ & Identity matrix of size $N\times N$ \\
%		$\mathbf{1}_N$ & Column vector with all $N$ elements being 1 \\
%		$\text{Tr}\left\lbrace\cdot\right\rbrace$ & Trace\\
%		$\otimes$ & Kronecker product \\
%		$\odot$ & Hadamard product \\
%		$||\cdot||_p$ & $\ell_p$ norm \\
%		$\left\Vert\cdot\right\Vert _{0}$ &  Number of non-zero elements in a vector \\
%		$|\cdot|$ & Cardinality of a set \\
%		$\left[\cdot\right]_n$ & The $n$-th element of a vector\\
%		$\mathbb{E}\left[\cdot\right]$ & Statistical expectation \\
%		$\nabla$ & Gradient operator\\
%		$\mathcal{N}\left(\cdot\right)$ & Gaussian distribution\\
%		%$\partial$ & Partial derivative operator\\		
%		\bottomrule
%	\end{tabular}
%\end{table}
Table~\ref{tbl:notations} summarizes some of the important notations in this paper.
%----------------------------------------------------------------
\begin{table}[t]	
%\captionsetup{labelfont={color=red},font={color=red}}
\caption{Glossary of notations}
        \label{tbl:notations}
		\vspace{-15pt}
		\footnotesize
        \begin{center}
		\begin{tabular}{ |p{1.5cm}||p{6.2cm}|}
			\multicolumn{2}{c}{} \\
    		\hline
			Symbol & Description \\
			\hline		
			$N$ & Total number of radars\\
            $N_c$ & Number of colocated MIMO radars\\
			$N_p$ & Number of phased array radars \\
			$N_m$ & Number of mechanical scanning radars \\
			$\varphi_{c}$ & Index set of colocated MIMO radars\\
			$\varphi_{p}$ & Index set of phased array radars \\
			$\varphi_{m}$ & Index set of mechanical scanning radars \\
			$\left(x_{i},y_{i}\right)$ & Coordinates of radar $i$\\
			$M_{i,q,k}$ &  number of measurements on target $q$ by radar $i$ in fusion interval $k$\\
			$t_{i,q,k}^{m}$ & arrival time of the $m$-th measurement\\
			$P_{i,q,k}^{m}$ & Power corresponding to receiving time $t_{i,q,k}^{m}$\\
			$T_{i,q,k}^{m}$ & Dwell time corresponding to receiving time $t_{i,q,k}^{m}$\\
			$T_{0}$ & Period of fusion interval\\
			$J$ & Number of communications macro users\\
			$L$ & Number of communications micro users\\
			$\beta^{j}$ & Transmission channel gain\\
			$\alpha_{j,i}^{r}$ & Interference channel gain from radar $i$ to user $j$\\
			$\alpha_{i,j}^{c}$ & Interference channel gain from userv $j$ to radar $i$\\			$\sigma_{c,j}^{2}$ & Communications noise power\\
            $\delta^{j}$ & Interference channel gain from macro user $j$\\
			$P_{c}^{j}$ & Power from user $j$\\
			$\Delta f$ & subchannel bandwidth \\
			$B$ & Total available bandwidth\\
			$F$ & Number of subchannels\\
            $F_{c}^{j}$ & Number of subchannels for user $j$\\
            $F_{r}^{i}$ & Number of subchannels for radar $i$\\
			$F_{i,j}^{o}$ & Number of overlapped subchannels between radar $i$ and user $j$\\
			$\boldsymbol{f}_{j,k}^{c}$ & Subchannel selection vector of user $j$\\
			$\boldsymbol{f}_{i,k}^{r}$ & Subchannel selection vector of radar $i$\\
            $\tilde{\alpha}_{i,j}^{c}$ & Fourier coefficients $y_{m,q}^{p}$ after focusing at frequency $\nu$\\
            $N$ & Number of Fourier coefficients in each channel\\
            $\tilde{\alpha}_{j,i}^{r}$ & Set of sampled Fourier coefficients per channel\\
			$\boldsymbol{s}_{i,q,k}^{m}$ & State of target $q$ from radar $i$ at $t_{i,q,k}^{m}$  \\
            $\boldsymbol{y}_{i,q,k}^{m}$ & Measurement of target $q$ from radar $i$ at $t_{i,q,k}^{m}$\\
            $x_{i,q,k}^{m} (y_{i,q,k}^{m})$ & Coordinate x (y) of target $q$ at time $t_{i,q,k}^{m}$\\
            $\dot{x}_{i,q,k}^{m} (\dot{y}_{i,q,k}^{m})$ & Velocity along x (y) axis of target $q$ at time $t_{i,q,k}^{m}$\\
            \hline
		\end{tabular}%	
        \end{center}
		%
		%	\vspace*{-5mm}
\end{table}		  
%----------------------------------------------------------------

\section{Network Configuration and System Model}\label{sec:model}
\begin{figure}[t]
\begin{centering}
\includegraphics[scale=0.5]{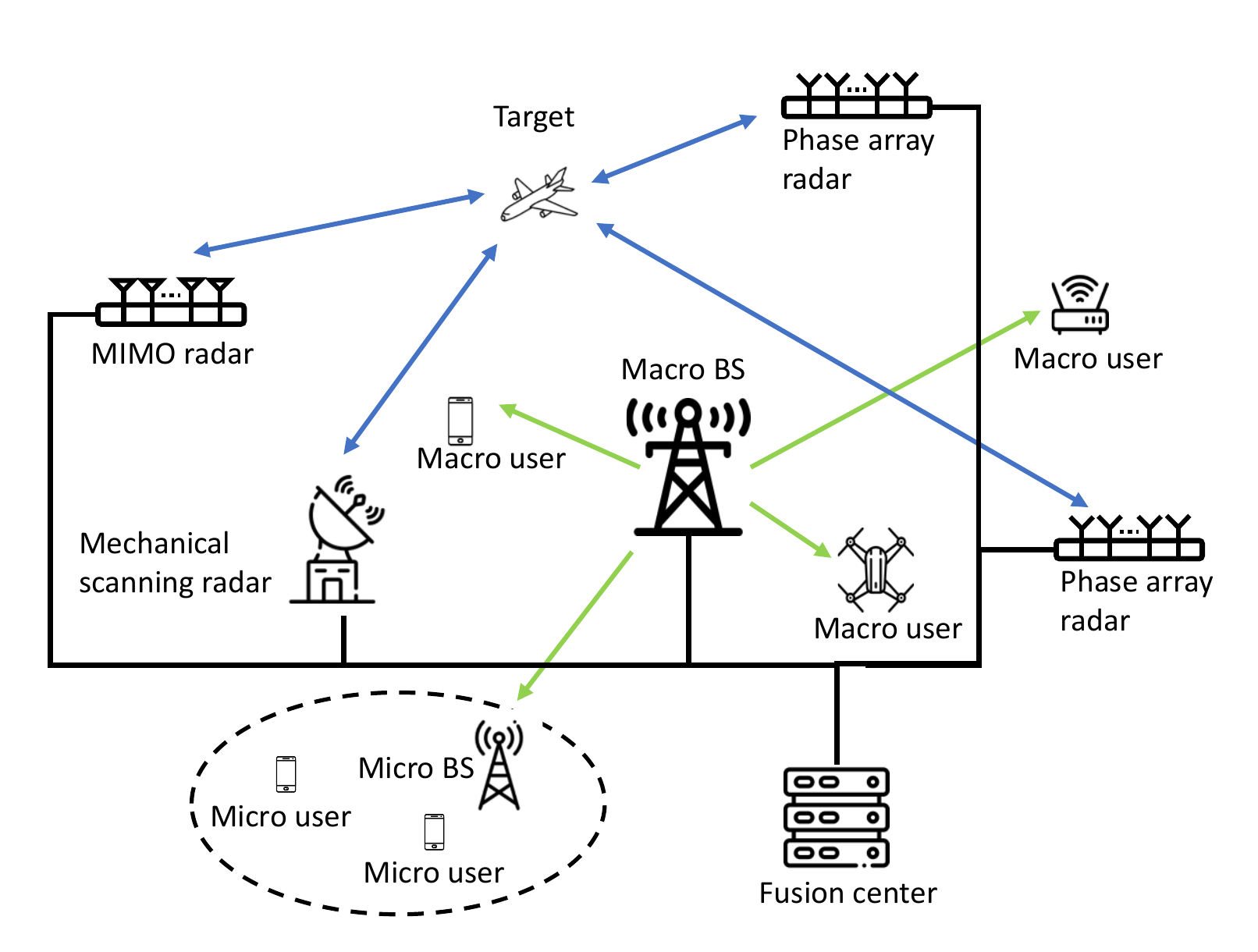}
\par\end{centering}
\caption{\label{fig:Sketch-of-the}An illustration of HRCN, wherein radar and cellular networks share the spectrum and their signals are source of mutual interference.}
\end{figure}

Consider a heterogeneously-distributed radar and communications network (HRCN) (Figure~\ref{fig:Sketch-of-the}), which employs different types of radars that simultaneously track multiple targets. The radars operate in the same spectrum as a multi-tier communications system that comprises a macro base station (BS) and multiple micro and macro users. The HRCN is equipped with a fusion center for signalling, synchronization, and resource allocation. Here, an efficient use of spectrum is possible provided the resources are well allocated amongst radar and communications entities. The HRCN is embedded with the following heterogeneities: 
\begin{description}
\item%Heterogeneity in network architecture: the HRCN is a cooperative network aiming to track multiple targets while providing wireless communications services. Thus, it consists of generic radar systems and cellular communications links. For the radar components, colocated MIMO radars, phased array radars and mechanical scanning radars also co-exist to provide a tracking service. In this work, we consider downlink transmissions from a single cell base station (BS).
[Heterogeneity in network architecture] The HRCN %is a cooperative network that aims to track multiple targets while providing communications services. Thus, it 
has both radar and cellular network components. The radar network comprises MIMO, phased array, and mechanically scanning radars that are heterogeneously distributed while tracking the same target. The multi-tier cellular communications network has macro and micro BSs along with multiple macro users.
\item%Heterogeneity in allocated resources: for both tracking and wireless communication, the performance is affected by some common factors such as transmit power, dwell time and bandwidth. These heterogeneous resources are, in general, required to be allocated in a well-designed manner to guarantee or optimize the overall performance. 
[Heterogeneity in allocated resources] Both tracking and communications QoS are affected by respective signal-to-interference-plus-noise ratios (SINRs), which further depend on the transmit power, radar's dwell time, and bandwidth usage. To mitigate the mutual interference and enhance the overall performance, these heterogeneous resources must be judiciously allocated.
\end{description}
%The sketch of the described HRCN and its application scenario is shown in Figure \ref{fig:Sketch-of-the}.  %The specific goal under this investigation is to exploit resource allocation to optimize the radar tracking performance while guaranteeing the communications throughput requirements. 
In the following, we describe each of these HRCN configurations in detail.

\subsection{Configurations of the HRCN}
\begin{figure*}[t]
\begin{centering}
\subfloat[\label{fig:Scheme-Colocated-MIMO-radar}Colocated MIMO radar.]{\begin{centering}
\includegraphics[scale=0.3]{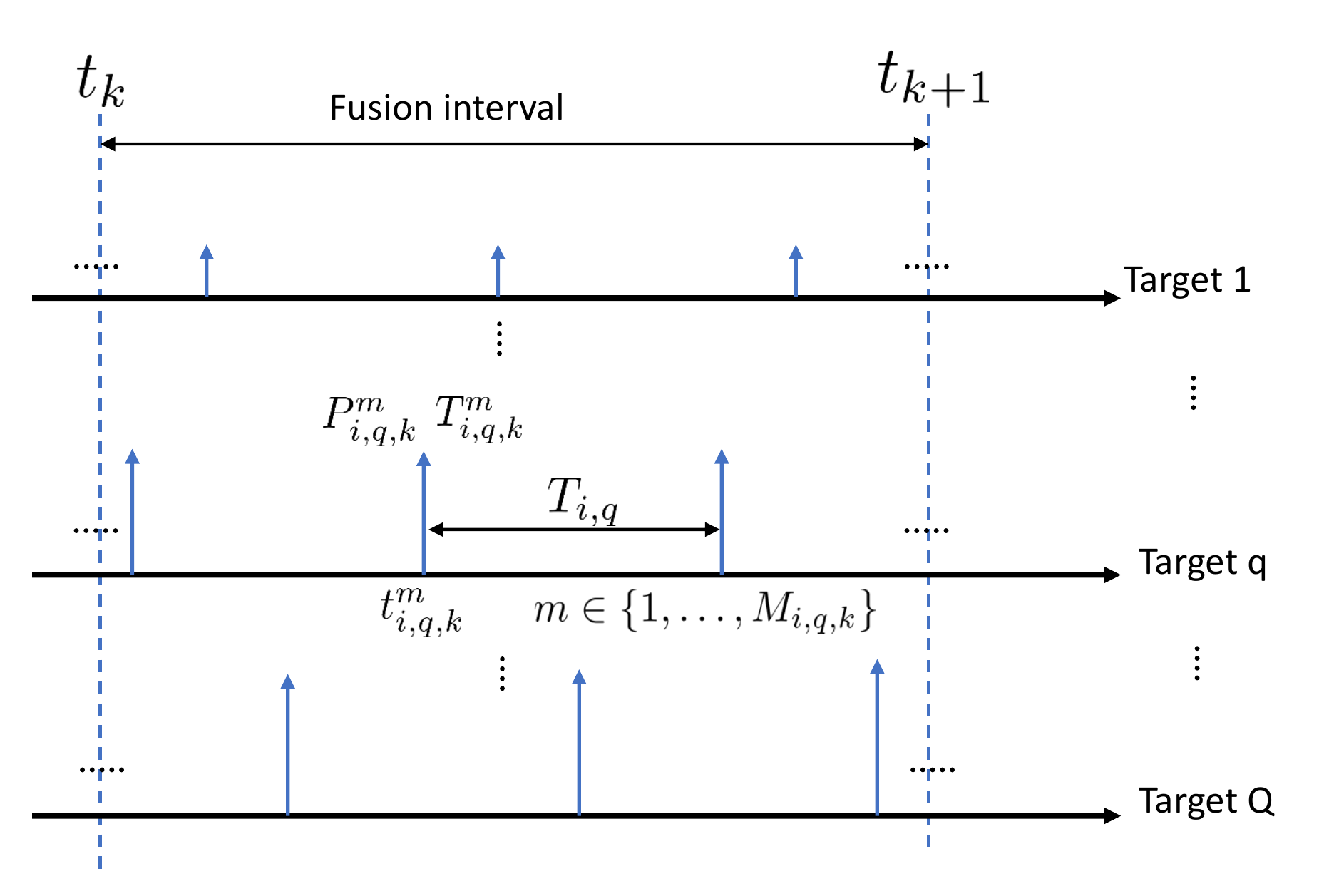}
\par\end{centering}
}\subfloat[\label{fig:Scheme-Phased-array-radar}Phased array radar.]{\begin{centering}
\includegraphics[scale=0.3]{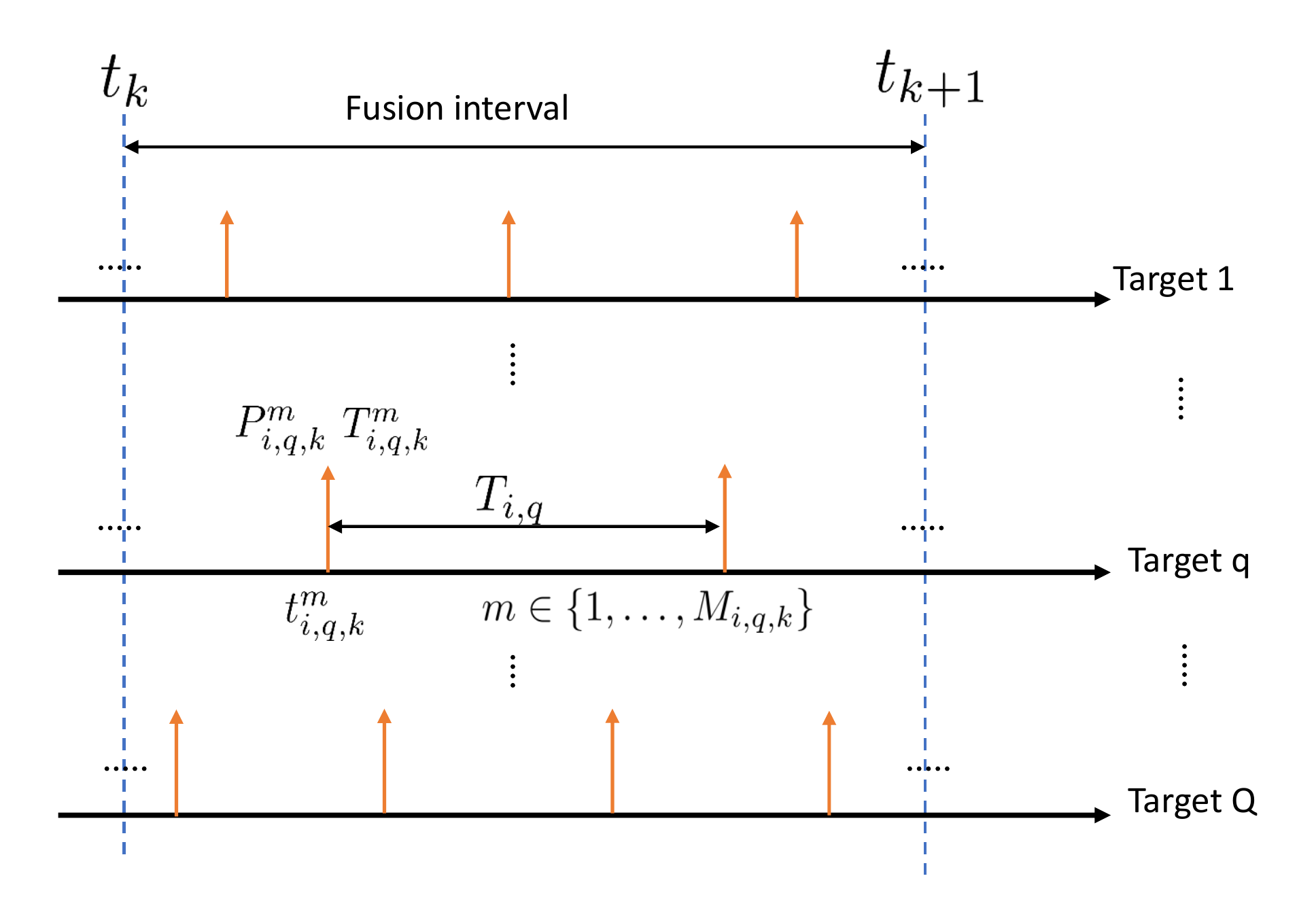}
\par\end{centering}
}\subfloat[\label{fig:Scheme-Mechanical-scanning-radar}Mechanical scanning radar.]{\begin{centering}
\includegraphics[scale=0.3]{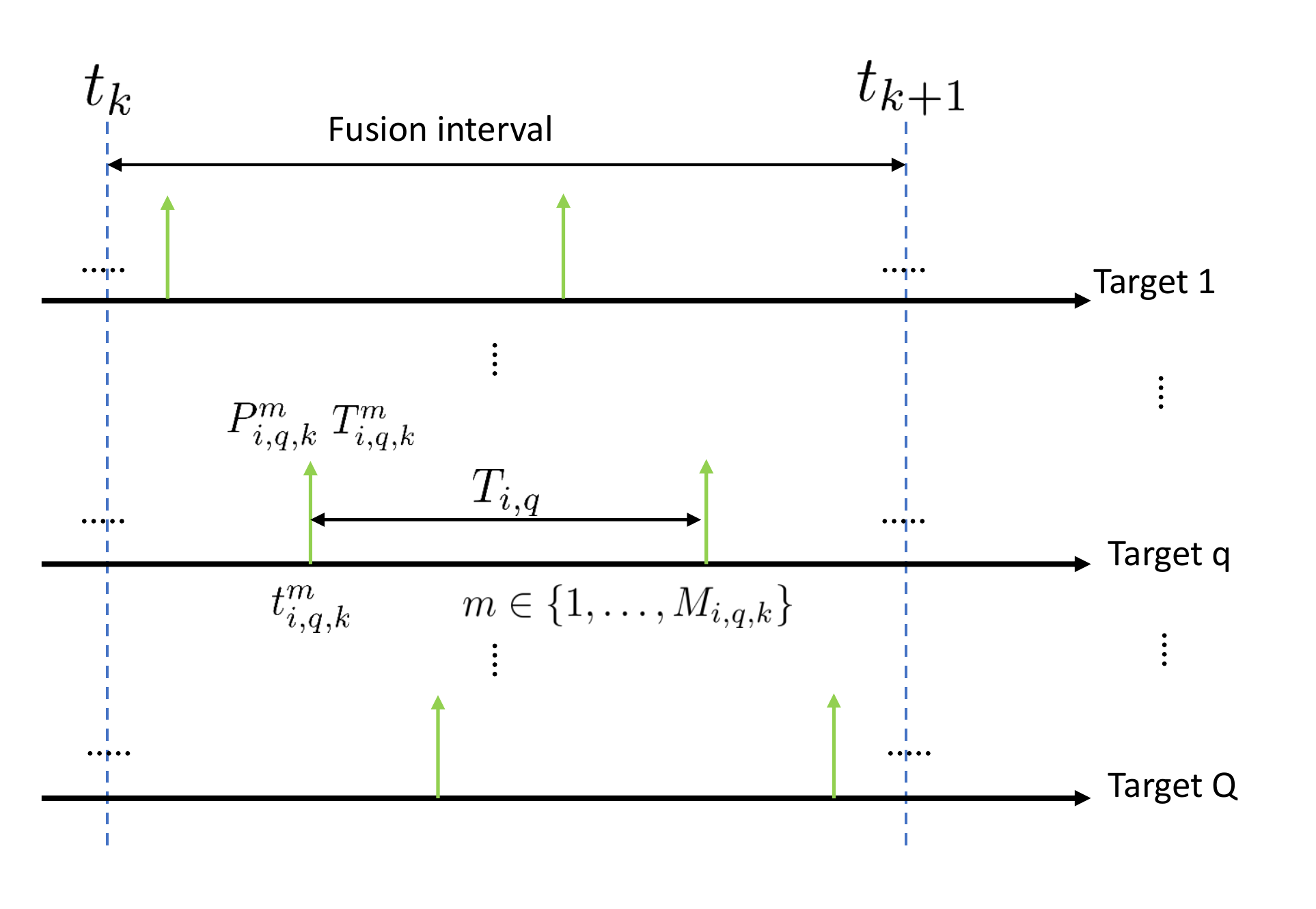}
\par\end{centering}}
\par\end{centering}
\caption{\label{fig:Operation-schemes}{Illustrations of scanning operations for (a) colocated MIMO, (b) phased array, and (c) mechanically scanning radars, where $M_{i,q,k}$ is the number of measurements collected from target $q$ by radar $i$ during the $k$-th fusion interval (from $t_{k}$ to $t_{k+1}$); $t_{i,q,k}^{m}$ is the arrival time of the $m$-th measurement for $m=1,\ldots,M_{i,q,k}$;
and $P_{i,q,k}^{m}$ and $T_{i,q,k}^{m}$ are,, respectively, the transmit power and dwell time corresponding to the measurement at time $t_{i,q,k}^{m}$.}}
\end{figure*}

The HRCN is a complex system with several operating units of radars and communications that require numerous system parameters to be optimized for multiple resources.
\paragraph{Radar set-up}
Consider $N$ radars in the HRCN to track $Q$ widely separated point-like targets in the surrounding environment. In particular, assume $N_{c}$ colocated MIMO radars, $N_{p}$ phased array radars, and $N_{m}$ mechanical scanning radars with $N=N_{c}+N_{p}+N_{m}$, located at different places with the coordinates $\left\{ \left(x_{i},y_{i}\right)\right\} _{i=1}^{N}$. Define the sets $\varphi_{c}\triangleq\left\{ 1,\ldots,N_{c}\right\} $, $\varphi_{p}\triangleq \left\{ N_{c}+1,\ldots,N_{c}+N_{p}\right\} $,
and $\varphi_{m} \triangleq \left\{ N_{c}+N_{p}+1,\ldots,N\right\} $ to index these radars. The  operational scheme of each of the radars are as follows:
\begin{description}
\item[{Colocated MIMO radar}] For $i\in\varphi_{c}$, the radar $i$ is capable of pointing multiple beams to illuminate multiple targets simultaneously.
Herein, the multiple targets are illuminated for the same dwell time but with different transmit powers. Hence, the revisit time intervals for the $Q$ targets are the same. 
\item[{Phased array radar}] For $i\in\varphi_{p}$, the radar $i$ steers the beam to illuminate multiple targets sequentially by configuring the phases of the transmit array. In this scheme, each target is illuminated with identical transmit power but with a different dwell time. Hence, the revisit time intervals for the $Q$ targets are usually different. 
\item[{Mechanical scanning radar}] For radars indexed by $\varphi_{m}$, the system parameters are usually fixed. The radar scans mechanically and illuminates all the targets sequentially with identical power and dwell time (and hence identical revisit times). 
\end{description}
The scanning operations of each radar are depicted in Figure~\ref{fig:Operation-schemes}, %(at the top of next page), 
where $M_{i,q,k}$ is the number of measurements collected from target $q$ by radar $i$ during the $k$-th fusion interval\footnote{Fusion interval \cite{olfati2005consensus,chen2003performance} is the period during which multiple sensors measure and estimate independently. The target state is determined by fusing all radar estimates.} (from $t_{k}$ to $t_{k+1}$); $t_{i,q,k}^{m}$ is the arrival time of the $m$-th measurement for $m=1,\ldots,M_{i,q,k}$; $P_{i,q,k}^{m}$ and $T_{i,q,k}^{m}$ denote, respectively, the transmit power and dwell time corresponding to the measurement at time $t_{i,q,k}^{m}$. Assume that the fusion interval is a constant denoted by $T_0$. 
%For example, during the $k$-th fusion interval, there are $M_{i,q,k}$  measurements from the $q$-th target for the $i$-th radar at the receiving time $t_{i,q,k}^{m}$ corresponding to the transmit power $P_{i,q,k}^{m}$ for $m=1,\ldots,M_{i,q,k}$.
The inter-measurement duration (Fig.~\ref{fig:Scheme-Colocated-MIMO-radar}) at $i$th MIMO radar for the $q$th target (revisit intervals) is denoted by a constant $T_{i,q}$. However, $P_{i,q,k}^{m}$ varies for $q=1,\ldots Q$.  %Figure~\ref{fig:Scheme-Phased-array-radar} indicates 
In case of phased-array radars (Fig.~\ref{fig:Scheme-Phased-array-radar}), $T_{i,q}$ are different but $P_{i,q,k}^{m}$ is identical for $q=1,\ldots Q$. Finally, for a mechanically scanning radar (Fig.~\ref{fig:Scheme-Mechanical-scanning-radar}), both $T_{i,q}$ and $P_{i,q,k}^{m}$ are the same for $q=1,\ldots Q$. 

\begin{figure}[t]
\begin{centering}
\includegraphics[scale=0.55]{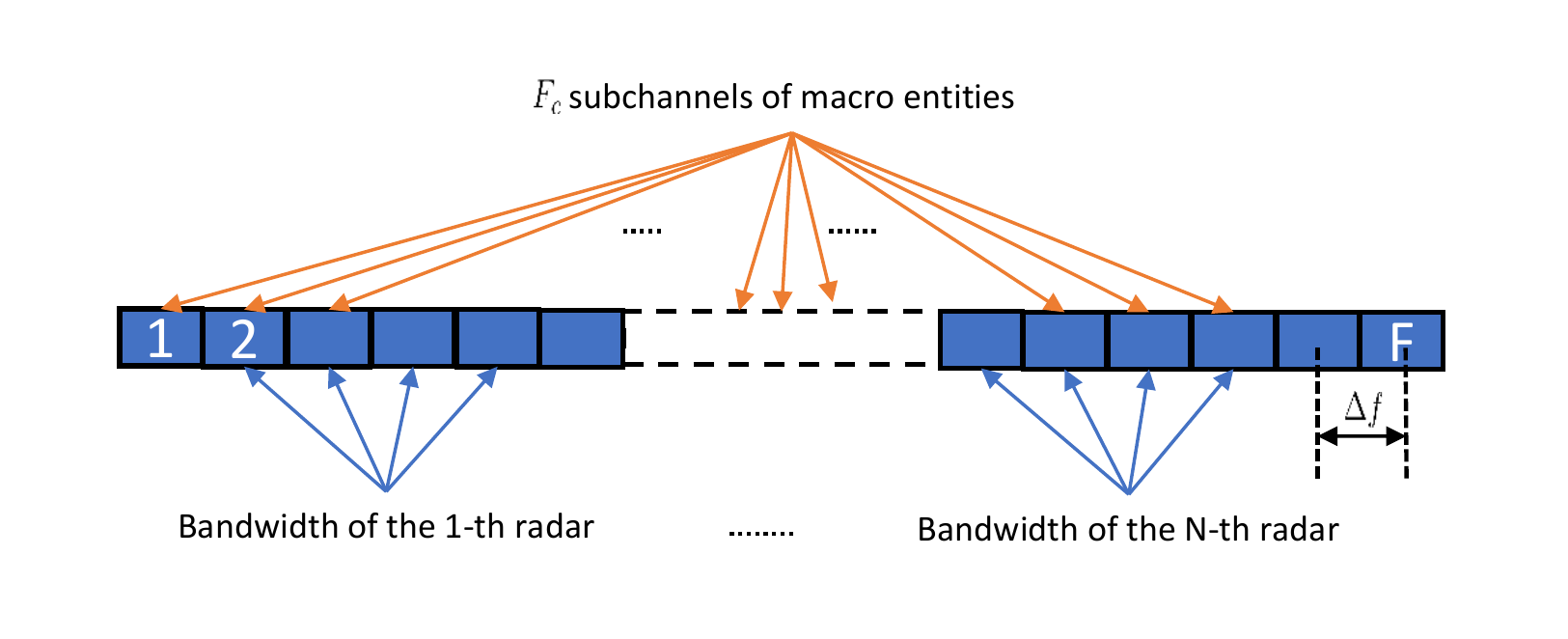}
\par\end{centering}
\caption{\label{fig:The-spectrum-usage}{Illustration of spectrum usage for the HRCN, where each blue block represents the subchannel or subcarriers. The occupied bandwidth of each radar is continuous and pre-allocated.}}
\end{figure}

\paragraph{Wireless communications configuration}
Consider a multi-tier heterogeneous network (HetNet) of communications (Fig.~\ref{fig:Sketch-of-the}) that consists of a macrocell (primary network) with multiple macro users and a microcell (secondary network). The primary network of the macro BS serves $J$ macro users and one micro BS. The secondary network is served by the micro BS with $L$ micro users. The co-existence of the multi-tier networks, frequency reuse, and precoding has been well-studied in literature \cite{olwal2016survey}. For the HRCN resource allocation, we make the following model to simplify the problem. %ing assumptions are made on the configuration of the communications network.
%The configurations and assumptions of this multi-tier HetNet are as follows:
%
\begin{description}
    \item[Communications downlinks]  The coverage areas of the macro and micro BSs are different. When reusing the frequencies, we assume that the macro downlinks cause interfere to only $L$ micro users. Similarly, the micro downlinks interfere within the coverage area of only micro BS. This is especially true when the micro network is power limited.
 \item[SINR of macro user] The macro BS employs orthogonal frequency-division multiple access (OFDMA) for downlink and allocates %orthogonal 
 non-interfering resource blocks. As a result, the mutual interference among the $J$ macro downlinks is non-existent. The SINR of the $j$-th macro user is\footnote{The radar illumination is irregular and non-continuous in a fusion interval. Hence, the commonly used transient SINR is unavailable. Instead, we use the average SINR as the communications metric. Additionally, the averaging period may be replaced by the airtime $T_c$ of the communications transmission, which has the same expression but with a scaling constant $T_c/T_0$.}
 \par\noindent\small
    \begin{equation}
        \text{SINR}_{j}=\frac{\beta^{j}P_{c}^{j}T_{0}}{\sum_{i=1}^{N}\sum_{q=1}^{Q}\sum_{m=1}^{M_{i,q,k}}\alpha_{j,i}^{r}P_{i,q,k}^{m}T_{i,q,k}^{m}+\sigma_{c,j}^{2}T_{0}},
        \label{eq:SINR_macro}
    \end{equation}\normalsize
    where $\beta^{j}$ and $P_{c}^{j}$ are the channel gain and allocated power to the $j$-th macro entity, respectively; $\alpha_{j,i}^{r}$ is the interference channel gain from the $i$-th radar; and $\sigma_{c,j}^{2}$ is the noise power at the receiver front-end of $j$th user.
    \item[SINR of micro user] To improve spectrum utilization, the micro BS may reuse the same band of the macro BS \cite{hasan2015distributed} and then apply the (non-interfering) OFDMA resource blocks amongst the $L$ micro users. This leads to negligible mutual interference among the $L$ micro users. Compared to the SINR of macro entities, the SINR for the $l$-th micro user becomes\par\noindent\small
    \begin{equation}
        \hspace*{-0.2in}\begin{aligned} & \small{\text{SINR}_{l}}= \small{\frac{\beta^{l}P_{c}^{l}T_{0}}{\delta^{j}P_{c}^{j}T_{0}+\sum_{i=1}^{N}\sum_{q=1}^{Q}\sum_{m=1}^{M_{i,q,k}}\alpha_{l,i}^{r}P_{i,q,k}^{m}T_{i,q,k}^{m}+\sigma_{c,l}^{2}T_{0}},}
        \label{eq:SINR_micro}
        \end{aligned}
    \end{equation}\normalsize
    where $\delta^{j}P_{c}^{j}T_{0}$ represents the interference from the $j$-th macro user of the same frequency; $\delta^{j}$ is the interference channel gain; and the other parameters are similar to their counterparts in \eqref{eq:SINR_macro}.
    \item[Codebook] We assume Gaussian codebook for all transmissions.
\end{description}
\paragraph{Configurations for shared bandwidth}
%Since the mm-wave cellular communications are considered in the HRCN, it enables efficient use of spectrum provided the resources are well allocated amongst radar and communications entities.
%the bandwidth within a cellular range provides another dimension of degree of freedom, which should be well allocated to further enhance the overall performance. 
Assume $B$ to be the total bandwidth available to the HRCN. The sub-carrier interval of OFDMA
$\Delta f$ and the spectrum is divided into $F$ sub-channels
with $B=\Delta f\left(F-1\right)$. The allocated bandwidth is configured as follows:
\begin{description}
\item[Bandwidth configuration for communications] To serve the $J$ macro entities,
the communications network requires  $F_{c}$ sub-channels  with $F_{c}=\sum_{j=1}^{J}F_{c}^{j}$,
where $F_{c}^{j}$ is the number of sub-channels for the $j$-th user. A selection vector $\boldsymbol{f}_{j,k}^{c}\in\mathbb{R}^{F}$, comprising of binary entries, represents the selected sub-channels for the $j$-th downlink in the $k$-th fusion interval. Similar operations are ascribed to the $L$ micro users. The bandwidth allocation (i.e. $\left\{ \boldsymbol{f}_{j,k}^{c}\right\} _{j=1}^{J}$)
may be reassigned for each fusion interval to meet the changing demands of the wireless communications.
\item[Bandwidth configuration for radars] The bandwidth required by the
$i$-th radar is $B_{i}=\Delta f\cdot F_{r}^{i}$. Similar to communications, a selection vector $\boldsymbol{f}_{i,k}^{r}\in\mathbb{R}^{F}$
represents the selected sub-channels for the $i$-th radar during the $k$th fusion interval. In this work, we fix the allocation to $N$ radar bands (i.e. the occupied sub-channels) implying a higher priority to radar performance than communications. This also simplifies the problem formulation because only communications bands need to be assigned. However, this may result in limited exploitation of the opportunities offered by a flexible HRCN.
\end{description}
Fig.~\ref{fig:The-spectrum-usage} illustrates the bandwidth shared by radars and communications, where some communications sub-channels  partially/completely overlap with those used by radars.
%
%\vspace*{-0.1in}
\paragraph{HRCN interference}
During a fusion interval, multiple radar measurements coexist with $J$ communications downlinks. Consequently, %the $J$ downlinks are affected by some radars, and the radars are also interfered constantly by the communications. 
we have the following cases of interference within the HRCN:
\begin{description}
\item[Intra-system interference] As mentioned earlier, there is no intra-tier interference arising from the non-interfering OFDMA resource-block allocation. But spectrum reuse results in cross-tier interference to the micro users. For $N$ radars, we do not impose any limitation on the bandwidth overlap. The interference among the $N$ radars is assumed to be negligible because all the constituent radars adopt directional beams. 
\item[Inter-system interference] This refers to the mutual interference between radars and communications. The coefficients accounting for the interference from the $j$-th macro downlink to the $i$-th radar and from the $i$-th radar to the $j$-th downlink are denoted by $\alpha_{i,j}^{c}$ and $\alpha_{j,i}^{r}$, respectively. Both $\alpha_{i,j}^{c}$ and $\alpha_{j,i}^{r}$ are determined by the spectral overlap. The in-band interference from communications to radar is proportional to the overlap \cite{cordill2013electromagnetic}. Thus, %$\alpha_{i,j}^{c}$ can be expressed as 
$\alpha_{i,j}^{c}=\tilde{\alpha}_{i,j}^{c}\left(\boldsymbol{f}_{i,k}^{r}\right)^{T}\boldsymbol{f}_{j,k}^{c}$, where $\tilde{\alpha}_{i,j}^{c}$ is a constant unrelated to the bandwidth\footnote{If the channel is frequency-selective \cite{rodriguez2018channel}, then $\alpha_{i,j}^{c}=\left(\boldsymbol{f}_{i,k}^{r}\right)^{T}\left(\tilde{\boldsymbol{\alpha}}_{i,j}^{c}\odot\boldsymbol{f}_{j,k}^{c}\right)$, where the vector $\tilde{\boldsymbol{\alpha}}_{i,j}^{c}$ represents frequency-selectivity.}. Similarly, %for the parameter $\alpha_{l,i}^{r}$, we have
$\alpha_{j,i}^{r}=\tilde{\alpha}_{j,i}^{r}\left(\boldsymbol{f}_{j,k}^{c}\right)^{T}\boldsymbol{f}_{i,k}^{r}$, where $\tilde{\alpha}_{j,i}^{r}$ is a constant unrelated to the bandwidth. The mutual interference between the $i$-th radar and the $l$-th micro downlink is analogously defined.
\end{description}
%
%\vspace*{-0.21in}
\subsection{Target tracking model}
During the $k$-th fusion interval, the state of target $q$ at time $t_{i,q,k}^{m}$ with $m\in\left\{ 1,\ldots,M_{i,q,k}\right\} $
is %defined as \par\noindent\small
%\begin{equation}
$\boldsymbol{s}_{i,q,k}^{m}=\left[x_{i,q,k}^{m},\dot{x}_{i,q,k}^{m},y_{i,q,k}^{m},\dot{y}_{i,q,k}^{m}\right]^{T}$, 
%\end{equation}\normalsize
where $\left(x_{i,q,k}^{m},y_{i,q,k}^{m}\right)$ and $\left(\dot{x}_{i,q,k}^{m},\dot{y}_{i,q,k}^{m}\right)$ denote target's
position and Doppler velocity, respectively, in Cartesian coordinate system. The state $\boldsymbol{s}_{t_{k+1}}^{q}$ of the target
$q$ at fusion time $t_{k+1}$ is
%\par\noindent\small
%\begin{equation}
$\boldsymbol{s}_{t_{k+1}}^{q}=\left[x_{t_{k+1}}^{q},\dot{x}_{t_{k+1}}^{q},y_{t_{k+1}}^{q},\dot{y}_{t_{k+1}}^{q}\right]^{T}$. 
%\end{equation}\normalsize
The state transition and measurement model is \par\noindent\small
\begin{equation}
\begin{cases}
\boldsymbol{s}_{t_{k+1}}^{q}=f\left(\boldsymbol{s}_{i,q,k}^{m},t_{k+1}-t_{i,q,k}^{m}\right)+\boldsymbol{\gamma}_{k}^{q},\\
\boldsymbol{y}_{i,q,k}^{m}=h\left(\boldsymbol{s}_{i,q,k}^{m},i\right)+\boldsymbol{w}_{i,q,k}^{m},
\end{cases}\label{eq:3}
\end{equation}\normalsize
where $\boldsymbol{s}_{i,q,k}^{m}$ represents the state of target
$q$ to radar $i$ at the receive time $t_{i,q,k}^{m}$ with $m\in\left\{ 1,\ldots,M_{i,q,k}\right\} $, 
$f\left(\cdot\right)$ is the transition function with $f\left(\boldsymbol{s}_{i,q,k}^{m},t_{k+1}-t_{i,q,k}^{m}\right)=\boldsymbol{F}_{k}^{q}\boldsymbol{s}_{i,q,k}^{m}$, $\boldsymbol{F}_{k}^{q}=\left[\boldsymbol{I}_{2}\otimes\boldsymbol{T}\right]$ with %\par\noindent\small
%\begin{equation}
$\boldsymbol{T}=\left[\begin{array}{cc}
1 & t_{k+1}-t_{i,q,k}^{m}\\
0 & 1
\end{array}\right]$, 
%\end{equation}\normalsize
$\boldsymbol{\gamma}_{k}^{q}\sim\mathcal{N}\left(\boldsymbol{0},\boldsymbol{\Gamma}_{k}^{q}\right)$
represents the process noise, $\boldsymbol{w}_{i,q,k}^{m}\sim\mathcal{N}\left(\boldsymbol{0},\boldsymbol{\Sigma}_{i,q,k}^{m}\right)$
represents the measurement error, and 
\par\noindent\small
\begin{equation}
\footnotesize{h\left(\boldsymbol{s}_{i,q,k}^{m},i\right)=\left[\begin{array}{c}
R_{i,q,k}^{m}\\
\theta_{i,q,k}^{m}\\
\nu_{i,q,k}^{m}
\end{array}\right]=\left[\begin{array}{c}
\sqrt{\left(x_{i,q,k}^{m}-x_{i}\right)^{2}+\left(y_{i,q,k}^{m}-y_{i}\right)^{2}}\\
\arctan\left[\frac{y_{i,q,k}^{m}-y_{i}}{x_{i,q,k}^{m}-x_{i}}\right]\\
\frac{\left(x_{i,q,k}^{m}-x_{i}\right)\dot{x}_{t_{k+1}}^{q}+\left(y_{i,q,k}^{m}-y_{i}\right)\dot{y}_{i,q,k}^{m}}{\sqrt{\left(x_{i,q,k}^{m}-x_{i}\right)^{2}+\left(y_{i,q,k}^{m}-y_{i}\right)^{2}}}
\end{array}\right],}
\label{eq_eq}
\end{equation}\normalsize
%$h\left(\cdot\right)$ 
is the measurement function given by \eqref{eq_eq} with $R_{i,q,k}^{m}$, $\theta_{i,q,k}^{m}$ and $\nu_{i,q,k}^{m}$
being the range, AoA and velocity, respectively.

Define $\boldsymbol{\Sigma}_{i,q,k}^{m}=\text{diag}\left(\sigma_{R_{i,q,k}^{m}}^{2},\sigma_{\theta_{i,q,k}^{m}}^{2},\sigma_{\nu_{i,q,k}^{m}}^{2}\right),$
where $\sigma_{R_{i,q,k}^{m}}^{2}$, $\sigma_{\theta_{i,q,k}^{m}}^{2}$
and $\sigma_{\nu_{i,q,k}^{m}}^{2}$ are the lower bounds on the MSE
error of the corresponding parameters in the subscripts. As per \cite{skolnik1960theoretical,peebles2007radar},\par\noindent\small
\begin{equation}
\begin{cases}
\sigma_{R_{i,q,k}^{m}}^{2}= & \frac{\sum_{j=1}^{J}\alpha_{i,j}^{c}P_{c,k}^{j}+\sigma_{r,i}^{2}}{P_{i,q,k}^{m}T_{i,q,k}^{m}}\eta_{i,q,k}^{m}\zeta_{i}^{-2}c_{R},\\
\sigma_{\theta_{i,q,k}^{m}}^{2}= & \frac{\sum_{j=1}^{J}\alpha_{i,j}^{c}P_{c,k}^{j}+\sigma_{r,i}^{2}}{P_{i,q,k}^{m}T_{i,q,k}^{m}}\eta_{i,q,k}^{m}B_{i}^{2}c_{\theta},\\
\sigma_{\nu_{i,q,k}^{m}}^{2}= & \frac{\sum_{j=1}^{J}\alpha_{i,j}^{c}P_{c,k}^{j}+\sigma_{r,i}^{2}}{P_{i,q,k}^{m}T_{i,q,k}^{m}}\eta_{i,q,k}^{m}\zeta_{i}^{2}c_{\nu},
\end{cases}
\end{equation}\normalsize
where $T_{i,q,k}^{m}$ is the dwell time\footnote{The dwell time is defined as the product of number of pulses for each measurement and pulsewidth \cite{peebles2007radar}.} of radar $i$ on target $q$,
$\alpha_{i,j}^{c}$ accounts for the interference coefficient from the $j$-th
communications downlink to the $i$-th radar, $\sigma_{r,i}^{2}$ is
the variance of the noise at the receiver of radar $i$, $\eta_{i,q,k}^{m}$
is the  radar cross-section of the target $q$ to radar $i$ at the receive time
$t_{i,q,k}^{m}$, $\zeta_{i}$ and $B_{i}$ are the transmit signal
bandwidth and the 3dB receive beam width of radar $i$, respectively,
and $c_{R}$, $c_{\theta}$ and $c_{\nu}$ are unrelated constants.
Therefore, we rewrite %$\boldsymbol{\Sigma}_{i,q,k}^{m}$ as
\par\noindent\small 
\begin{equation}
\boldsymbol{\Sigma}_{i,q,k}^{m}=\frac{\sum_{j=1}^{J}\alpha_{i,j}^{c}P_{c,k}^{j}+\sigma_{r,i}^{2}}{P_{i,q,k}^{m}T_{i,q,k}^{m}}\boldsymbol{C}_{i,q,k}^{m},\label{eq:7}
\end{equation}\normalsize
where $\boldsymbol{C}_{i,q,k}^{m}$ is the matrix that contains the above-mentioned parameters.
%
%\vspace*{-0.1in}
\section{Problem Formulation}\label{sec:problem}
The HRCN configurations presented above indicate that the constituent
radars operate in an asynchronous manner. To circumvent the demanding
synchronization requirement, we adopt the concept of composition measure (CM), i.e., the CM estimate will serve as the input measurements to
the tracking filter \cite{klein2016tracking}. Thus, the estimation accuracy of the CM will determine the tracking performance. It can be gauged from \eqref{eq:3}-\eqref{eq:7} that the Cram\'{e}r-Rao bound  (CRB) of CM would depend on resources allocated to different entities. Further, the overall filter estimates also depend on these allocations. Here, contrary to \cite{yan2020optimal}, our HRCN performance involves the additional dimension of communications which influences resource allocation and impacts CRB derivations. %Based on the CM and its CRB, we formulate the resource allocation problem in this section. %
%\vspace*{-0.1in}
\subsection{Composition measure and Bayesian CRB}
%
%Towards addressing the asynchronous operation, 
A composition measure (CM) $\hat{\boldsymbol{s}}_{t_{k+1}}^{q}$ for each target $q$ at the fusion time $t_{k+1}$ is obtained based on all the measurements during the $k$-th fusion interval $\boldsymbol{y}_{k}^{q}$. The CM ${\hat{\boldsymbol{s}}_{t_{k+1}}^{q}}$ is an estimate of the true state $\boldsymbol{s}_{t_{k+1}}^{q}$ and shown to be statistically efficient even for small sampling sizes \cite{osborne2013statistical}. In particular, \par\noindent\small
\begin{align}
    \boldsymbol{y}_{k}^{q}&=\left[\left(\boldsymbol{y}_{1,q,k}^{1}\right)^{T},\ldots,\left(\boldsymbol{y}_{1,q,k}^{M_{1,q,k}}\right)^{T},\ldots,\left(\boldsymbol{y}_{N,q,k}^{M_{N,q,k}}\right)^{T}\right]^{T}, \label{eqn:All_meas}\\
%\end{equation}
%
% we  construct
%a composition measure (CM) for each target $q$ at the fusion time $t_{k+1}$ as,
%which will be used as the measurement in the Kalman filter. At the
%fusion time $t_{k+1}$, the CM for target $q$ is denoted by 
%
%\begin{equation}
    \hat{\boldsymbol{s}}_{t_{k+1}}^{q}&=\left[\widehat{x}_{t_{k+1}}^{q},\hat{\dot{x}}_{t_{k+1}}^{q},\hat{y}_{t_{k+1}}^{q},\hat{\dot{y}}_{t_{k+1}}^{q}\right]^{T}. \label{eqn:CM}
\end{align}\normalsize
%
%Further, based on the measurements during the $k$-th fusion interval,
%we define the stacked measurement vector $\boldsymbol{y}_{k}^{q}$ as 
%
%\begin{equation}
%\boldsymbol{y}_{k}^{q}=\left[\left(\boldsymbol{y}_{1,q,k}^{1}\right)^{T},\ldots,\left(\boldsymbol{y}_{1,q,k}^{M_{1,q,k}}\right)^{T},\ldots,\left(\boldsymbol{y}_{N,q,k}^{M_{N,q,k}}\right)^{T}\right]^{T}.
%\end{equation}
%
%Towards obtaining this CM, we first exploit the fact that
The measurements are independent and the probability density function\par\noindent\small
\begin{equation}
\label{eq:PDF}
%p\left(\boldsymbol{y}_{k}^{q}|\hat{\boldsymbol{s}}_{t_{k+1}}^{q}\right)
p\left(\boldsymbol{y}_{k}^{q}|{\boldsymbol{s}_{t_{k+1}}^{q}}\right)
=\prod_{i=1}^{N}\prod_{m}^{M_{i,q,k}}\mathcal{N}\left(h\left(\boldsymbol{s}_{i,q,k}^{m},i\right),\boldsymbol{\Sigma}_{i,q,k}^{m}\right),
\end{equation}\normalsize
where $\boldsymbol{s}_{i,q,k}^{m}$ is predicted based on $\boldsymbol{s}_{t_{k+1}}^{q}=f\left(\boldsymbol{s}_{i,q,k}^{m},t_{k+1}-t_{i,q,k}^{m}\right)$.
From the maximum likelihood (ML) estimate, the CM is %can be obtain by
\par\noindent\small
\begin{equation}
%\hat{\boldsymbol{s}}_{t_{k+1}}^{q}=\underset{\hat{\boldsymbol{s}}_{t_{k+1}}^{q}}{\arg\max}\left[\ln\left(p\left(\boldsymbol{y}_{k}^{q}|\hat{\boldsymbol{s}}_{t_{k+1}}^{q}\right)\right)\right].\label{eq:9}
%
\hat{\boldsymbol{s}}_{t_{k+1}}^{q}=\underset{{\boldsymbol{s}_{t_{k+1}}^{q}}}{\arg\max}\left[\ln\left(p\left(\boldsymbol{y}_{k}^{q}|{\boldsymbol{s}_{t_{k+1}}^{q}}\right)\right)\right].\label{eq:9}
\end{equation}\normalsize
This requires solving a nonlinear least-square problem and the iterated least squares algorithm \cite{bar2004estimation} is usually applied to obtain the ML estimate. Following \cite{yan2020optimal}, we approximate the CM covariance matrix by the CRB of $\boldsymbol{s}_{t_{k+1}}^{q}$ at ${\hat{\boldsymbol{s}}_{t_{k+1}}^{q}}$. The following Lemma~\ref{lem:CRB} states the Fisher information matrix (FIM) of the true state.
\begin{lem}\label{lem:CRB}
The FIM $\boldsymbol{J}_{\boldsymbol{y}_{k}^{q}}\left(\boldsymbol{s}_{t_{k+1}}^{q}\right)$ of the target state $\boldsymbol{s}_{t_{k+1}}^{q}$ is
\begin{equation}
\begin{aligned} & \boldsymbol{J}_{\boldsymbol{y}_{k}^{q}}\left(\boldsymbol{s}_{t_{k+1}}^{q}\right)\\
= & \sum_{i=1}^{N}\sum_{m}^{M_{i,q,k}}\frac{P_{i,q,k}^{m}T_{i,q,k}^{m}}{\sum_{j=1}^{J}\alpha_{i,j}^{c}P_{c,k}^{j}+\sigma_{r,i}^{2}}\boldsymbol{H}_{i,q,k}^{mT}\left(\boldsymbol{C}_{i,q,k}^{m}\right)^{-1}\boldsymbol{H}_{i,q,k}^{m},
\end{aligned}
\label{eq:FIM}
\end{equation}\normalsize
where $\boldsymbol{H}_{i,q,k}^{m}$ is the Jacobian of $h\left(\boldsymbol{s}_{i,q,k}^{m},i\right)$ on $\boldsymbol{s}_{t_{k+1}}^{q}$ with $\boldsymbol{s}_{t_{k+1}}^{q} = f\left(\boldsymbol{s}_{i,q,k}^{m},t_{k+1}-t_{i,q,k}^{m}\right)$. 
\end{lem}
\begin{IEEEproof}
    See Appendix \ref{Appendix-0}.
\end{IEEEproof}

The CM $\hat{\boldsymbol{s}}_{t_{k+1}}^{q}$ and its approximate covariance matrix
$\boldsymbol{J}_{\boldsymbol{y}_{k}^{q}}^{-1}\left(\hat{\boldsymbol{s}}_{t_{k+1}}^{q}\right)$
are the input to the tracking filter. We denote the filtered estimate
by $\tilde{\boldsymbol{s}}_{t_{k+1}}^{q}$, which is function of $\hat{\boldsymbol{s}}_{t_{k+1}}^{q}$.
The mean-squared error (MSE) of $\boldsymbol{s}_{t_{k+1}}^{q}$ satisfies
the inequality\par\noindent\small
\begin{equation}
\mathbb{E}_{\boldsymbol{y}_{k}^{q}}\left[\left(\tilde{\boldsymbol{s}}_{t_{k+1}}^{q}-\boldsymbol{s}_{t_{k+1}}^{q}\right)\left(\tilde{\boldsymbol{s}}_{t_{k+1}}^{q}-\boldsymbol{s}_{t_{k+1}}^{q}\right)^{T}\right]\succeq\left[\boldsymbol{B}\left(\boldsymbol{s}_{t_{k+1}}^{q}\right)\right]^{-1},
\end{equation}\normalsize
where $\left[\boldsymbol{B}\left(\boldsymbol{s}_{t_{k+1}}^{q}\right)\right]^{-1}$
is the Bayesian CRB matrix, and $\boldsymbol{B}\left(\boldsymbol{s}_{t_{k+1}}^{q}\right)$
is the corresponding FIM that is adapted from \cite{yan2014prior} by further incorporating the interference from communications system. These steps produce,\par\noindent\small
\begin{align}
\boldsymbol{B}\left(\boldsymbol{s}_{t_{k+1}}^{q}\right)&=\sum_{i=1}^{N}\sum_{m}^{M_{i,q,k}}\frac{P_{i,q,k}^{m}T_{i,q,k}^{m}}{\sum_{j=1}^{J}\alpha_{i,j}^{c}P_{c,k}^{j}+\sigma_{r,i}^{2}}\hat{\boldsymbol{C}}_{i,q,k}^{m}
+\tilde{\boldsymbol{\Gamma}}_{k}^{q},\\
%\end{equation}
%where
%\begin{eqnarray}
    \hat{\boldsymbol{C}}_{i,q,k}^{m}&=\hat{\boldsymbol{H}}_{i,q,k}^{mT}\left(\boldsymbol{C}_{i,q,k}^{m}\right)^{-1}\hat{\boldsymbol{H}}_{i,q,k}^{m}, \\
%\end{equation}
%\begin{equation}
\tilde{\boldsymbol{\Gamma}}_{k}^{q}&=\left[\boldsymbol{\Gamma}_{k}^{q}+\boldsymbol{F}_{k}^{q}\boldsymbol{B}^{-1}\left(\boldsymbol{s}_{t_{k}}^{q}\right)\boldsymbol{F}_{k}^{qT}\right]^{-1},
\end{align}\normalsize
where $\hat{\boldsymbol{H}}_{i,q,k}^{m}$ is the Jacobian of the measurement at $\bar{\boldsymbol{s}}_{t_{k+1|k}}^{q}$ that is the predicted estimate of the true target state. 
\subsection{Resource allocation}
To avoid frequent reconfigurations of the system, the allocated power and dwell time from radar $i$ on for target $q$ are constant over multiple measurements of a fusion interval, i.e.,\par\noindent\small
\begin{equation}
\begin{cases}
P_{i,q,k}=P_{i,q,k}^{1}=\dots=P_{i,q,k}^{M_{i,q,k}}, & \forall q=1,\ldots,Q\\
T_{i,q,k}=T_{i,q,k}^{1}=\dots=T_{i,q,k}^{M_{i,q,k}}, & \forall q=1,\ldots,Q.
\end{cases}
\end{equation}\normalsize
Recall that the subchannel selection vectors of $j$-th user %, $\boldsymbol{f}_{j,k}^{c}$, 
and $i$-th radar  %$\boldsymbol{f}_{i,k}^{r}$ 
are, respectively,\par\noindent\small
\begin{eqnarray}
%&\hspace*{-0.1in}
\boldsymbol{f}_{j,k}^{c}&=\left[0,\ldots,0,1,\ldots,1,0,\ldots,0\right]^{T}\in\mathbb{R}^{F\times1},\;\forall j, \\
%\end{equation}
%and
%\begin{equation}
%&\hspace*{-0.1in}
\boldsymbol{f}_{i,k}^{r}&=\left[0,\ldots,0,1,\ldots,1,0,\ldots,0\right]^{T}\in\mathbb{R}^{F\times1},\;\forall i,
%\end{equation}
\end{eqnarray}\normalsize
where $\boldsymbol{f}_{i,k}^{r}$ is predefined and fixed over fusion
intervals. The vectors $\left\{ \boldsymbol{f}_{j,k}^{c}\right\} $
satisfy the following constraints,\par\noindent\small
\begin{equation}
\mathcal{F}_{c}^{k}=\begin{cases}
\boldsymbol{1}_{F}^{T}\boldsymbol{f}_{j,k}^{c}=F_{min}^{c}\le F\\
\left(\boldsymbol{f}_{j,k}^{c}\right)^{T}\boldsymbol{f}_{\hat{j},k}^{c}=0,\forall j\neq\hat{j}\\
\left\Vert{f}_{j,k}^{c}\right\Vert _{0}=1, \forall j,
\end{cases}
\label{eq:F_constraints}
\end{equation}\normalsize

Based on the above definitions, the bandwidth overlap of the $i$-th
radar and the $j$-th user is simply $\left(\boldsymbol{f}_{i,k}^{r}\right)^{T}\boldsymbol{f}_{j,k}^{c},\forall i=1,\ldots N,j=1,\ldots,J$.
This yields % the parameters accounting for the mutual interference between the radars and the communications links can be expressed as
\par\noindent\small
\begin{equation}
\begin{cases}
\alpha_{i,j}^{c}=\tilde{\alpha}_{i,j}^{c}\left(\boldsymbol{f}_{i,k}^{r}\right)^{T}\boldsymbol{f}_{j,k}^{c}\\
\alpha_{j,i}^{r}=\tilde{\alpha}_{j,i}^{r}\left(\boldsymbol{f}_{i,k}^{r}\right)^{T}\boldsymbol{f}_{j,k}^{c}.
\end{cases}
\end{equation}\normalsize

%Using the aforementioned notations, 
The FIM of the CMs becomes %can be further expressed as
\par\noindent\small
\begin{equation}
   \boldsymbol{B}\left(\boldsymbol{s}_{t_{k+1}}^{q}\right)=
   \sum_{i=1}^{N}\frac{P_{i,q,k}T_{i,q,k}}{\sum_{j=1}^{J}\tilde{\alpha}_{i,j}^{c}\left(\boldsymbol{f}_{i,k}^{r}\right)^{T}\boldsymbol{f}_{j,k}^{c}P_{c,k}^{j}+\sigma_{r,i}^{2}}\tilde{\boldsymbol{C}}_{i,q,k}+\tilde{\boldsymbol{\Gamma}}_{k}^{q},
\label{eq:B_definition}
\end{equation}\normalsize
where $\tilde{\boldsymbol{C}}_{i,q,k}=\sum_{m}^{M_{i,q,k}}\hat{\boldsymbol{C}}_{i,q,k}^{m}$. Hereafter, we use  $\boldsymbol{B}\left(\boldsymbol{s}_{t_{k+1}}^{q}\right)$ and $\boldsymbol{B}_{q}$ interchangeably.
%The corresponding CRB matrix is thereby $\left[\boldsymbol{B}\left(\boldsymbol{s}_{t_{k+1}}^{q}\right)\right]^{-1}$.
Note that the diagonal elements of $\left[\boldsymbol{B}\left(\boldsymbol{s}_{t_{k+1}}^{q}\right)\right]^{-1}$
are not homogeneous because of different units for distance and
velocity. For fair optimization, we normalize them to $\boldsymbol{\Lambda}\boldsymbol{B}^{-1}\left(\boldsymbol{s}_{t_{k+1}}^{q}\right)\boldsymbol{\Lambda}^{T}$, 
where %$\boldsymbol{\Lambda}$ is the normalized matrix given by
%\par\noindent\small
%\begin{equation}
$\boldsymbol{\Lambda}=\text{diag}\left(\boldsymbol{I}_{2}\otimes\left[\begin{array}{cc}
1 & 0\\
0 & T_{0}
\end{array}\right]\right)$ 
%\end{equation}\normalsize
%
is the normalized matrix.

%Towards considering a scalar objective representative of the performance of all involved targets, we 
Define the following metric of estimation accuracy \par\noindent\small
\begin{equation}
g\left(\left\{ P_{c,k}^{j}\right\} ,\left\{ P_{i,q,k}\right\} ,\left\{ T_{i,q,k}\right\} ,\left\{ \boldsymbol{f}_{j,k}^{c}\right\} \right)=\sum_{q=1}^{Q}\frac{1}{\text{Tr}\left(\boldsymbol{\Lambda}\boldsymbol{B}^{-1}_{q}\boldsymbol{\Lambda}^{T}\right)},
\end{equation}\normalsize
where $\text{Tr}\left(\boldsymbol{\Lambda}\boldsymbol{B}^{-1}\left(\boldsymbol{s}_{t_{k+1}}^{q}\right)\boldsymbol{\Lambda}^{T}\right)$
accounts for the total MSE error of $\boldsymbol{s}_{t_{k+1}}^{q}$. Therefore, maximizing this metric will lead to reduction in the MSE for each target estimate. 

Based on \eqref{eq:SINR_macro}, the $j$-th macro throughput, $\forall j=1,\ldots,J$, is %\par\noindent\small
%\vspace{-2mm}
%\begin{equation}
%\footnotesize{\begin{aligned} &
$r\left(\left\{P_{c,k}^{j}\right\} ,\left\{ P_{i,q,k}\right\} ,\left\{ T_{i,q,k}\right\} ,\left\{ \boldsymbol{f}_{j,k}^{c}\right\} \right)
=  \text{\ensuremath{\log}}\left(1+\frac{\beta^{j}P_{c,k}^{j}T_{0}}{\sum_{i=1}^{N}\sum_{q=1}^{Q}M_{i,q,k}\tilde{\alpha}_{j,i}^{r}P_{i,q,k}T_{i,q,k}\boldsymbol{f}_{i,k}^{rT}\boldsymbol{f}_{j,k}^{c}+\sigma_{c,j}^{2}T_{0}}\right)$. 
%\end{aligned}}
%\vspace{2mm}
%\end{equation}\normalsize
Similarly, the downlink throughput expression for the $l$-th micro users, $\forall l=1,\ldots,L$, is based on \eqref{eq:SINR_micro}. 

It is natural to consider both macro and micro throughput in resource allocation for HRCN system. However, this results in a centralized solution to the multi-tier cellular network, which is computationally expensive because of the high cross-tier signaling overhead \cite{hasan2015distributed}. Therefore, assuming that the microcell resource is allocated independently within the micro BS, we consider the joint allocation over the radars and the macro BS of the multi-tier cellular network. Without loss of generality, our proposed algorithm is also applicable when the micro user throughput is included with appropriate constraints because of the linearity of interference terms in the denominator of \eqref{eq:SINR_micro}. This leads to a two-stage optimization procedure, which could be one-shot or used in a loop; we omit this here and focus on only macroscopic problem.

Define the optimization variables %$\boldsymbol{z}_{k}$ and $\boldsymbol{F}_{k}^{c}$ as
\par\noindent\small
%
%\begin{equation}
\begin{align}
%\vspace*{-0.1in}
%&&\begin{aligned}
\boldsymbol{z}_{k}= & \left[P_{1,1,k},\ldots,P_{N_{c},Q,k},T_{N_{c}+1,1,k},\ldots,T_{N_{c}+N_{p},Q,k},\right.\\
 & \left.P_{c,k}^{1},\ldots,P_{c,k}^{J}\right]^{T}\\
%\end{aligned} \\
%\end{equation}
%and
%\begin{equation}
\boldsymbol{F}_{k}^{c}&=\left[\boldsymbol{f}_{1,k}^{c},\ldots,\boldsymbol{f}_{J,k}^{c}\right],
%\end{equation}
\end{align}\normalsize
The resource allocation requires solving the optimization %problem is formulated as 
\par\noindent\small
\begin{equation}
\begin{aligned} & \underset{\boldsymbol{z}_{k},\boldsymbol{F}_{k}^{c}}{\text{maximize}} &  & g\left(\boldsymbol{z}_{k},\boldsymbol{F}_{k}^{c}\right)\\
 & \text{subject to} &  & r\left(\boldsymbol{z}_{k},\boldsymbol{F}_{k}^{c}\right)\ge\epsilon_{k}^{j},\forall j\\
 &  &  & \sum_{q=1}^{Q}M_{i,q,k}P_{i,q,k}\le P_{total}^{i},\forall i\in\varphi_{c}\\
 &  &  & \sum_{q=1}^{Q}M_{i,q,k}T_{i,q,k}\le T_{total}^{i},\forall i\in\varphi_{p}\\
 &  &  & \sum_{j=1}^{J}P_{c,k}^{j}\le P_{total}^{c}\\
 &  &  & \boldsymbol{f}_{j,k}^{c}\in\mathcal{F}_{c}^{k},\forall j.
\end{aligned}
\label{eq:RA opt problem}
\end{equation}\normalsize
%Before addressing the problem, we have following observations:
%\begin{itemize}
%\item 

In \eqref{eq:RA opt problem}, the first constraint represents the throughput requirements on the $J$ macro downlinks. The second and third constraints, respectively, account for radar's power and dwell time budgets. The last two constraints represent the budgets of communications power and bandwidth, respectively. This is a general formulation that subsumes the problem in \cite{yan2020optimal} as a special case when only radar's power and dwell time are considered. For the sake of simplicity, we considered only the macro users. However, the problem can be easily extended to micro users and, importantly, our proposed ANCHOR algorithm is still applicable. Overall, the problem is highly challenging because the objective function as well as some constraints are non-convex; variables $\boldsymbol{z}_{k}$ and $\boldsymbol{F}_{k}^{c}$ are coupled; and constraints on $\boldsymbol{F}_{k}^{c}$ imply a mixed-integer programming. % will be involved.

\section{Optimization Method for Resource Allocation}\label{sec:algo}
We now develop an algorithm to solve \eqref{eq:RA opt problem} of the $k$-th fusion interval and, for notational simplicity, we omit the subscript $k$ hereafter. % unless specified otherwise. 
We employ the alternating optimization to decouple the optimization for $\boldsymbol{z}$ and $\boldsymbol{F}^{c}$; this corresponds to an alternating update of the binary frequency allocation and the continuous power and dwell time allocation.

\subsection{Frequency allocation}% via global optimal search}
For a fixed $\boldsymbol{z}_{\ell}$ at the $\ell$-th iteration, the subproblem of optimizing $\boldsymbol{F}^{c}$ in problem \eqref{eq:RA opt problem} is \par\noindent\small
\begin{equation}
    \begin{aligned} & \underset{\boldsymbol{F}^{c}}{\text{maximize}} &  & \sum_{q=1}^{Q}\frac{1}{\text{Tr}\left(\boldsymbol{\Lambda}\boldsymbol{B}_{q}^{-1}\boldsymbol{\Lambda}^{T}\right)}\\
     & \text{subject to} &  & \boldsymbol{a}_{j}^{T}\boldsymbol{f}_{j}^{c}\le\tilde{\epsilon}^{j},\forall j\\
     &  &  & \boldsymbol{f}_{j}^{c}\in\mathcal{F}_{c},\forall j=1,\ldots,J
    \end{aligned}
    \label{eq:subprob_F}
\end{equation}\normalsize
with \par\noindent\small
\begin{equation}
    \begin{cases}
        \boldsymbol{B}_{q}=\sum_{i=1}^{N}\frac{\tilde{\boldsymbol{C}}_{i,q,k}}{\sum_{j=1}^{J}\boldsymbol{b}_{i,j}^{T}\boldsymbol{f}_{j}^{c}+\tilde{\sigma}_{r,i}^{2}}+\tilde{\boldsymbol{\Gamma}}_{k}^{q}\\
        \boldsymbol{b}_{i,j}=\frac{\tilde{\alpha}_{i,j}^{c}P_{c,k}^{j}}{P_{i,q,k}T_{i,q,k}}\boldsymbol{f}_{i,k}^{r}\\
        \boldsymbol{a}_{j}=\sum_{i=1}^{N}\tilde{\alpha}_{j,i}^{r}\left(\sum_{q=1}^{Q}M_{i,q}P_{i,q}T_{i,q}\right)\boldsymbol{f}_{i}^{r}\\
        \tilde{\sigma}_{r,i}^{2}=\frac{\sigma_{r,i}^{2}}{P_{i,q,k}T_{i,q,k}}\\
        \tilde{\epsilon}^{j}=\frac{\left(\beta^{j}P_{c,k}^{j}-\sigma_{c,j}^{2}\left(e^{\epsilon^{j}}-1\right)\right)T_{0}}{e^{\epsilon^{j}}-1}.
    \end{cases}
    \label{eq:constraint_subprob_F}
\end{equation}\normalsize

Considering the special structure of the constraint set $\mathcal{F}_{c}$ specified in \eqref{eq:F_constraints}, define  %\par\noindent\small
%\begin{equation}
    $\boldsymbol{f}_{j}^{c}=\boldsymbol{s}_{j}\varotimes\boldsymbol{1}_{F_{min}^{c}},\forall j=1,\ldots J$,
%\end{equation}\normalsize
where $\boldsymbol{s}_{j}$ is a vector of length  $N_{f}=\frac{F}{F_{min}^{c}}$ (clearly, $N_{f}$ is an integer). Thus, problem \eqref{eq:subprob_F} is equivalently recast into\par\noindent\small
\begin{equation}
    \begin{aligned}  
    & \underset{\left\{ \boldsymbol{s}_{j}\right\} }{\text{maximize}} &  & \sum_{q=1}^{Q}\frac{1}{\text{Tr}\left(\boldsymbol{\Lambda}\boldsymbol{B}_{q}^{-1}\boldsymbol{\Lambda}^{T}\right)}\\
     & \text{subject to} &  & \tilde{\boldsymbol{a}}_{j}^{T}\boldsymbol{s}_{j}\le\tilde{\epsilon}^{j},\forall j\\
     &  &  & \boldsymbol{s}_{j}\in\left\{ 0,1\right\} ^{N_{f}\times1}\\
     &  &  & \boldsymbol{1}^{T}\boldsymbol{s}_{j}=1,\forall j\\
     &  &  & \boldsymbol{s}_{j}^{T}\boldsymbol{s}_{\hat{j}}=0,\forall j\neq\hat{j}\in\{1,\ldots,J\},
    \end{aligned}
    \label{eq:prob_NCMIP}
\end{equation}\normalsize
where $\boldsymbol{B}_{q}$ in \eqref{eq:constraint_subprob_F} is recast as $\boldsymbol{B}_{q}=\sum_{i=1}^{N}\frac{\tilde{\boldsymbol{C}}_{i,q,k}}{\sum_{j=1}^{J}\tilde{\boldsymbol{b}}_{i,j}^{T}\boldsymbol{s}_{j}+\tilde{\sigma}_{r,i}^{2}}+\tilde{\boldsymbol{\Gamma}}_{k}^{q}$, $\tilde{\boldsymbol{b}}_{i,j}=\left(\boldsymbol{I}_{N_{f}}\varotimes\boldsymbol{1}_{F_{min}^{c}}^{T}\right)\boldsymbol{b}_{i,j}$, and $\tilde{\boldsymbol{a}}_{j}=\left(\boldsymbol{I}_{N_{f}}\varotimes\boldsymbol{1}_{F_{min}^{c}}^{T}\right)\boldsymbol{a}_{j}$. %Once problem \eqref{eq:prob_NCMIP} is solved with 
Denote the optimal solution of \eqref{eq:subprob_F} and \eqref{eq:prob_NCMIP} by $\left(\boldsymbol{f}_{j}^{c}\right)^{\star}=\boldsymbol{s}_{j}^{\star}\varotimes\boldsymbol{1}_{F_{min}^{c}}$ and $\boldsymbol{s}_{j}^{\star}$, respectively. 

%It is worth mentioning that 
The problem in \eqref{eq:prob_NCMIP} can be interpreted as an assignment problem with a nonlinear objective function. The constraints in \eqref{eq:prob_NCMIP} indicate that the problem is to assign an exclusive location to the only non-zero element, i.e. 1, for each $\boldsymbol{s}_{j}$ so that the objective value is maximized. For the classical assignment problem, Hungarian method~\cite{kuhn1955hungarian} is a classical approach that yields the global optimal solution with a cubic complexity. However, it can only solve for linear objective function. Although many subsequent works extend it to the nonlinear cases, most of them are heuristic and only available to bilinear, quadratic, biquadratic, and cubic objectives \cite{pardalos2013nonlinear}. The objective function of problem \eqref{eq:prob_NCMIP} does not belong to any of the above-mentioned classes. Consequently, prior approaches to the nonlinear assignment problems are inapplicable here. 

Since the number of feasible assignments is finite, an exhaustive search over the whole feasible space can always yield the optimal solution in polynomial time. However, for large $N_{f}$, this approach suffers from heavy computations. Here, simulated annealing method offers a trade-off between computational load and optimality while solving \eqref{eq:prob_NCMIP}. In traditional simulated annealing, computational time and complexity are dictated by the exploration/exploitation mechanism and the termination criteria potentially leading to global optimum. However, in the current setting, fewer and faster computations are preferred and hence, in the sequel, we adapt the classical simulated annealing to account for these constraints.

\paragraph{State Transition}
Note that in problem \eqref{eq:prob_NCMIP}, each $\boldsymbol{s}_{j}$ should satisfy the constraints\par\noindent\small
\begin{equation}
    \begin{cases}
    \tilde{\boldsymbol{a}}_{j}^{T}\boldsymbol{s}_{j}\le\tilde{\epsilon}^{j},\forall j\\
    \boldsymbol{s}_{j}\in\left\{ 0,1\right\} ^{N_{f}\times1}.
    \end{cases}
\end{equation}\normalsize
The corresponding $\hat{\boldsymbol{s}}_{j}\in\mathbb{R}^{N_{f}\times1},\forall j=1,\ldots,J$ is %constructed according to
\par\noindent\small
\begin{equation}
    \left[\hat{\boldsymbol{s}}_{j}\right]_{n}=\begin{cases}
    0, & \left[\tilde{\boldsymbol{a}}_{j}\right]_{n}>\tilde{\epsilon}^{j}\\
    1, & \left[\tilde{\boldsymbol{a}}_{j}\right]_{n}\le\tilde{\epsilon}^{j},
    \end{cases}
    \label{eq:49}
\end{equation}\normalsize
where $\left[\cdot\right]_{n}$ represents the $n$-th element of a vector. Thus, a feasible $\boldsymbol{s}_{j}$ is generated by selecting one nonzero position from $\hat{\boldsymbol{s}}_{j}$. Through this generation approach, the original searching space is reduced. %ompressed to a smaller region. 
Accordingly, we propose state transform (i.e. moving from the current $\left\{ \boldsymbol{s}_{j}\right\}$ to a neighbor) in Algorithm \ref{alg:SA_trans}. This problem-specific state transition enables faster neighbourhood discovery within the feasible set.%, potentially leading to enhanced performance.

\begin{algorithm}[H]		 	
	\caption{Scheme of state transition}	 	
	\label{alg:SA_trans}
\begin{algorithmic}[1]	 			 		
		\Require $\mathcal{J}=\left\{ 1,\ldots,J\right\}$, $\left\{ \boldsymbol{s}_{j}\right\} $, $\left\{ \hat{\boldsymbol{s}}_{j}\right\} $
		\Ensure $\left\{ \boldsymbol{s}_{j}\right\} $
		\State{Randomly select $\hat{j}\in\mathcal{J}$}	
		\State{$\mathcal{N}\leftarrow\left\{ n|\left[\hat{\boldsymbol{s}}_{\hat{j}}\right]_{n}>0\right\} $}
		\While{$\mathcal{N}\ne\emptyset$}
			\State{Randomly select $\hat{n}\in\mathcal{N}$} as the nonzero location of $\boldsymbol{s}_{\hat{j}}$
			\State{$\mathcal{N}\leftarrow\mathcal{N\backslash}\hat{n}$}
			\If{$\boldsymbol{s}_{\hat{j}}^{T}\boldsymbol{s}_{j}=0,\forall j\in\mathcal{J}\backslash\hat{j}$ }
				\State{Exit}
			\EndIf
		\EndWhile
		\State{Define $\bar{j}$ be the index associated with $\boldsymbol{s}_{\hat{j}}^{T}\boldsymbol{s}_{\bar{j}}\neq0$}
		\State{$\hat{j}\leftarrow\bar{j}$ and go to Line 3}
\end{algorithmic}
\end{algorithm}

\paragraph{%Easing the 
Optimality requirement}
Denote the state transition in Algorithm \ref{alg:SA_trans} by $\mathcal{H}\left(\cdot\right)$ and combine it with the simulated annealing algorithm summarized in Algorithm \ref{alg:Alg_subprob_F}. Note that the simulated annealing , as a global optimization approach, has the theoretical guarantee of convergence in the probabilistic sense \cite{henderson2003theory}. %It has to be admitted that the convergence to the global optimum is in the sense of probability, and thereby 
However, the convergence cannot be guaranteed for a specific run/instance. Algorithm \ref{alg:Alg_subprob_F} is a nested component of the alternating optimization framework. So, it is viable to terminate the iterations of the algorithm as long as the objective function of problem \eqref{eq:subprob_F} improves over the previous iteration. This mitigates the demanding requirement on the simulated-annealing-based algorithm to find a global optimum while offering the designer a flexibility in the selection of the termination criterion.
\begin{algorithm}[H]		 	
	\caption{Global optimal search for frequency allocation}
	\label{alg:Alg_subprob_F}
    \begin{algorithmic}[1]	 			 		
		\Require $\left\{ \hat{\boldsymbol{b}}_{j}\right\},\left\{\delta_{j}\right\},\left\{\hat{\boldsymbol{a}}_{i,j}\right\} ,T_\text{max},T_\text{min},\delta T$
		\Ensure $\left\{ \boldsymbol{s}_{j}\right\} $  
		\State{$T\leftarrow T_\text{max}$}	
		\While{$T \ge T_\text{min}$}
            \State{$\boldsymbol{B}_{q}\leftarrow\sum_{i=1}^{N}\frac{\tilde{\boldsymbol{C}}_{i,q,k}}{\sum_{j=1}^{J}\tilde{\boldsymbol{b}}_{i,j}^{T}\boldsymbol{s}^{'}_{j}+\tilde{\sigma}_{r,i}^{2}}+\tilde{\boldsymbol{\Gamma}}_{k}^{q}$}
            \State{$\text{OBJ}_{1}\leftarrow\sum_{q=1}^{Q}\frac{1}{\text{Tr}\left(\boldsymbol{\Lambda}\boldsymbol{B}_{q}^{-1}\boldsymbol{\Lambda}^{T}\right)}$}
            \State{$\left\{\boldsymbol{s}_{j}^{\prime}\right\}\leftarrow\mathcal{H}\left(\left\{ \boldsymbol{s}_{j}\right\} \right)$}
            \State{$\boldsymbol{B}_{q}\leftarrow\sum_{i=1}^{N}\frac{\tilde{\boldsymbol{C}}_{i,q,k}}{\sum_{j=1}^{J}\tilde{\boldsymbol{b}}_{i,j}^{T}\boldsymbol{s}^{'}_{j}+\tilde{\sigma}_{r,i}^{2}}+\tilde{\boldsymbol{\Gamma}}_{k}^{q}$}
            \State{$\text{OBJ}_{2}\leftarrow\sum_{q=1}^{Q}\frac{1}{\text{Tr}\left(\boldsymbol{\Lambda}\boldsymbol{B}_{q}^{-1}\boldsymbol{\Lambda}^{T}\right)}$}
            \State{$\Delta\leftarrow\text{OBJ}_{2}-\text{OBJ}_{1}$}
            \If{$\Delta>0$}
              \State{$\left\{ \boldsymbol{s}_{j}\right\}\leftarrow\left\{ \boldsymbol{s}_{j}^{\prime}\right\} $}
            \ElsIf{$e^{\frac{\Delta}{T}}>\text{rand}\left(0,1\right)$}   
            	\State{$\left\{ \boldsymbol{s}_{j}\right\}\leftarrow\left\{ \boldsymbol{s}_{j}^{\prime}\right\} $}
            \EndIf
            \State{$T \leftarrow T-\delta T$}
        \EndWhile
    \end{algorithmic}
\end{algorithm}

\subsection{Power and time allocation}
To allocate power and time-slots, we adopt the alternating ascent-descent method.
\subsubsection{Maximin reformulation}
For a fixed $\boldsymbol{F}_{\ell+1}^{c}$, the subproblem of optimizing $\boldsymbol{z}$ is \par\noindent\small
\begin{equation}
    \begin{aligned} & \underset{\boldsymbol{z}}{\text{maximize}} &  & g\left(\boldsymbol{z},\boldsymbol{F}_{\ell+1}^{c}\right)\\
     & \text{subject to} &  & r\left(\boldsymbol{z},\boldsymbol{F}_{\ell+1}^{c}\right)\ge\epsilon^{j},\forall j;\; \sum_{q=1}^{Q}M_{i,q}P_{i,q}\le P_{total}^{i},\forall i\in\varphi_{c}\\ 
     & & &\sum_{q=1}^{Q}M_{i,q}T_{i,q}\le T_{total}^{i},\forall i\in\varphi_{p}; \; \sum_{j=1}^{J}P_{c}^{j}\le P_{total}^{c}.
    \end{aligned}
\label{eq:subpro_z}
\end{equation}\normalsize
The following Lemma~\ref{lemma1} casts \eqref{eq:subpro_z} to a more convenient form. 
\begin{lem}\label{lemma1}
Denote a set of slack variables by $\left\{ \boldsymbol{V}_{q}\right\} _{q=1}^{Q}$. The problem \eqref{eq:RA opt problem} is equivalent to \par\noindent\small
\begin{equation}
    \begin{alignedat}{2} & \begin{aligned}\underset{\boldsymbol{z}}{\mathrm{max}} & \underset{\left\{ \boldsymbol{V}_{q}\right\} }{\mathrm{min}}\end{aligned}
     &  & \sum_{q=1}^{Q}\mathrm{Tr}\left(\boldsymbol{V}_{q}^{T}\tilde{\boldsymbol{\Lambda}}^{T}\boldsymbol{B}_{q}\tilde{\boldsymbol{\Lambda}}\boldsymbol{V}_{q}\right)\\
     & \mathrm{subject}\:\mathrm{to} &  & \mathrm{Tr}\left(\boldsymbol{V}_{q}\right)=1,\forall q\\
     &  &  & r\left(\boldsymbol{z},\boldsymbol{F}_{\ell+1}^{c}\right)\ge\epsilon^{j},\forall j; \; \sum_{q=1}^{Q}M_{i,q}P_{i,q}\le P_{total}^{i},\forall i\in\varphi_{c}\\
     &  &  & \sum_{q=1}^{Q}M_{i,q}T_{i,q}\le T_{total}^{i},\forall i\in\varphi_{p}; \; \sum_{j=1}^{J}P_{c}^{j}\le P_{total}^{c},
    \end{alignedat}
    \label{eq:subpro_z_maximin}
\end{equation}\normalsize
where $\tilde{\boldsymbol{\Lambda}}=\boldsymbol{\Lambda}^{-1}$.
\end{lem}
\begin{IEEEproof}
	See Appendix \ref{Appendix-A}.
\end{IEEEproof}

Clearly, through this reformulation, we get rid of the inverse operation on $\boldsymbol{B}_{q}$. %We now solve the maximin problem in \eqref{eq:subpro_z_maximin}.

\subsubsection{Alternating descent-ascent method}
For problem \eqref{eq:subpro_z_maximin}, we resort to the alternating ascent-decent method, which has been widely applied to solve maximin problems\footnote{For a minimax problem, the counterpart is usually called alternating descent-ascent method.} and has several variants or extensions \cite{nouiehed2019solving,lu2020hybrid}. The convergence of these methods can be guaranteed under some mild conditions. For more details, we refer the interested reader to \cite{razaviyayn2013unified,razaviyayn2020nonconvex}. 

At the $n$-th iteration of the alternating ascent-descent method, given a fixed $\boldsymbol{z}^{n}$, the inner minimization problem of $\left\{\boldsymbol{V}_{q}\right\}$ has the closed form solution\par\noindent\small
\begin{equation}
    \boldsymbol{V}_{q}^{n+1}=\frac{\left(\tilde{\boldsymbol{\Lambda}}^{T}\boldsymbol{B}_{q}^{\ell,n}\tilde{\boldsymbol{\Lambda}}\right)^{-1}}{\text{Tr}\left[\left(\tilde{\boldsymbol{\Lambda}}^{T}\boldsymbol{B}_{q}^{\ell,n}\tilde{\boldsymbol{\Lambda}}\right)^{-1}\right]},\;\forall q=1,\ldots,Q,
\end{equation}\normalsize
where $\boldsymbol{B}_{q}^{\ell,n}$ is obtained by substituting  $\boldsymbol{F}_{\ell+1}^{c}$ and $\boldsymbol{z}^{n}$ into \eqref{eq:B_definition}.

For the fixed $\left\{\boldsymbol{V}_{q}^{n+1}\right\}$, we have\par\noindent\small
\begin{equation}
    \begin{aligned} & \sum_{q=1}^{Q}\text{Tr}\left[\left(\tilde{\boldsymbol{\Lambda}}\boldsymbol{V}_{\ell+1}^{q}\right)^{T}\boldsymbol{B}_{q}\left(\tilde{\boldsymbol{\Lambda}}\boldsymbol{V}_{\ell+1}^{q}\right)\right]\\
    = & \sum_{i=1}^{N}\frac{\sum_{q=1}^{Q}\omega_{i,q}^{n}P_{i,q}T_{i,q}}{\left(\boldsymbol{f}_{i}^{r}\right)^{T}\left(\sum_{j=1}^{J}\tilde{\alpha}_{i,j}^{c}\boldsymbol{f}_{j,\ell}^{c}P_{c}^{j}\right)+\sigma_{r,i}^{2}}+\textrm{constant},
    \end{aligned}
\end{equation}\normalsize
where $\omega_{i,q}^{n}=\text{Tr}\left[\left(\tilde{\boldsymbol{\Lambda}}\boldsymbol{V}_{q}^{n+1}\right)^{T}\tilde{\boldsymbol{C}}_{i,q}\left(\tilde{\boldsymbol{\Lambda}}\boldsymbol{V}_{q}^{n+1}\right)\right]\ge0$, and %$const.$ is a 
the constant term is unrelated to $\boldsymbol{z}$. Thus, the maximization problem of $\boldsymbol{z}$ becomes\par\noindent\small
\begin{equation}
    \begin{aligned} & \underset{\boldsymbol{z}}{\text{maximize}} &  & \sum_{i=1}^{N}\frac{\sum_{q=1}^{Q}\omega_{i,q}^{n}P_{i,q}T_{i,q}}{\left(\boldsymbol{f}_{i}^{r}\right)^{T}\left(\sum_{j=1}^{J}\tilde{\alpha}_{i,j}^{c}\boldsymbol{f}_{j,\ell+1}^{c}P_{c}^{j}\right)+\sigma_{r,i}^{2}}\\
     & \text{subject to} &  & r\left(\boldsymbol{z},\boldsymbol{F}_{\ell+1}^{c}\right)\ge\epsilon^{j},\forall j=1,\ldots,J\\
     &  &  & \sum_{q=1}^{Q}M_{i,q}P_{i,q}\le P_{total}^{i},\forall i\in\varphi_{c}\\
     &  &  & \sum_{q=1}^{Q}M_{i,q}T_{i,q}\le T_{total}^{i},\forall i\in\varphi_{p}\\
     &  &  & \sum_{j=1}^{J}P_{c}^{j}\le P_{total}^{c}.
    \end{aligned}
    \label{eq:subpro_z_outer_max}
\end{equation}\normalsize
Rewrite the objective function in \eqref{eq:subpro_z_outer_max}  as\par\noindent\small
\begin{equation}
    \sum_{i=1}^{N}\frac{\sum_{q=1}^{Q}\omega_{i,q}^{n}P_{i,q}T_{i,q}}{\left(\boldsymbol{f}_{i}^{r}\right)^{T}\left(\sum_{j=1}^{J}\tilde{\alpha}_{i,j}^{c}\boldsymbol{f}_{j,\ell+1}^{c}P_{c}^{j}\right)+\sigma_{r,i}^{2}}=\sum_{i=1}^{N}\frac{\boldsymbol{c}_{i}^{T}\boldsymbol{z}_{k}+d_{i}}{\boldsymbol{e}_{i}^{T}\boldsymbol{z}_{k}+\sigma_{r,i}^{2}},
\end{equation}\normalsize
where\par\noindent\small
\begin{equation}
    \boldsymbol{c}_{i}=\begin{cases}
    [\underset{\left(i-1\right)Q}{\underbrace{0,\ldots,0}},\omega_{i,1}^{n}T_{i,1},\ldots,\omega_{i,Q}^{n}T_{i,Q},\underset{\left(N_{c}+N_{p}-i\right)Q+J}{\underbrace{0,\ldots,0}}], & i\in\varphi_{c}\\{}
    [\underset{\left(i-1\right)Q}{\underbrace{0,\ldots,0}},\omega_{i,1}^{n}P_{i,1},\ldots,\omega_{i,Q}^{n}P_{i,Q},\underset{\left(N_{c}+N_{p}-i\right)Q+J}{\underbrace{0,\ldots,0}}], & i\in\varphi_{p}\\{}
    [\underset{N_{c}Q}{\underbrace{0,\ldots,0}},\underset{N_{p}Q}{\underbrace{0,\ldots,0}},\underset{J}{\underbrace{0,\ldots,0}}] & i\in\varphi_{m},
    \end{cases}
    \label{eq:c_subpro_z}
\end{equation}
\begin{align}
    d_{i}&=\begin{cases}
    0, & i\in\varphi_{c}\\
    0, & i\in\varphi_{p}\\
    \sum_{q=1}^{Q}\omega_{i,q}^{\ell}P_{i,q}T_{i,q}, & i\in\varphi_{m},
    \end{cases}\label{eq:d_subpro_z}\\
%\end{equation}
%\begin{equation}
    %\begin{aligned}
    \boldsymbol{e}_{i} &= [\underset{N_{c}Q}{\underbrace{0,\ldots,0}},\underset{N_{p}Q}{\underbrace{0,\ldots,0}}\\
     & \tilde{\alpha}_{i,1}^{c}P_{c}^{1}\boldsymbol{f}_{i}^{rT}\boldsymbol{f}_{1,\ell+1}^{c},\ldots,\tilde{\alpha}_{i,J}^{c}P_{c}^{J}\boldsymbol{f}_{i}^{rT}\boldsymbol{f}_{J,\ell+1}^{c}],\forall i.
    %\end{aligned}
    \label{eq:e_subpro_z}
%\end{equation}
\end{align}
\normalsize
The constraints on $\boldsymbol{z}$ of problem \eqref{eq:subpro_z_outer_max} expressed compactly are %\par\noindent\small
%\begin{equation}
    $\boldsymbol{A}\boldsymbol{z}\le\boldsymbol{b}$,
%\end{equation}\normalsize
where \par\noindent\small
\begin{equation}
    \begin{aligned}\boldsymbol{b}= & \left[P_{total}^{1},\ldots P_{total}^{N_{c}},T_{total}^{N_{c}+1},\ldots T_{total}^{N_{c}+N_{p}},P_{total}^{c},\right.\\
     & \left.-\sigma_{c,1}^{2}T_{0},\ldots,-\sigma_{c,J}^{2}T_{0}\right]^{T},
    \end{aligned}
    \label{eq:b_subpro_z}
\end{equation}\normalsize
and \par\noindent\small
\begin{equation}
    \boldsymbol{A}=\left[\begin{array}{c}
    \begin{array}{cccc}
    \boldsymbol{m}_{1} & \boldsymbol{0} & \cdots & \boldsymbol{0}\\
    \boldsymbol{0} & \ddots & \ddots & \vdots\\
    \vdots & \ddots & \boldsymbol{m}_{N_{c}+N_{p}} & \boldsymbol{0}\\
    \boldsymbol{0} & \cdots & \boldsymbol{0} & \boldsymbol{1}_{J}^{T}
    \end{array}\\
    \begin{array}{ccc}
    \boldsymbol{n}_{\varphi_{c}}^{1} & \boldsymbol{n}_{\varphi_{p}}^{1} & \boldsymbol{n}_{c}^{1}\\
    \vdots & \vdots & \vdots\\
    \boldsymbol{n}_{\varphi_{c}}^{J} & \boldsymbol{n}_{\varphi_{p}}^{J} & \boldsymbol{n}_{c}^{J}
    \end{array}
    \end{array}\right]
    \label{eq:A_subpro_z}
\end{equation}\normalsize
with\par\noindent\small
\begin{align}
    \boldsymbol{m}_{i}&=\left[M_{i,1},\ldots,M_{i,Q}\right],\forall i=1,\ldots,N_{c}+N_{p},\\
%\end{equation}
%\begin{equation}
    \boldsymbol{n}_{c}^{j}&=\left[\underset{j-1}{\underbrace{0,\ldots,0}},\frac{\beta^{j}T_{0}}{1-e^{\epsilon^{j}}},\underset{J-j}{\underbrace{0,\ldots,0}}\right],\forall j,\\
%    \end{equation}
%    \begin{equation}
    \boldsymbol{n}_{\varphi_{c}}^{j}&=\left[M_{1,1}\tilde{\alpha}_{j,1}^{r}\boldsymbol{f}_{1}^{rT}\boldsymbol{f}_{j,\ell+1}^{c}T_{1,1},\right.\nonumber\\
     & \left.\ldots,M_{N_{c},Q}\tilde{\alpha}_{j,N_{c}}^{r}\boldsymbol{f}_{N_{c}}^{rT}\boldsymbol{f}_{j,\ell+1}^{c}T_{N_{c},Q}\right],\\
%\end{equation}
%\begin{equation}
    %\begin{aligned}
    \boldsymbol{n}_{\varphi_{p}}^{j}&= \left[M_{N_{c}+1,1}\tilde{\alpha}_{j,N_{c}+1}^{r}\boldsymbol{f}_{N_{c}+1}^{rT}\boldsymbol{f}_{j,\ell+1}^{c}P_{N_{c}+1,1},\right.\nonumber\\
     & \left.\ldots,M_{N_{c}+N_{p},1}\tilde{\alpha}_{j,N_{c}+N_{p}}^{r}\boldsymbol{f}_{N_{c}+N_{p}}^{rT}\boldsymbol{f}_{j,\ell+1}^{c}P_{N_{c}+N_{p},1}\right].
    %\end{aligned}
\end{align}\normalsize
Therefore, problem \eqref{eq:subpro_z_outer_max} is equivalently recast into \par\noindent\small
\begin{equation}
    \begin{aligned} & \underset{\boldsymbol{z}}{\text{maximize}} &  & f\left(\boldsymbol{z}\right)\triangleq\sum_{i=1}^{N}\frac{\boldsymbol{c}_{i}^{T}\boldsymbol{z}+d_{i}}{\boldsymbol{e}_{i}^{T}\boldsymbol{z}+\sigma_{r,i}^{2}}\\
     & \text{subject to} &  & \boldsymbol{A}\boldsymbol{z}\le\boldsymbol{b},\;%\\
     %&  &  & 
     \boldsymbol{z}\ge\boldsymbol{0},
    \end{aligned}
    \label{eq:linFrac_z}
\end{equation}\normalsize
which maximizes a sum of linear fractional function over a convex set. 

Several methods have been proposed in the
literature to solve this fractional programming problem \cite{benson2004global,falk2014optimizing,phuong2003unified,shen2018fractional}. However, in order to be within the framework of the alternating
ascent-descent method, an incremental update on the variable $\boldsymbol{z}$ is necessary. In other words, a global solution to problem
\eqref{eq:linFrac_z} might violate the alternating ascent-descent method and its theoretical guarantees. Therefore, within the alternating ascent-descent framework, the update rule of $\boldsymbol{z}$ is simply %\par\noindent\small
%\begin{equation}
    $\boldsymbol{z}^{n+1}=\mathcal{P}_{\mathcal{Z}}\left(\boldsymbol{z}^{n}+\eta\nabla_{z}f\left(\boldsymbol{z}^{n}\right)\right)$, 
%\end{equation}\normalsize
where $\eta$ is the stepsize parameter, \par\noindent\small
\begin{equation}
    \nabla_{z}f\left(\boldsymbol{z}^{n}\right)=\sum_{i=1}^{N}\frac{\left(\boldsymbol{e}_{i}^{T}\boldsymbol{z}^{n}+\sigma_{r,i}^{2}\right)\boldsymbol{c}_{i}-\left(\boldsymbol{c}_{i}^{T}\boldsymbol{z}^{n}+d_{i}\right)\boldsymbol{e}_{i}}{\left(\boldsymbol{e}_{i}^{T}\boldsymbol{z}^{n}+\sigma_{r,i}^{2}\right)^{2}},
\end{equation}\normalsize
and $\mathcal{P}_{\mathcal{Z}}\left(\cdot\right)$ represents the projection to the convex set $\mathcal{Z}=\left\{\boldsymbol{z}|\boldsymbol{A}\boldsymbol{z}\le\boldsymbol{b},\boldsymbol{z}\ge\boldsymbol{0}\right\}$, which actually involves solving a convex quadratic problem.

\subsubsection{Computational complexity and convergence analysis}
The proposed method for the power and time allocation is summarized in Algorithm \ref{alg:Alg_subprob_z}. The overall computational complexity of Algorithm \ref{alg:Alg_subprob_z} is linear with the number of iterations. 
\begin{algorithm}[H]		 	
	\caption{Alternating ascent-descent method to power and dwell time allocation}	 	
	\label{alg:Alg_subprob_z}
\begin{algorithmic}[1]	 			 		
		\Require $\left\{ \boldsymbol{B}_{q}\right\} ,\left\{ \tilde{\boldsymbol{C}}_{i,q}\right\}, \left\{ P_{total}^{i}\right\} ,\left\{ T_{total}^{i}\right\} ,P_{total}^{c}, \left\{ M_{i,q}\right\} , \boldsymbol{F}_{\ell}^{c}$
		\Ensure Allocation vector $\boldsymbol{z}$  
		\State{$n\leftarrow 0$}
		\State{initialize a feasible $\boldsymbol{z}^{n}$}	
		\Repeat
			\State{calculate $\boldsymbol{B}_{q}^{\ell,n}$ by (\ref{eq:B_definition})}
			\State{$\boldsymbol{V}_{q}^{n+1}\leftarrow\left(\tilde{\boldsymbol{\Lambda}}^{T}\boldsymbol{B}_{q}^{\ell,n}\tilde{\boldsymbol{\Lambda}}\right)^{-1}/\text{Tr}\left[\left(\tilde{\boldsymbol{\Lambda}}^{T}\boldsymbol{B}_{q}^{\ell,n}\tilde{\boldsymbol{\Lambda}}\right)^{-1}\right]$}
			\State{$\omega_{i,q}^{n}\leftarrow\text{Tr}\left[\left(\tilde{\boldsymbol{\Lambda}}\boldsymbol{V}_{q}^{n+1}\right)^{T}\tilde{\boldsymbol{C}}_{i,q}\left(\tilde{\boldsymbol{\Lambda}}\boldsymbol{V}_{q}^{n+1}\right)\right]$}
			\State{Calculate $\boldsymbol{c}_{i},d_{i},\boldsymbol{e}_{i},\boldsymbol{b}, \boldsymbol{A}$ by (\ref{eq:c_subpro_z}), (\ref{eq:d_subpro_z}), (\ref{eq:e_subpro_z}), (\ref{eq:b_subpro_z}) and (\ref{eq:A_subpro_z})}
			\State{$\nabla_{z}f\left(\boldsymbol{z}^{n}\right)\leftarrow\sum_{i=1}^{N}\frac{\left(\boldsymbol{e}_{i}^{T}\boldsymbol{z}^{n}+\sigma_{r,i}^{2}\right)\boldsymbol{c}_{i}-\left(\boldsymbol{c}_{i}^{T}\boldsymbol{z}^{n}+d_{i}\right)\boldsymbol{e}_{i}}{\left(\boldsymbol{e}_{i}^{T}\boldsymbol{z}^{n}+\sigma_{r,i}^{2}\right)^{2}}$}
			\State{$\boldsymbol{z}^{n+1}\leftarrow\mathcal{P}_{\mathcal{Z}}\left(\boldsymbol{z}^{n}+\eta\nabla_{z}f\left(\boldsymbol{z}^{n}\right)\right)$}
			\State{$n\leftarrow n+1$}
		\Until convergence
\end{algorithmic}
\end{algorithm}

For the convenience of analysis, we focus on the deterministic cost on a per-iteration basis, which mainly comes from the following sources: $\boldsymbol{V}_{q}^{n+1}$,
$\omega_{i,q}^{n}$ and $\boldsymbol{z}^{n+1}$. The computational cost of $\left\{\boldsymbol{V}_{q}^{n+1}\right\} $ is linear with
$\mathcal{O}\left(Q\right)$ due to the matrix multiplication and inverse of $\left(\tilde{\boldsymbol{\Lambda}}^{T}\boldsymbol{B}_{q}^{\ell,n}\tilde{\boldsymbol{\Lambda}}\right)$
is a constant. Similarly, the computational of $\left\{ \omega_{i,q}^{n}\right\} $
is also linear with $\mathcal{O}\left(QN\right)$. The update rule of $\boldsymbol{z}^{n+1}$ involves the projection operator $\mathcal{P}_{\mathcal{Z}}\left(\cdot\right)$,
which is, in fact, to solve a convex quadratic problem. The MOSEK optimization package will reformulate the problem into epigraph form by introducing an additional slack variable. Consequently, there is a second order cone constraint and the remaining constraints are linear. The computational complexity of solving the reformulated problem is therefore upper bound by $\mathcal{O}\left(\left(N_{c}Q+N_{p}Q+J\right)^{3.5}\right)$, the same order as second-order cone programming. The convergence of Algorithm \ref{alg:Alg_subprob_z} is stated by the following theorem.
\begin{lem}\label{lemma2}
Assume $\left\{ \boldsymbol{z}^{n}\right\} $ to be the generated sequence of Algorithm~\ref{alg:Alg_subprob_z}. Then, every limit point of the sequence $\left\{ \boldsymbol{z}^{n}\right\} $is the stationary point of problem \eqref{eq:subpro_z}.
\end{lem}
\begin{IEEEproof}
	See Appendix \ref{Appendix-B}.
\end{IEEEproof}
%Building on the above derivations, 
The convergence of the combined algorithm ANCHOR to solve problem \eqref{eq:RA opt problem} is stated in the following Lemma~\ref{theorem3}.
\begin{thm}\label{theorem3}
Assume that Algorithm \ref{alg:Alg_subprob_F} outputs an (sub-)optimal solution, then the sequence $\left\{g\left(\boldsymbol{z}_{\ell},\boldsymbol{F}_{\ell}^{c}\right)\right\} $ is non-decreasing and converges to a finite value, where $\left(\boldsymbol{z}_{\ell},\boldsymbol{F}_{\ell}^{c}\right)$ is generated by Algorithm \ref{alg:Alg_RA} at the $\ell$-th iteration.
\end{thm}
\begin{IEEEproof}
	See Appendix \ref{Appendix-C}.
\end{IEEEproof}
Algorithm \ref{alg:Alg_RA} summarizes all steps of ANCHOR.
\begin{algorithm}[H]		 	
	\caption{\textit{A}lter\textit{n}ating allo\textit{c}ation of \textit{h}eterogene\textit{o}us \textit{r}esources (ANCHOR)}	 	
	\label{alg:Alg_RA}	 	
	\begin{algorithmic}[1]		 		
		\Require $\varphi_{c},\varphi_{p},\varphi_{m}, \left\{ P_{total}^{i}\right\}, \left\{ T_{total}^{i}\right\} ,P_{total}^{c},\left\{ M_{i,q}\right\}$
		\Ensure Resource allocation vector $\mathbf{z}$  	
		\State{$\ell\leftarrow 0$}
		\State{initialize a feasible $\boldsymbol{z}_{\ell}$}
		\Repeat
		\State{update $\boldsymbol{F}_{\ell+1}^{c}$ via Algorithm \ref{alg:Alg_subprob_F}}
		\State{update $\boldsymbol{z}_{\ell+1}$ via Algorithm \ref{alg:Alg_subprob_z}} 
		\State{$\ell\leftarrow\ell+1$}	 		
		\Until convergence
	\end{algorithmic}	 
\end{algorithm}
%
%\subsection{The HRCN Workflow for Resource Allocation}
\subsection{Algorithmic workflow}
\begin{figure}[t]
\begin{centering}
\includegraphics[scale=0.37]{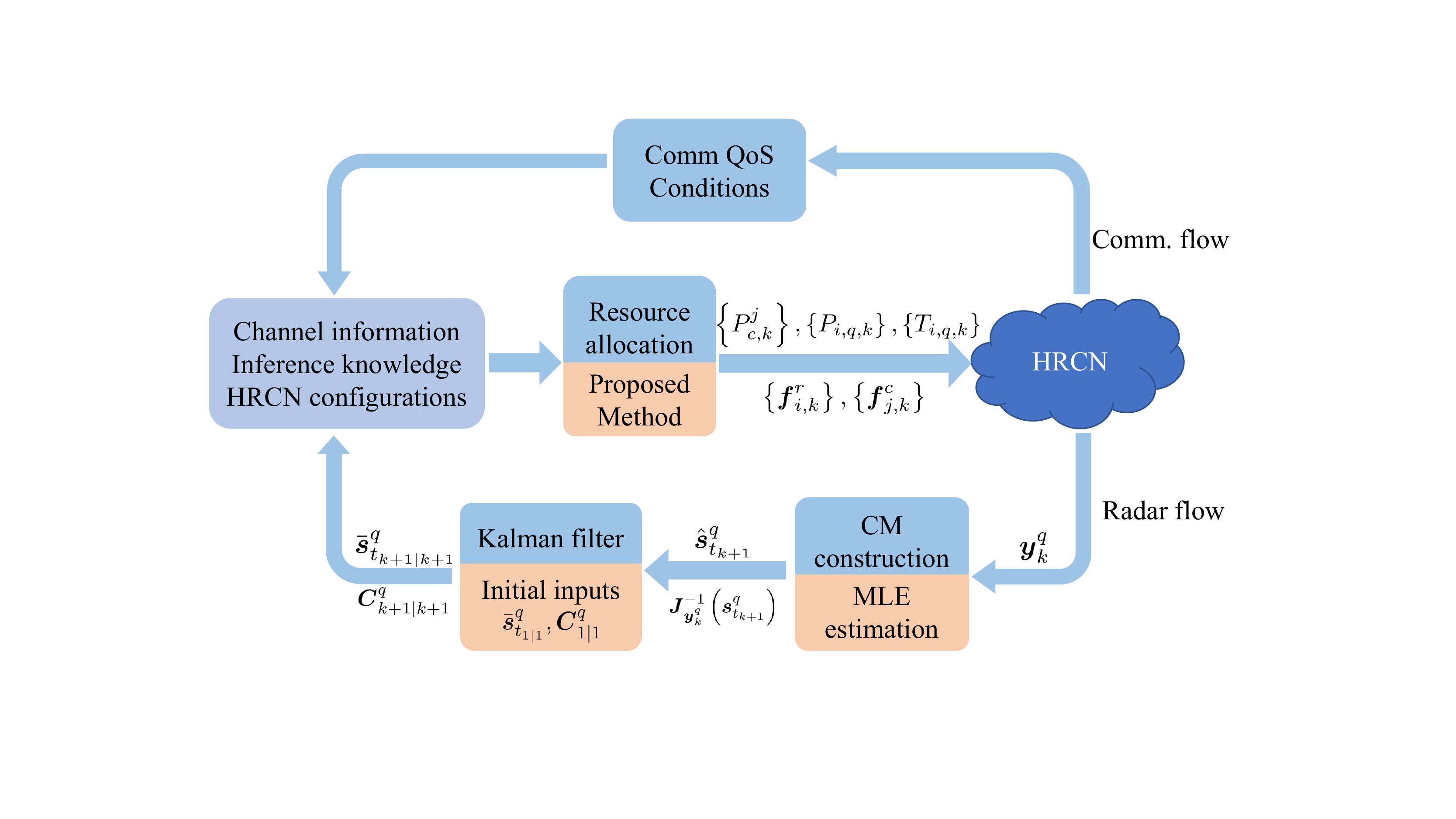}
\par\end{centering}
\caption{\label{fig:The-RA-workflow}{Resource allocation procedure for HRCN. The symbols represent either the inputs or outputs of the related modules.}}
\end{figure}

The ANCHOR algorithm is applied to allocate resources for a fusion interval. It is expected that the optimized resource allocation will improve the accuracy of the CM $\hat{\boldsymbol{s}}_{t_{k+1}}^{q}$
and its covariance matrix $\boldsymbol{J}_{\boldsymbol{y}_{k}^{q}}^{-1}\left(\boldsymbol{s}_{t_{k+1}}^{q}\right)$, which further enhances the tracking performance based on Kalman filter. At the $k$-th fusion interval, $\bar{\boldsymbol{s}}_{t_{k|k}}^{q}$ is the filtered estimate of the target state $\boldsymbol{s}_{t_{k}}^{q}$ of the last fusion
interval and assumed to be known, and $\hat{\boldsymbol{s}}_{t_{k+1}}^{q}$
is the CM estimated based on the measurements $\boldsymbol{y}_{k}^{q}$.
Based on the system model as equation \eqref{eq:3}, we have \par\noindent\small
\begin{equation}
\begin{cases}
\bar{\boldsymbol{s}}_{t_{k+1|k}}^{q}=f\left(\bar{\boldsymbol{s}}_{t_{k|k}}^{q},T_{0}\right)\\
\boldsymbol{C}_{k+1|k}^{q}=\boldsymbol{\Gamma}_{k}^{q}+\boldsymbol{F}_{k}^{q}\boldsymbol{C}_{k|k}^{q}\boldsymbol{F}_{k}^{q}\\
\bar{\boldsymbol{s}}_{t_{k+1|k+1}}^{q}=\bar{\boldsymbol{s}}_{t_{k+1|k}}^{q}+\boldsymbol{K}_{k+1}^{q}\left(\hat{\boldsymbol{s}}_{t_{k+1}}^{q}-\bar{\boldsymbol{s}}_{t_{k+1|k}}^{q}\right)\\
\boldsymbol{C}_{k+1|k+1}^{q}=\left(\boldsymbol{I}-\boldsymbol{K}_{k+1}^{q}\right)\boldsymbol{C}_{k+1|k}^{q},
\end{cases}\label{eq:kalmanUpdate}
\end{equation}\normalsize
where $\bar{\boldsymbol{s}}_{t_{k+1|k}}^{q}$ and $\bar{\boldsymbol{s}}_{t_{k+1|k+1}}^{q}$
are the predicted and filtered estimate of the true target state,
respectively, and $\boldsymbol{C}_{k+1|k}^{q}$ and $\boldsymbol{C}_{k+1|k+1}^{q}$
are the corresponding covariance matrices, and $\boldsymbol{K}_{k+1}^{q}$
is Kalman gain with \par\noindent\small
\begin{equation}
\boldsymbol{K}_{k+1}^{q}=\boldsymbol{C}_{k+1|k}^{q}\left[\boldsymbol{C}_{k+1|k}^{q}+\boldsymbol{J}_{\boldsymbol{y}_{k}^{q}}^{-1}\left(\boldsymbol{s}_{t_{k+1}}^{q}\right)\right]^{-1}.
\end{equation}\normalsize

The filtered estimate $\bar{\boldsymbol{s}}_{t_{k+1|k+1}}^{q}$ is the prior knowledge of the Kalman filter for the next fusion interval. Note that the initial state $\bar{\boldsymbol{s}}_{t_{1|1}}^{q}$
and its covariance matrix $\boldsymbol{C}_{1|1}^{q}$ are usually
estimated. The workflow of the resource allocation for the HRCN is summarized
in Fig.~\ref{fig:The-RA-workflow}. We observe that the entire resource allocation procedure proceeds in an iterative closed-loop as the Kalman filtering process. As the fusion interval elapses, the resource allocation iteratively improves the Bayesian CRB matrix of Kalman filtering thereby leading to an enhanced tracking performance in the sense of smaller estimation covariance.
\section{Numerical Experiments}\label{sec:simu}
\begin{figure}[t]
\begin{centering}
\includegraphics[scale=0.6]{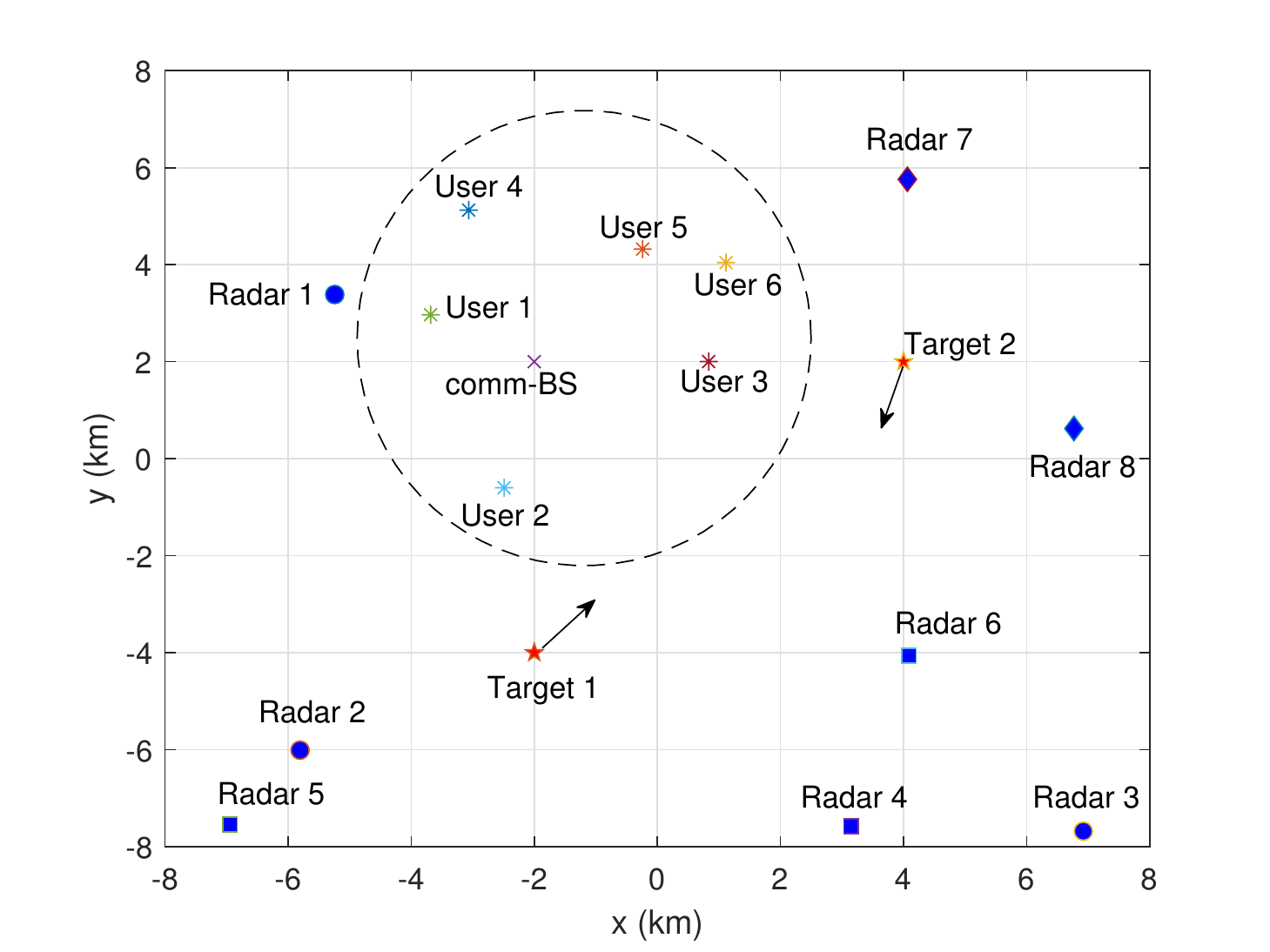}
\par\end{centering}
\caption{\label{fig:Test-scenario-of}{Test scenario of the HRCN with three types of radars, which coexist with communications BS and users,  tracking two moving targets simultaneously.
}}
\end{figure}

\begin{figure}[t]
\begin{centering}
\includegraphics[scale=0.63]{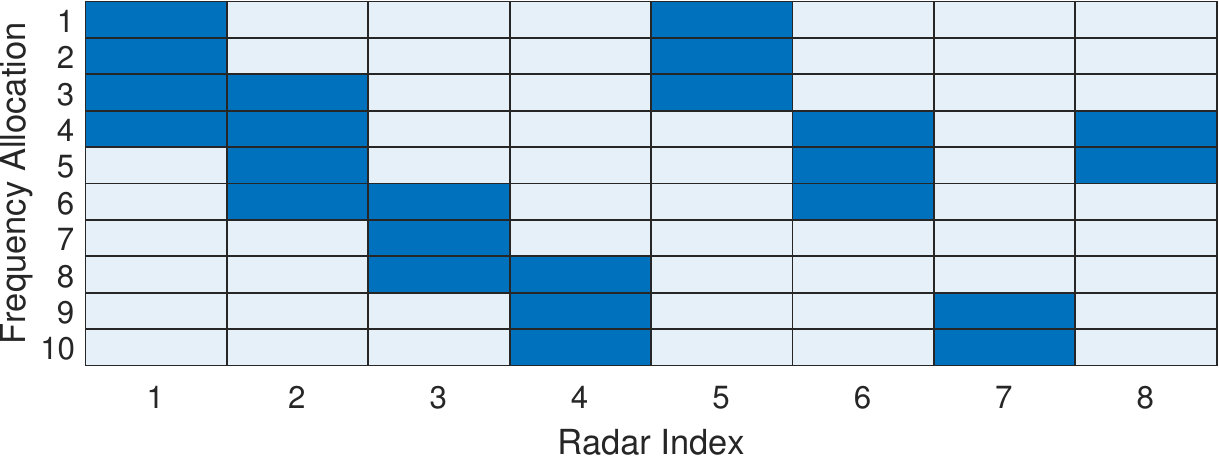}
\par\end{centering}
\caption{\label{fig:Frequency-Radar} {Preassigned frequency allocations for the 8 radars, where each block represents the bandwidth $\Delta f$.}}
\end{figure}

We conducted extensive numerical experiments to evaluate the performance of our proposed resource allocation algorithm ANCHOR for the HRCN scenario. Throughout the experiments,  unless specified otherwise, we employ the following parameter settings:
\begin{enumerate}
    \item HRCN scenario settings: %The test scenario is depicted in Figure \ref{fig:Test-scenario-of}. 
    We consider a typical outdoor scenario (Fig.~\ref{fig:Test-scenario-of}), where target 1 and 2 start from the locations $\left(-2\text{ km},-4\text{ km}\right)$ and $\left(4\text{ km},2\text{ km}\right)$ and move with the velocities $\left(50\text{ m/s},50\text{ m/s}\right)$ and $\left(-25\text{ m/s},-50\text{ m/s}\right)$, respectively. We deploy heterogeneously-distributed radars to track the two moving targets. Specifically, we set $N_{cr}=3$ colocated MIMO radars (MMRs), $N_{pr}=3$ phased array radars (PARs) and $N_{mr}=2$ mechanical scanning radars (MSRs), and their locations are randomly generated in the region of interest. The duration of a single fusion interval is $T_{0}=10$ s. With respect to the staring time $t=0$ of a fusion interval, the initial and revisit time of these radars are provided in Table \ref{tab:Initial-sampling-time}. For the HetNet, we consider only the downlink transmission from a BS to $J=6$ devices, which are randomly placed within the BS coverage area. For the shared bandwidth between radars and communications, assume that the total bandwidth is $B=400$ MHz available, for example, in X-band and higher spectral bands) and the unit interval is $\triangle f=4$ MHz (i.e. $B/\triangle f=100$ frequencies to be assigned). Fig.~\ref{fig:Frequency-Radar} demonstrates pre-assigned radar frequencies, where each block represents a $40$ MHz frequency band. 
    \item Algorithmic parameter settings: %In the algorithm of the global optimal search for frequency allocation, 
    We set $T_\text{max}=1000$, $T_\text{min}=0.1$ and $\delta T=1$. For the alternating ascent-descent method to power and dwell time allocation, both power and dwell time are initialized using the uniform allocation (i.e. $P_{i,q,k}={P_{total}^i}/{N_c}$, $P_{c,k}^{j}={P_{total}^c}/{J}$ and $T_{i,q,k}={T_{total}^i}/{N_p}$), and the stopping criterion is the increase of the objective value is less than $1\times10^{-4}$. The ANCHOR algorithm stops when the objective value increase is less than $1\times10^{-4}$ or the number of iterations is beyond 100.
    \item System parameter settings: All interference channel gains\footnote{We assume block fading for the interference channel gain, wherein it is a constant over a fusion interval. However, it may be appropriately changed depending on the actual operating band \cite{meijerink2014physical,gustafson2013mm,zhang2010channel}. This also applies to other interference channel gains.} and noise power are obtained by squaring the value generated from the Gaussian distribution. In the Kalman filter, justified by the fact that the tracking is along a series of fusion intervals, the initial state $\bar{\boldsymbol{s}}_{t_{1|1}}^{q}$ is set as the true target state with Gaussian noise of standard deviation 5\% of the true value. The corresponding initial covariance matrix $\boldsymbol{C}_{1|1}^{q}$ is usually set to the estimated covariance in the previous fusion interval. However, this information is not available to us and hence we set it to be 10-times scaled version of the covariance matrix used in the state model.% which represents a rough initial guess and may lead to inaccurate tacking performance at the first few tracking rounds.
\end{enumerate}
%Unless otherwise specified, the above experiment settings are the same for all simulations.

\begin{table}[t]
\caption{\label{tab:Initial-sampling-time}Initial settings}% of sampling time and revisit intervals for all radars}
\centering{}%
\tabcolsep=0.16cm
\begin{tabular}{cccccccccc}
\toprule 
\multirow{2}{*}{Radar} & Type & \multicolumn{3}{c}{MMR} & \multicolumn{3}{c}{PAR} & \multicolumn{2}{c}{MSR}\tabularnewline
\cmidrule{2-10} \cmidrule{3-10} \cmidrule{4-10} \cmidrule{5-10} \cmidrule{6-10} \cmidrule{7-10} \cmidrule{8-10} \cmidrule{9-10} \cmidrule{10-10} 
 & Index & 1 & 2 & 3 & 4 & 5 & 6 & 7 & 8\tabularnewline
\midrule 
\multirow{2}{*}{Target 1} & Initial time (s) & 2 & 2.5 & 3 & 2.3 & 3 & 3.2 & 3.2 & 2.1\tabularnewline
\cmidrule{2-10} \cmidrule{3-10} \cmidrule{4-10} \cmidrule{5-10} \cmidrule{6-10} \cmidrule{7-10} \cmidrule{8-10} \cmidrule{9-10} \cmidrule{10-10} 
 & revisit interval (s) & 2 & 2 & 2 & 3 & 2 & 2 & 2.5 & 2.5\tabularnewline
\midrule 
\multirow{2}{*}{Target 2} & Initial time (s) & 2 & 2.5 & 3 & 3 & 3.5 & 3.8 & 4 & 3.5\tabularnewline
\cmidrule{2-10} \cmidrule{3-10} \cmidrule{4-10} \cmidrule{5-10} \cmidrule{6-10} \cmidrule{7-10} \cmidrule{8-10} \cmidrule{9-10} \cmidrule{10-10} 
 & revisit interval (s) & 2 & 2 & 2 & 3 & 2 & 2 & 2.5 & 2.5\tabularnewline
\bottomrule
\end{tabular}
\end{table}

\subsection{ANCHOR performance within a fusion interval}
\begin{figure}[t]
\begin{centering}
\includegraphics[scale=0.6]{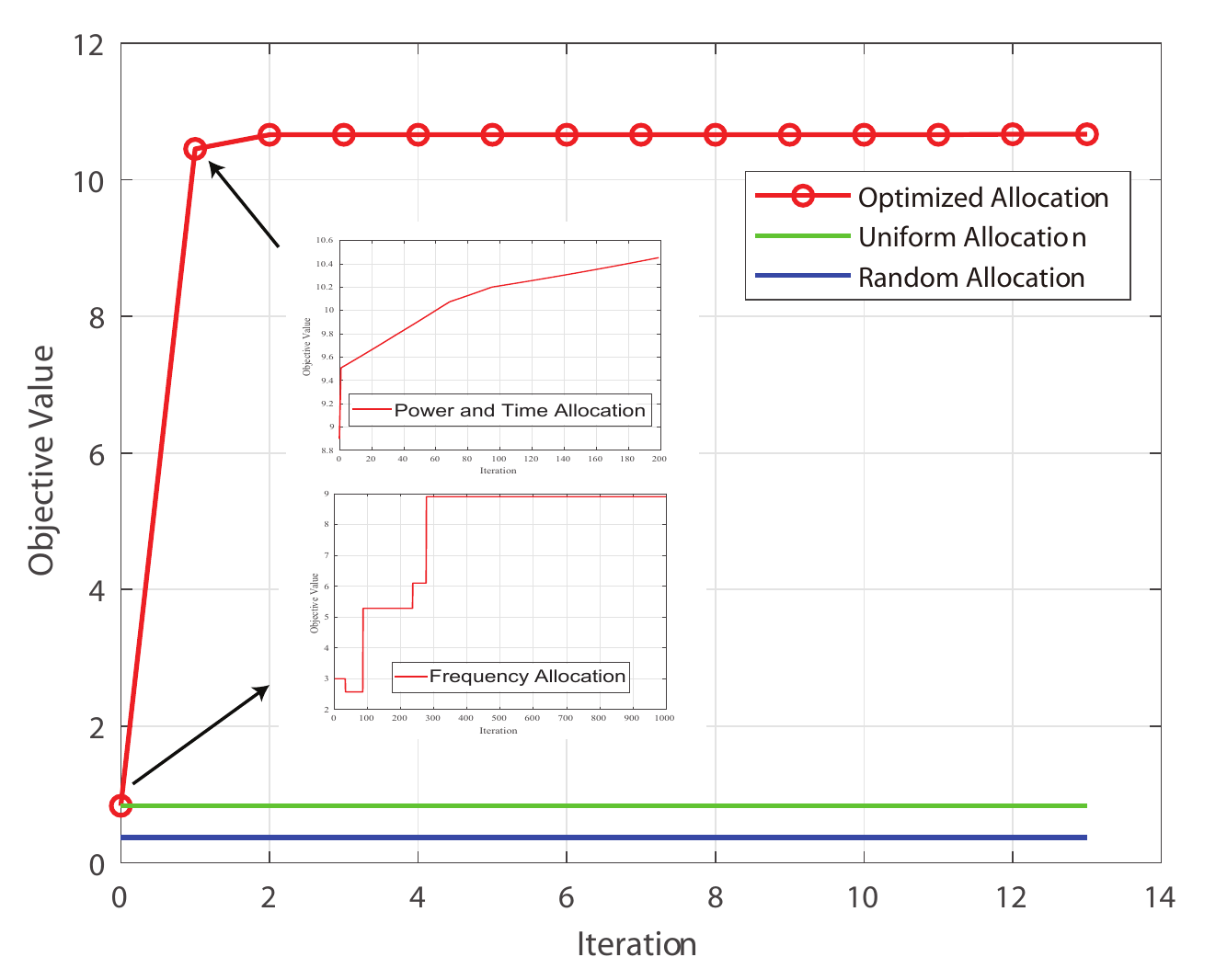}
\par\end{centering}
\caption{\label{fig:obj_improve_single} Comparison of allocation methods in terms of the improvement in the objective value; the insets depict performance improvement with iterations for allocating power-time (top) and frequencies (bottom) for a given outer iteration index.}
\end{figure}

We first explore the performance of ANCHOR within a fusion interval and subsequently consider its performance across intervals.
\paragraph{Convergence and optimized objective function}  We assess the performance in the $k$-th fusion interval (i.e. from $t_{k}$ to $t_{k+1}$). Note that the Bayesian CRB matrix $\boldsymbol{B}_{k}$ and
the Kalman estimation covariance matrix $\boldsymbol{C}_{k|k}$ should be available as the inputs to the Bayesian tracking scheme in the $k$-th interval. We initialize them with an estimate ( i.e. an amplified version of the state covariance matrix), as is typical in Kalman filtering given that they will converge quickly after the initial phase~\cite{kalman1960new}. Recall that the tracking performance depends on the improvement on the Bayesian CRB, which is evaluated by the objective function $\sum_{q=1}^{Q}\frac{1}{\text{Tr}\left(\boldsymbol{\Lambda}\boldsymbol{B}^{-1}\left(\boldsymbol{s}_{t_{k+1}}^{q}\right)\boldsymbol{\Lambda}^{T}\right)}$. Therefore, we compare the improvements of the objective value under three resource allocation schemes for the $k$-th fusion interval in Figure \ref{fig:obj_improve_single}. For random allocation, we randomly generate the allocated resources satisfying 
the constraints of problem (\ref{eq:RA opt problem}). The uniform allocation distributes the resources equally. The optimized allocation is designed by our proposed algorithm ANCHOR, in which the  uniform allocation serves as initialization. Given ANCHOR is based on the alternating optimization method solving two subproblems, we also demonstrate the two nested iterations of the ANCHOR for the first outer iteration. We note that, based on the two nested iterations, the ANCHOR can increase the objective value monotonically as the outer iteration index increases. Compared to the other two allocations, the converged objective value of the optimized allocation is much larger.

\begin{figure*}[t]
\centering
%\subfloat[\label{fig:Power-allocation-MMR}Power allocation for MIMO radars.]{\begin{centering}
%\includegraphics[scale=0.6]{Figures/PowerAllocation_MMR_SingleFusion_20210812}
%\par\end{centering}
%}
%\par\end{centering}
%\vspace{-2mm}
%\begin{centering}
%\subfloat[\label{fig:Dwell-time-allocation-PAR}Dwell time allocation for phased
%array radars.]{\begin{centering}
%\includegraphics[scale=0.6]{Figures/TimeAllocation_PAR_SingleFusion_20210812}
%\par\end{centering}
%}
%\par\end{centering}
\includegraphics[width=0.9\textwidth]{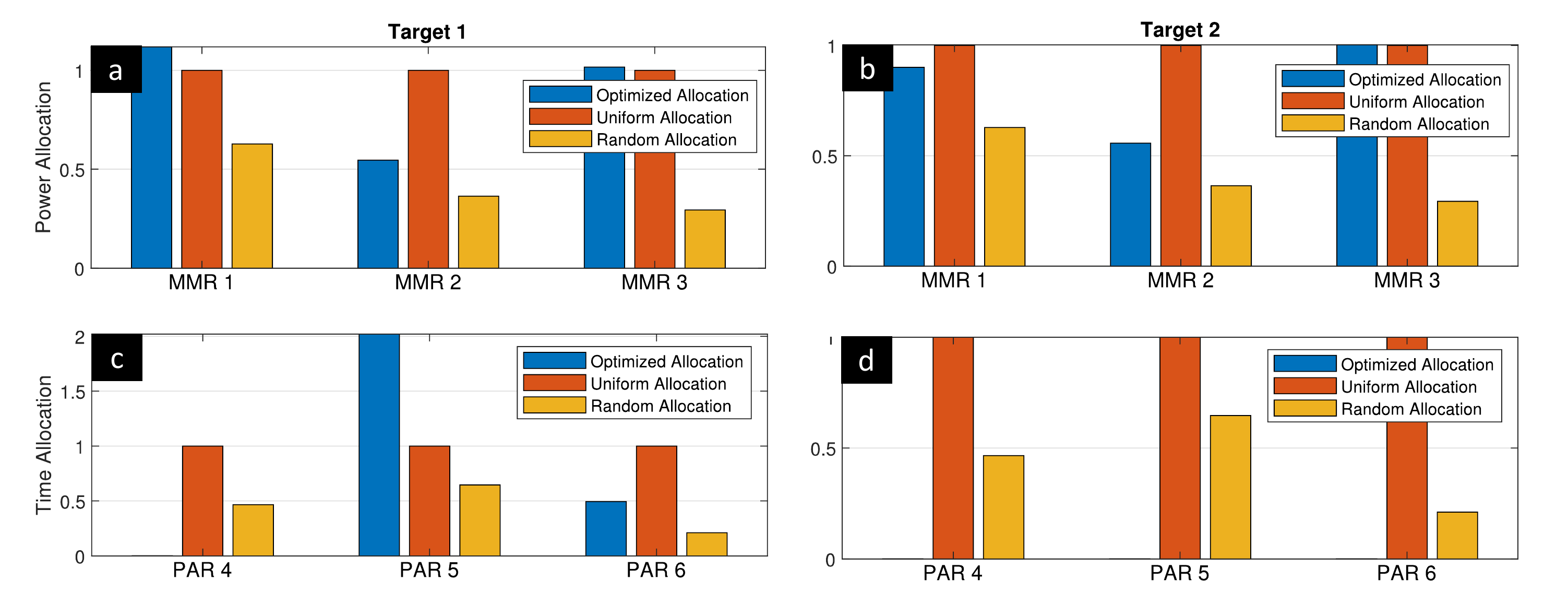}
\caption{{Power allocation in MIMO radars for (a) Target 1 and (b) Target 2, where all powers under three allocation schemes are normalized by the counterpart of uniform allocation for each radar. Time allocation in phased array radar for (c) target 1 and (d) target 2, where the allocated time is normalized similarly as in uniform allocation.}}
\label{fig:allocated_resources_single} 
\end{figure*}

\paragraph{Radar resource allocation} Fig.~\ref{fig:allocated_resources_single} demonstrates the allocated radar resources for three methods (random, uniform, and ANCHOR). All values are normalized by the counterparts of the uniform allocation. Note that for the PARs with the optimized allocation, only target $1$
is tracked by using radar $5$ and $6$. In fact, the designed dwell time $T_{4,1,k},T_{4,2,k},T_{5,2,k}$ and $T_{6,2,k}$ are close to zero numerically, which physically means that the corresponding radars
stop scanning at some given time. While it is intuitive that the tracking performance is improved by increasing the power and dwell time as they potentially increase the SINR of the receiving
measurements. However, in an interference scenario, we cannot increase all resources while satisfying various constraints and maintaining lower mutual interference. Thus, a trade-off is achieved via the optimized allocation (Fig.~\ref{fig:allocated_resources_single}).

\begin{figure}[t]
\begin{centering}
\subfloat[\label{fig:Communication-power-allocation}{Allocation of communications power (left) and throughput margin (the difference between the achieved and required throughput) (right)}.]{\begin{centering}
\includegraphics[scale=0.3]{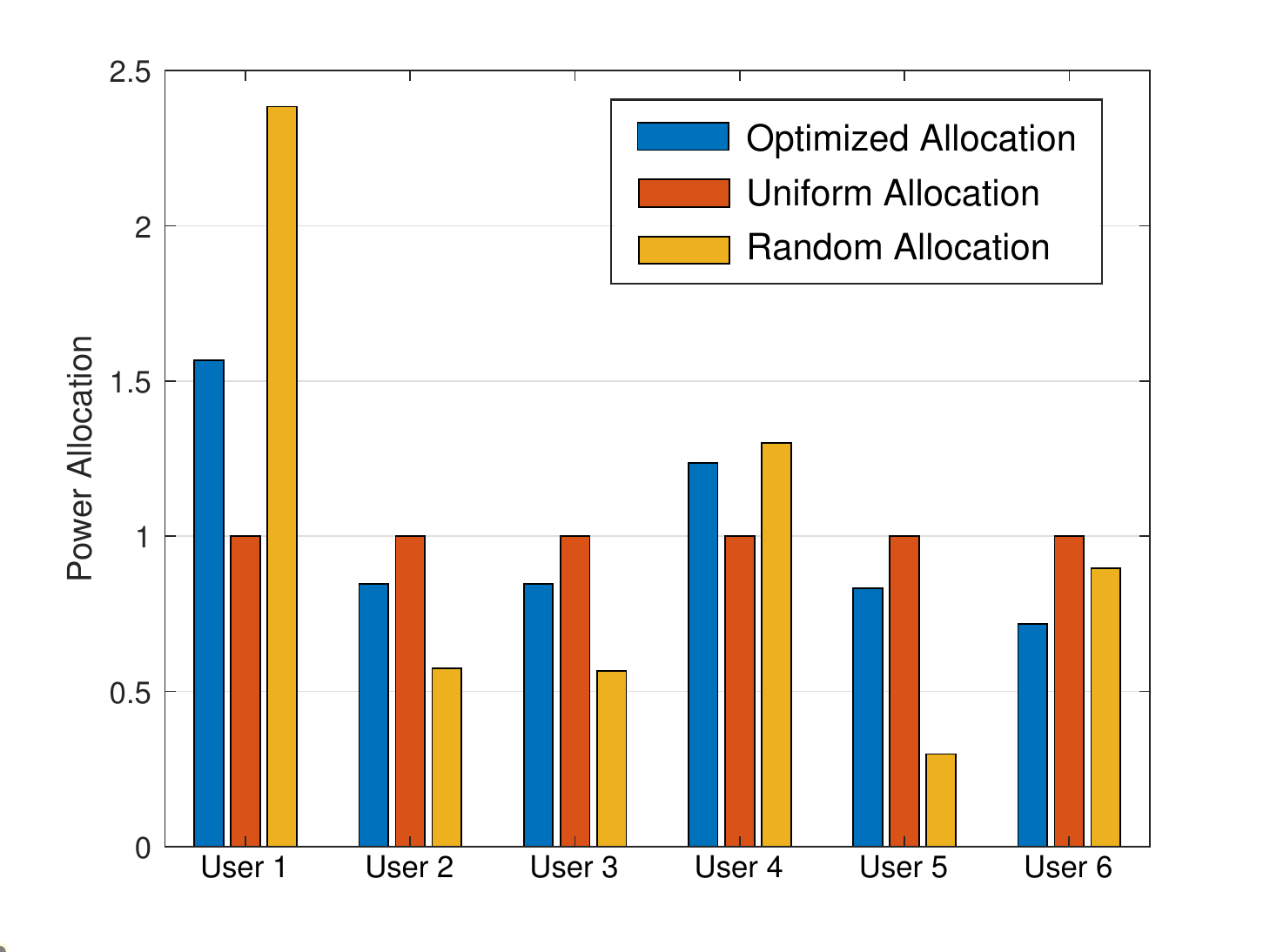}\includegraphics[scale=0.3]{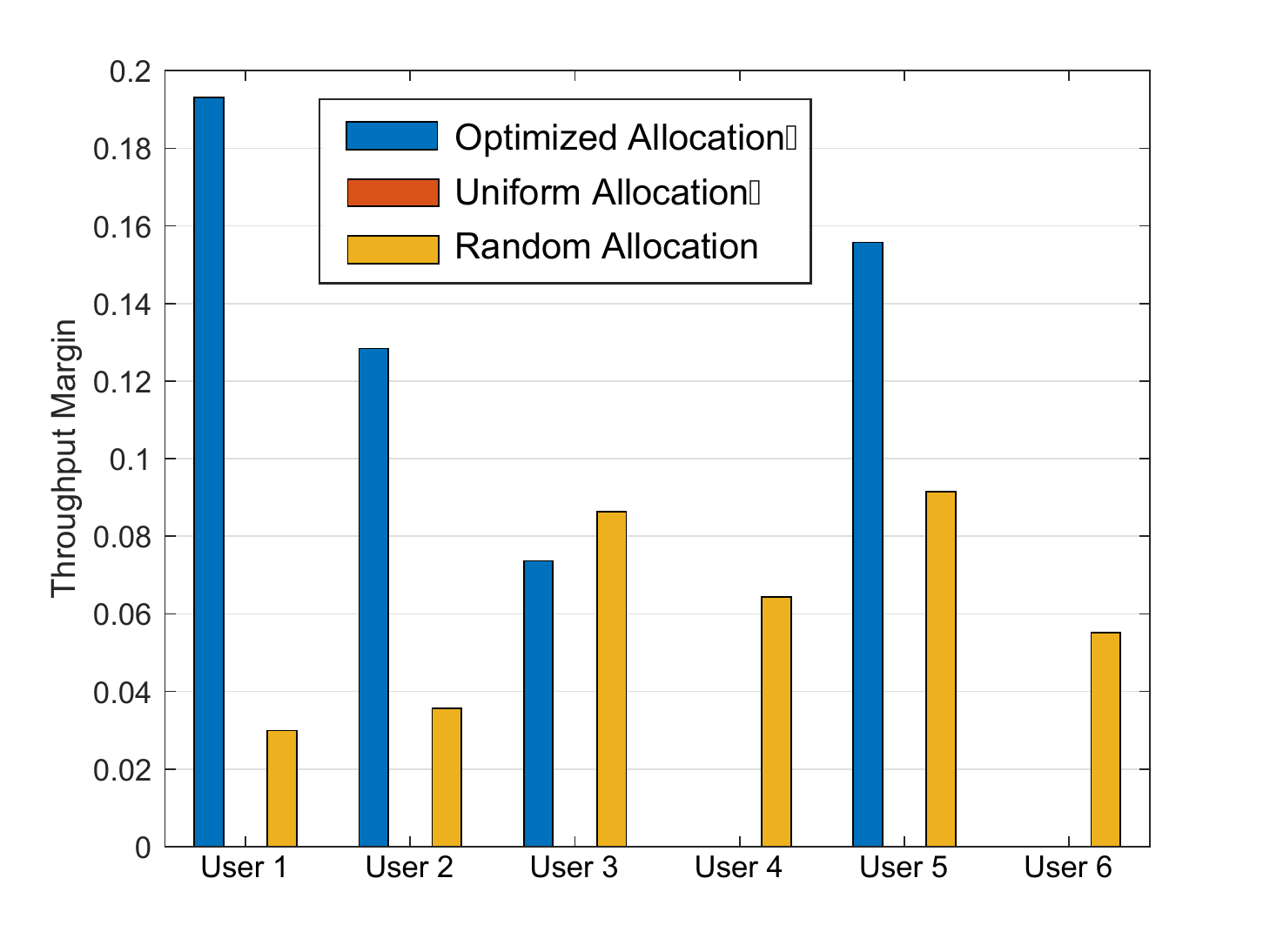}
\par\end{centering}
}
\par\end{centering}
\begin{centering}
\subfloat[Frequency allocation to communications system.]{\begin{centering}
\includegraphics[scale=0.32]{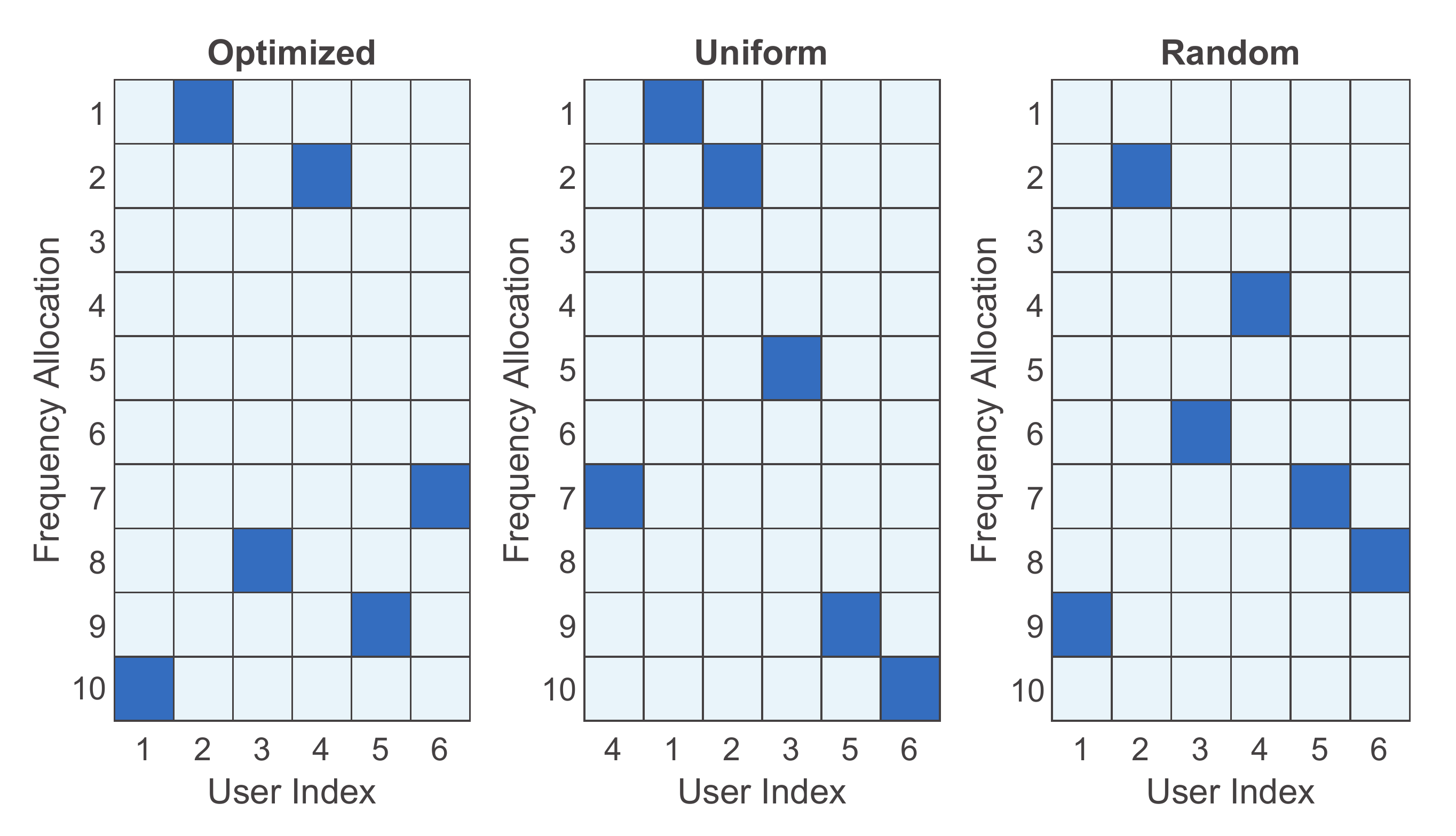}
\par\end{centering}
}
\par\end{centering}
\caption{\label{fig:QoS_comm_single}Resource allocation for communications downlinks.}
\end{figure}

\paragraph{Communications resource allocation} Fig.~\ref{fig:QoS_comm_single} shows the power and frequency allocation for the communications downlink transmissions, where the throughput margin is defined as the difference between the achieved and required throughput. Note that the constraint parameters are set to satisfy the uniform allocation in the simulations. Consequently, the throughput margins of the uniform allocation in Fig.~\ref{fig:Communication-power-allocation} are all zero among the users. Interestingly, all users in the random allocation have margins to their throughput thresholds. Compared to the random allocation which only needs to satisfy the constraints of the optimization problem,
the ANCHOR allocation optimizes the Bayesian CRB while being feasible to the constraint set. In practice, if the throughput margin is necessary, it is always achieved equivalently by increasing the throughput threshold for the optimized allocation probably at cost of a slight degeneration on tracking performance. 

\paragraph{State estimation performance} For the three allocations, we compare the tracking performance in Table \ref{tab:Radar-RMSE}, where
the calibrated root mean square error (CRMSE) is \par\noindent\small
\begin{equation}
\text{CRMSE}=\sum_{q=1}^{Q}\sqrt{\frac{1}{N_{t}}\left(\sum_{n=1}^{N_{t}}\left\Vert \boldsymbol{\Lambda}\left(\tilde{\boldsymbol{s}}_{t_{k+1}}^{q\left(n\right)}-\boldsymbol{s}_{t_{k+1}}^{q}\right)\right\Vert ^{2}\right)},\label{eq:CRMSE}
\end{equation}\normalsize
with $N_{t}$ being the number of the Monte Carlo trials, and the estimated target state is averaged over the $N_{t}$ states. The true target states at time $t_{k+1}$ are given by $\boldsymbol{s}_{t_{k+1}}^{1}=\left[-1.5\text{ km},50\text{ m/s},-3.5\text{ km},50\text{ m/s}\right]^{T}$
and $\boldsymbol{s}_{t_{k+1}}^{2}=\left[3.75\text{ km},-25\text{ m/s},1.5\text{ km},-50\text{ m/s}\right]^{T}$. Although the averaged target states are estimated well for all the three allocations, the optimized allocation achieves the smallest CRMSE on average, which reflects the estimation stability over the $500$ trials.

\begin{table}[t]
\caption{\label{tab:Radar-RMSE}RMSE and estimated target states over 500 Monte
Carlo trials}

\centering{}%
\tabcolsep=0.18cm
\begin{tabular}{cccc}
\toprule 
Allocation & CRMSE (m) & Target & Estimated Target State\tabularnewline
\midrule
\multirow{2}{*}{Optimized} & \multirow{2}{*}{12.8830} & 1 & $\left[-1.5,50,-3.5,50\right]^{T}$\tabularnewline
\cmidrule{3-4} \cmidrule{4-4} 
 &  & 2 & $\left[3.7497,-25.1,1.499,50\right]^{T}$\tabularnewline
\midrule 
\multirow{2}{*}{Uniform} & \multirow{2}{*}{87.3946} & 1 & $\left[-1.4998,50,-3.5001,50\right]^{T}$\tabularnewline
\cmidrule{3-4} \cmidrule{4-4} 
 &  & 2 & $\left[-3.75,25,-1.5,50\right]^{T}$\tabularnewline
\midrule 
\multirow{2}{*}{Random} & \multirow{2}{*}{113.5374} & 1 & $\left[-1.4997,50.1,-3.5002,49.9\right]^{T}$\tabularnewline
\cmidrule{3-4} \cmidrule{4-4} 
 &  & 2 & $\left[3.75,-25,1.5001,-50\right]^{T}$\tabularnewline
\bottomrule
\end{tabular}
\end{table}

\begin{figure}[t]
\begin{centering}
\includegraphics[scale=0.6]{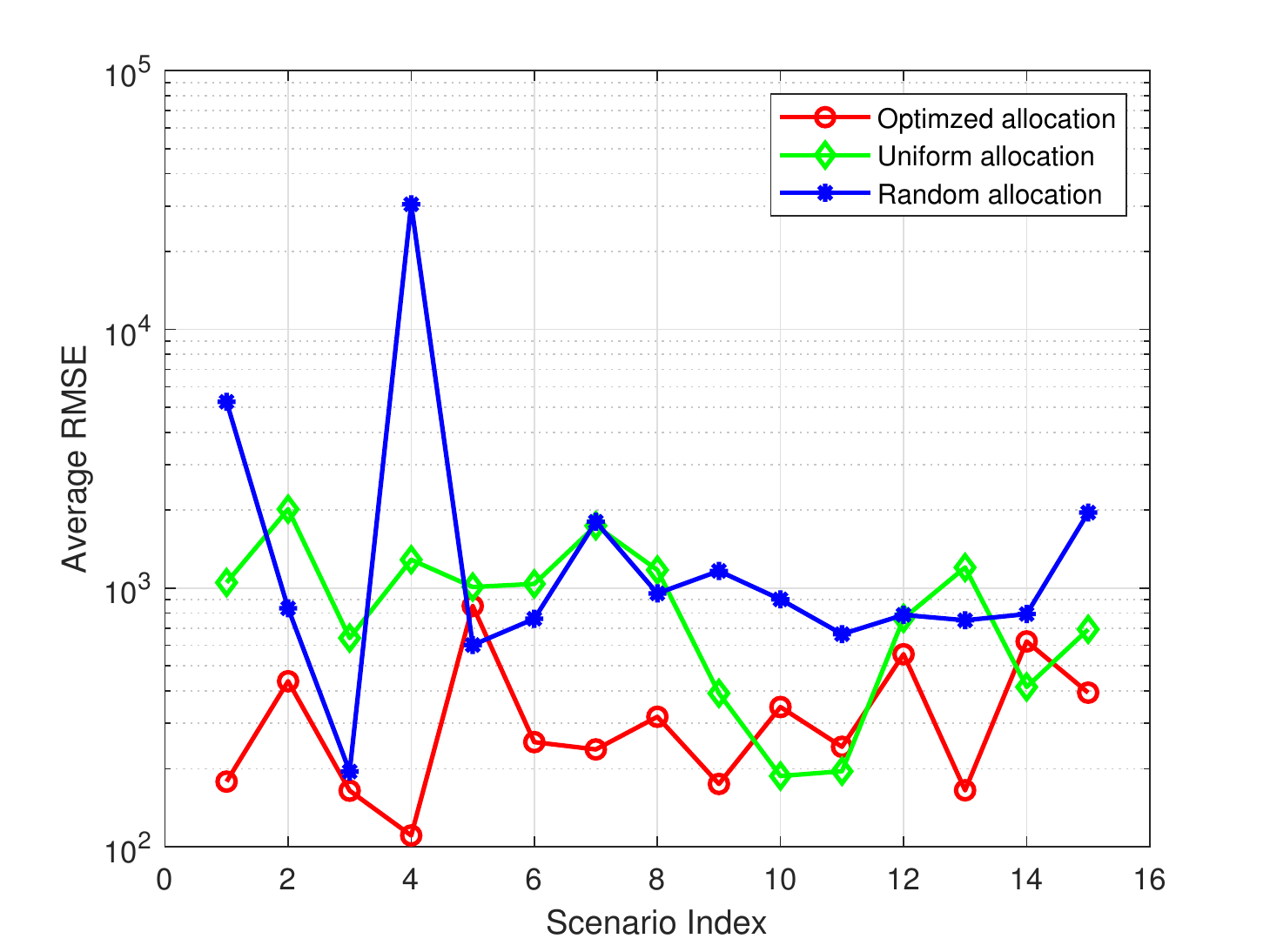}
\par\end{centering}
\caption{\label{fig:RMSE-over-20}Average CRMSE of the first fusion interval
over 15 different testing scenarios.}
\end{figure}

\paragraph{Performance across scenarios} To explore the statistical stability of the proposed allocation algorithm, we demonstrate the CRMSE performance over $15$ different testing scenarios. In each scenario, the radars are randomly placed to track two targets with randomly generated initial states. The BS is also randomly placed and the communications users are within the BS coverage. The system parameters are initialized correspondingly by the same distributions used in the previous experiments. Fig.~\ref{fig:RMSE-over-20} shows the CRMSE of a single fusion interval, where the initial $\boldsymbol{B}_{k|k}$ and $\boldsymbol{C}_{k|k}$ are randomly generated for different scenarios. Each CRMSE value is averaged over $200$ Monte Carlo trials. Clearly, over these scenarios, the average CRMSEs for the optimized allocation are roughly between $100$ m and $800$ m, which are further improved in the subsequent fusion intervals.

\subsection{ANCHOR performance across fusion intervals}
\begin{figure}[t]
\begin{centering}
\includegraphics[scale=0.3]{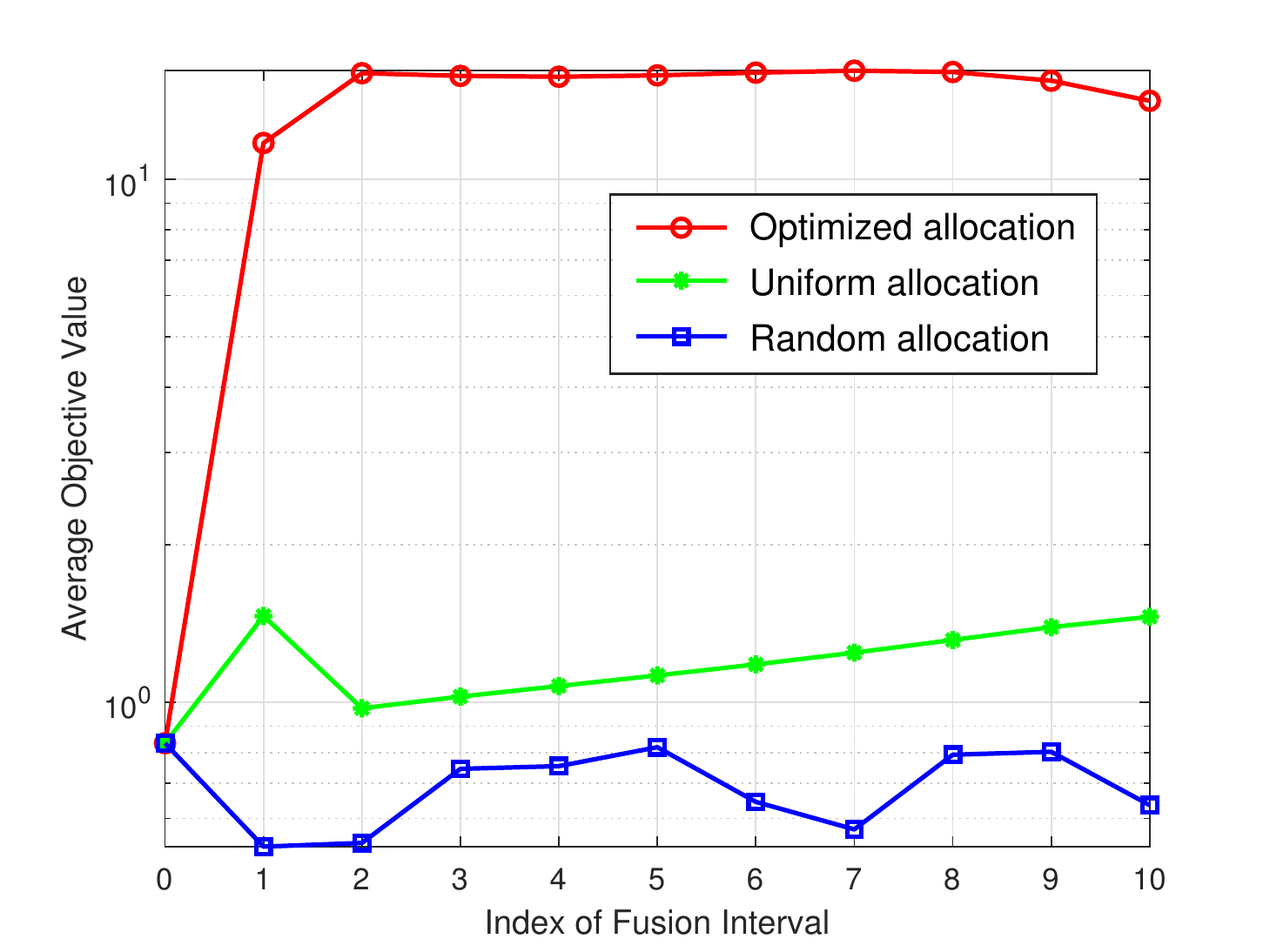}\includegraphics[scale=0.3]{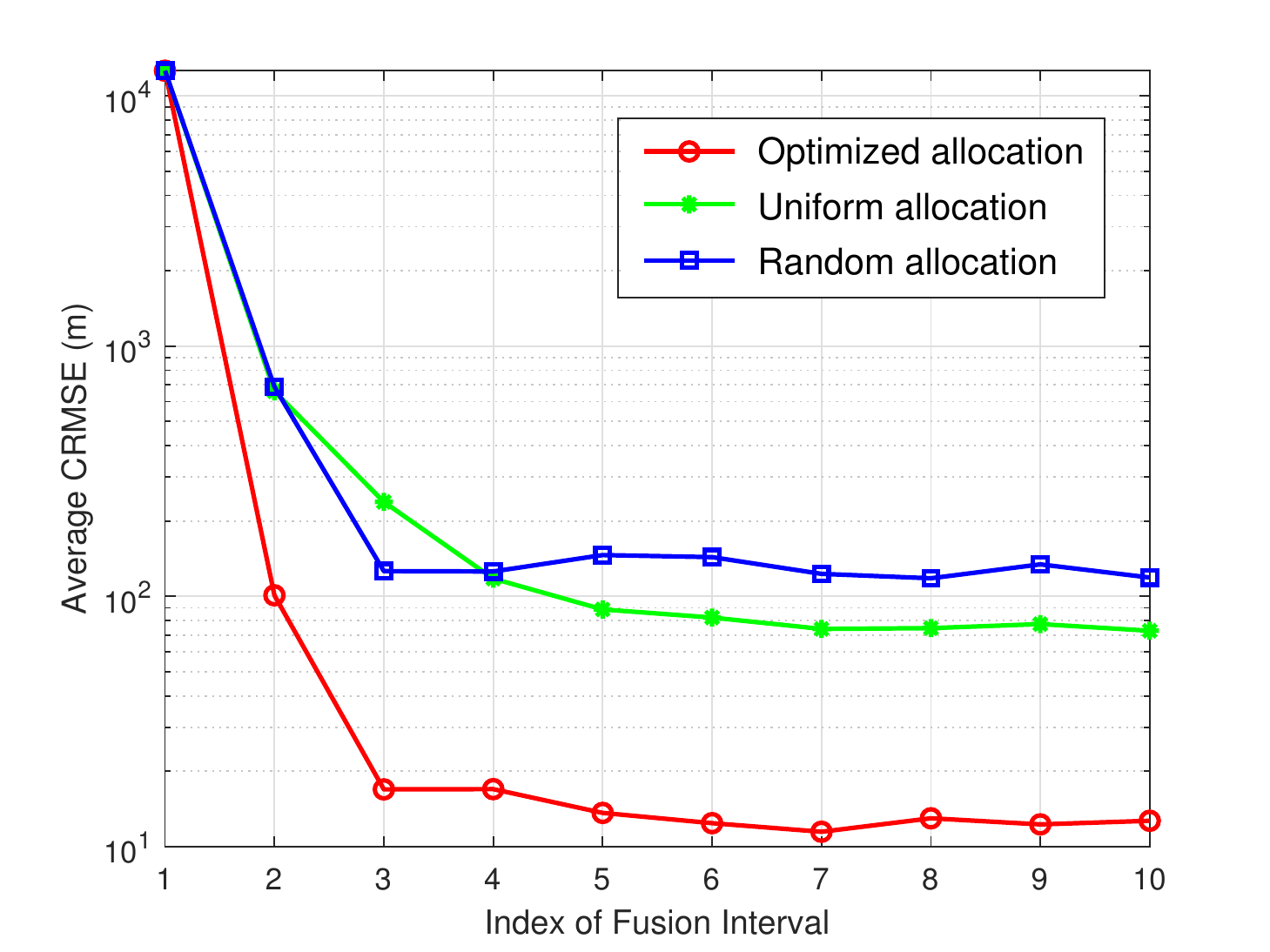}
\par\end{centering}
\caption{\label{fig:Average-multipleInterval}Average performance over multiple fusion intervals.  Left: Average CRB; right: Average CRMSE.}
\end{figure}

We conducted the experiments over $10$ consecutive fusion intervals indexing from $k=1$ to $k=10$. Based on the allocated resources, we update $\left\{ \boldsymbol{B}_{k+1}^{q}\right\} _{q=1}^{Q}$
and $\left\{ \boldsymbol{C}_{k+1|k+1}\right\} _{q=1}^{Q}$ of the $k$-th fusion interval as the inputs to the next interval. To evaluate the efficacy of the consecutive tracking scheme, we initialized $\boldsymbol{B}_{1}$
and $\boldsymbol{C}_{1|1}$ with a significantly amplified version of the state covariance matrix as a rough guess on the initial state. The tracking estimation is conducted based on \eqref{eq:kalmanUpdate}
and the performance is averaged over $200$ Monte Carlo trials. The trials are independent and each of them run the $10$ consecutive fusion intervals completely. Fig.~\ref{fig:Average-multipleInterval} demonstrates the averaged values over multiple fusion intervals. Starting with the same values, the optimized allocation improves the objective value significantly compared to the other two allocations. Besides, for the random allocation, it is expected that the objective value fluctuates over the different intervals. 

\begin{figure}[t]
\begin{centering}
\includegraphics[scale=0.3]{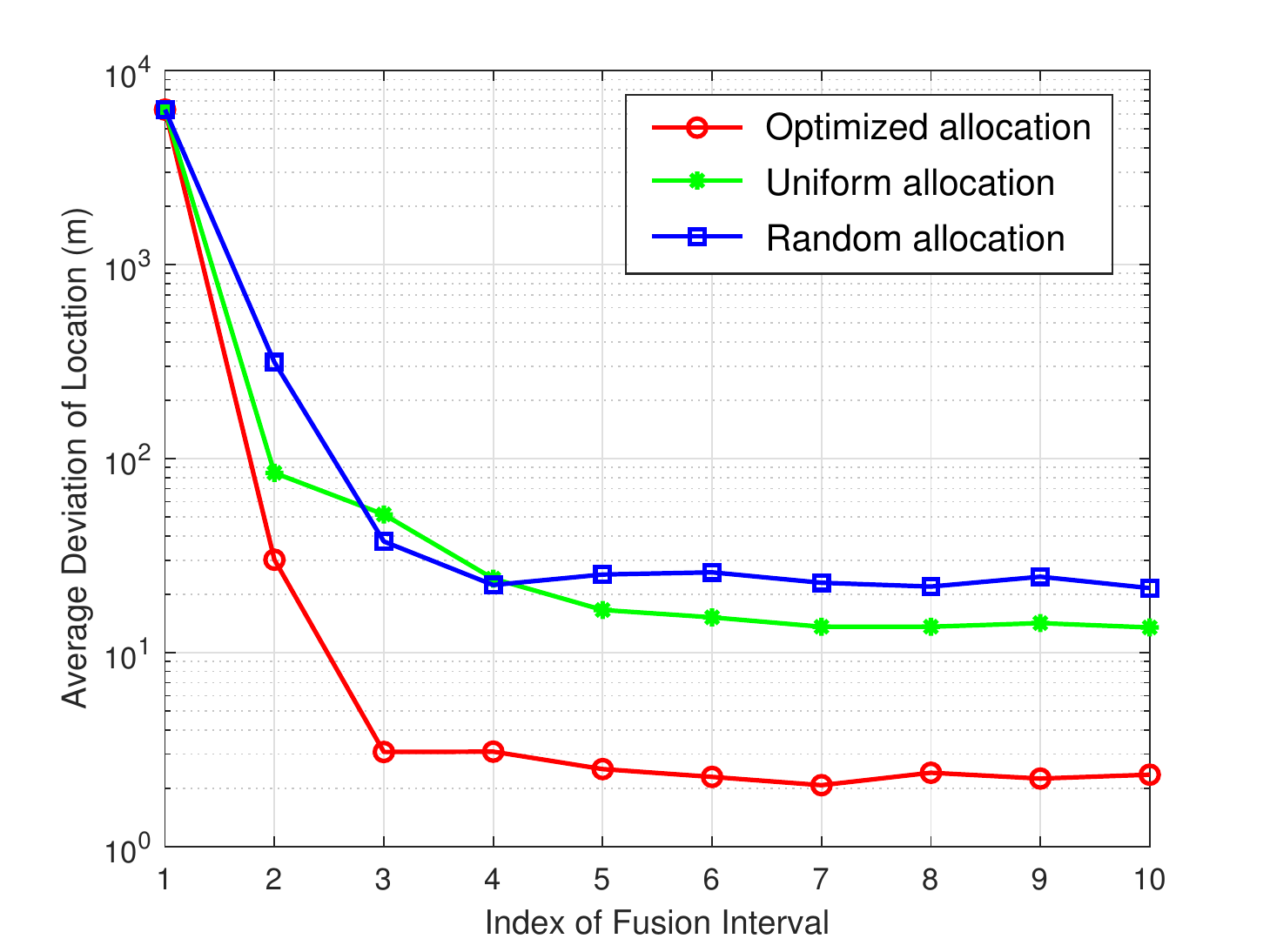}\includegraphics[scale=0.3]{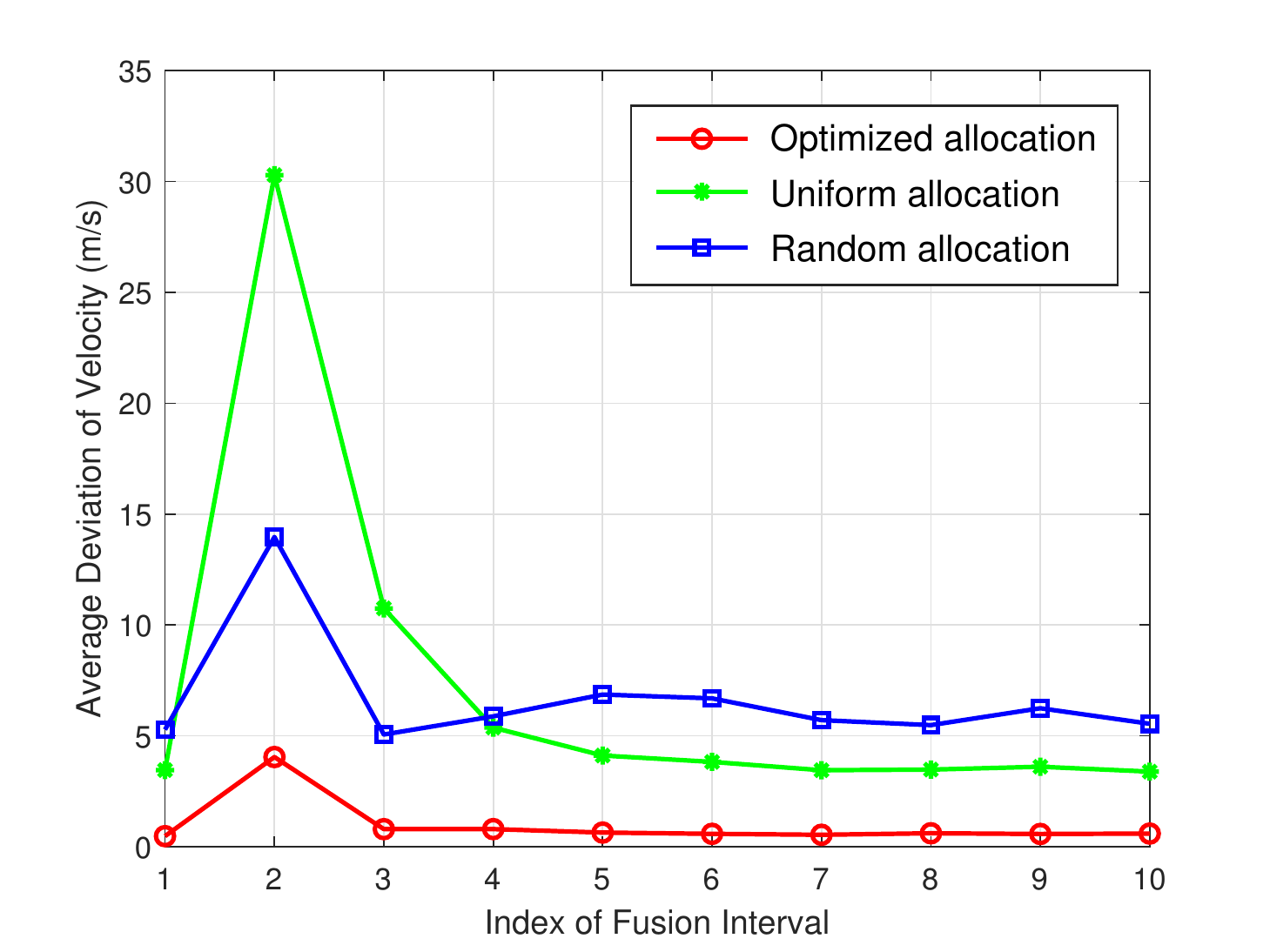}
\par\end{centering}
\caption{\label{fig:Average-tracking-deviations}Average deviations of target location and velocity.}
\end{figure}

\begin{comment}
\begin{figure}[t]
\begin{centering}
\includegraphics[scale=0.6]{Figures/ObjSumInvTr_vs_FusionIdx_MultiFusionMultiRA_20210814}
\par\end{centering}
\caption{\label{fig:CRB_multiple}Average CRB-based metric over multiple fusion intervals.}
\end{figure}

\begin{figure}[t]
\begin{centering}
\includegraphics[scale=0.6]{Figures/AvgCRMSE_vs_FusionIdx_MultiFusionMultiRA_20210814}
\par\end{centering}
\caption{\label{fig:RMSE_multiple}Average CRMSE over multiple fusion intervals.}
\end{figure}
\end{comment}

In terms of the average CRB, the average CRMSE performance over 200 Monte Carlo trials is demonstrated in Fig.~\ref{fig:Average-multipleInterval}. The CRMSE values of the three allocations at the first fusion interval is extremely large and close to each other. Recall that we set $\boldsymbol{B}_{1}$ and $\boldsymbol{C}_{1|1}$ with large values to represent a rough initial guess, and thereby we infer that the first tracking performance is heavily influenced by this "bad" initial guess and the improvement brought by the optimized allocation is negligible. Although the first tracking performance is not usable, all three CRMSE curves monotonically decrease with the fusion interval. This validates the efficacy of the Bayesian tracking scheme. The optimized allocation achieves the smallest CRMSE over all fusion intervals.  

Fig.~\ref{fig:Average-tracking-deviations} shows the tracking performance by inspecting the average deviations of location and velocity over all the intervals. Each average deviation is obtained by calculating the RMSE similar to \eqref{eq:CRMSE} without matrix $\boldsymbol{\varLambda}$. Compared to the uniform and random allocations, the optimized allocation consistently performs the best over all fusion intervals for both the location and velocity derivations; this is consistent with the result in Fig.~\ref{fig:Average-multipleInterval} indicating the average CRMSE of the optimized allocation as being the lowest. In addition, it is interesting to note that the velocity derivation has a relatively large fluctuation at the second fusion interval compared to the monotonicity of the location derivation. To explain this phenomenon, recall that each radar can only observe the radial velocity in our system model and thereby an accurate estimation on target velocity relies on the information fusion from all radars. Each radar is expected to provide some ``unique'' information about the velocity, which requires a proper placement of the radars with respect to the tracking target. Otherwise, the velocity ambiguity cannot be reduced substantially and consequently leads to unsatisfying velocity estimation.
\section{Summary}
\label{sec:summ}
We considered spectrum sharing between heterogeneously-distributed JRC with the goal of tracking multiple radar targets while maintaining the throughput levels for the communications downlinks. Within the asynchronous multi-target tracking framework, we proposed a Bayesian CRB-based metric for optimization, which further depends on the system resources (i.e. radar and communications power, dwell time, and shared bandwidth). The resulting resource allocation problem was non-convex and involved discrete and continuous variables. We solved this through our proposed ANCHOR algorithm based on the alternating optimization framework with guaranteed monotonicity. The frequency and power-time were allocated alternately. Our numerical experiments illustrated the key algorithmic and system aspects of resource allocation. We demonstrated that, compared to the trivial uniform and random allocations, the system performance is significantly improved by properly allocating the accessible heterogeneous resources of both radar and communications through ANCHOR algorithm. This study is helpful in meeting challenges of next-generation wireless systems where heterogeneous networks are envisaged such as different radio access technologies (RATs) and cell-free massive MIMO systems \cite{flores2021rate}.
%\appendices{}

\appendix
\subsection{Proof of Lemma \ref{lem:CRB}}\label{Appendix-0}
By definition, the FIM $\boldsymbol{J}_{\boldsymbol{y}_{k}^{q}}\left(\boldsymbol{s}_{t_{k+1}}^{q}\right)$ is
%\begin{equation}
%	\begin{aligned} &
$\boldsymbol{J}_{\boldsymbol{y}_{k}^{q}}\left(\boldsymbol{s}_{t_{k+1}}^{q}\right)%\\
    %= & 
    =\mathbb{E}\left\{ \left[\nabla_{\boldsymbol{s}_{t_{k+1}}^{q}}\ln p\left(\boldsymbol{y}_{k}^{q}|\boldsymbol{s}_{t_{k+1}}^{q}\right)\right]\left[\nabla_{\boldsymbol{s}_{t_{k+1}}^{q}}\ln p\left(\boldsymbol{y}_{k}^{q}|\boldsymbol{s}_{t_{k+1}}^{q}\right)\right]^{T}\right\}$. 
%    \end{aligned}
%	\label{eq:new-1}
%\end{equation}
Since the measurement $\boldsymbol{y}_{k}^{q}$ follows a normal distribution with 
%\begin{equation}
    $p\left(\boldsymbol{y}_{k}^{q}|\boldsymbol{s}_{t_{k+1}}^{q}\right)=\prod_{i=1}^{N}\prod_{m}^{M_{i,q,k}}\mathcal{N}\left(h\left(\boldsymbol{s}_{i,q,k}^{m},i\right),\boldsymbol{\Sigma}_{i,q,k}^{m}\right)$, 
%\end{equation} 
we have
\begin{equation}
    \begin{aligned} & \nabla_{\boldsymbol{s}_{t_{k+1}}^{q}}\ln p\left(\boldsymbol{y}_{k}^{q}|\boldsymbol{s}_{t_{k+1}}^{q}\right)\\
    = & \sum_{i=1}^{N}\sum_{m=1}^{M_{i,q,k}}\boldsymbol{H}_{i,q,k}^{mT}\left(\boldsymbol{\Sigma}_{i,q,k}^{m}\right)^{-1}\left(\boldsymbol{y}_{k}^{q}-h\left(\boldsymbol{s}_{i,q,k}^{m},i\right)\right),
\end{aligned}
\label{eq:new-2}
\end{equation}
where $\boldsymbol{H}_{i,q,k}^{m}$ is the Jacobian matrix of $h\left(\boldsymbol{s}_{i,q,k}^{m},i\right)$ on $\boldsymbol{s}_{t_{k+1}}^{q}$ based on $\boldsymbol{s}_{t_{k+1}}^{q} = f\left(\boldsymbol{s}_{i,q,k}^{m},t_{k+1}-t_{i,q,k}^{m}\right)$. 

We have 
\begin{equation}
\begin{aligned} & \mathbb{E}\left[\left(\boldsymbol{y}_{k}^{q}-h\left(\boldsymbol{s}_{i,q,k}^{m},i\right)\right)\left(\boldsymbol{y}_{k}^{q}-h\left(\boldsymbol{s}_{i',q',k}^{m},i\right)\right)^{T}\right]\\
= & \begin{cases}
\boldsymbol{\Sigma}_{i,q,k}^{m} & i=i',p=p'\\
\boldsymbol{0} & \text{otherwise}.
\end{cases}
\end{aligned}
\end{equation}
Hence, substituting \eqref{eq:new-2} into $\boldsymbol{J}_{\boldsymbol{y}_{k}^{q}}$ %\eqref{eq:new-1} 
yields 
%\begin{equation}
%\begin{aligned} & 
$\boldsymbol{J}_{\boldsymbol{y}_{k}^{q}}\left(\boldsymbol{s}_{t_{k+1}}^{q}\right)%\\
= %& 
\sum_{i=1}^{N}\sum_{m}^{M_{i,q,k}}\frac{P_{i,q,k}^{m}T_{i,q,k}^{m}}{\sum_{j=1}^{J}\alpha_{i,j}^{c}P_{c,k}^{j}+\sigma_{r,i}^{2}}\boldsymbol{H}_{i,q,k}^{mT}\left(\boldsymbol{C}_{i,q,k}^{m}\right)^{-1}\boldsymbol{H}_{i,q,k}^{m}$. 
%\end{aligned}
%\end{equation}
This completes the proof.

\subsection{Proof of Lemma \ref{lemma1}}\label{Appendix-A}
The inner minimization problem w.r.t. $\left\{ \boldsymbol{V}_{q}\right\} $ is \par\noindent\small
	\begin{equation}
		\begin{aligned} & \underset{\left\{ \boldsymbol{V}_{q}\right\} }{\text{minimize}} &  & \sum_{q=1}^{Q}\text{Tr}\left(\boldsymbol{V}_{q}^{T}\tilde{\boldsymbol{\Lambda}}^{T}\boldsymbol{B}_{q}\tilde{\boldsymbol{\Lambda}}\boldsymbol{V}_{q}\right)\\
			& \text{subject to} &  & \text{Tr}\left(\boldsymbol{V}_{q}\right)=1,\forall q=1,\ldots Q,
		\end{aligned}
	\end{equation}\normalsize
	which is further decoupled into $Q$ subproblems as \par\noindent\small
	\begin{equation}
		\begin{aligned} & \underset{\boldsymbol{V}_{q}}{\text{minimize}} &  & \text{Tr}\left(\boldsymbol{V}_{q}^{T}\tilde{\boldsymbol{\Lambda}}^{T}\boldsymbol{B}_{q}\tilde{\boldsymbol{\Lambda}}\boldsymbol{V}_{q}\right)\\
			& \text{subject to} &  & \text{Tr}\left(\boldsymbol{V}_{q}\right)=1.
		\end{aligned}
		\label{eq:32}
	\end{equation}\normalsize
	Problem \eqref{eq:32} has the Lagrangian %\par\noindent\small
	%\begin{equation}
		$\mathcal{L}\left(\boldsymbol{V}_{q},\lambda_{q}\right)=\text{Tr}\left(\boldsymbol{V}_{q}^{T}\tilde{\boldsymbol{\Lambda}}^{T}\boldsymbol{B}_{q}\tilde{\boldsymbol{\Lambda}}\boldsymbol{V}_{q}\right)+\lambda_{q}\left(\text{Tr}\left(\boldsymbol{V}_{q}\right)-1\right)$. 
	%\end{equation}\normalsize
	Following the Lagrangian method yields \par\noindent\small
	\begin{equation}
		\begin{cases}
			\frac{\partial\mathcal{L}}{\partial\boldsymbol{V}^{q}}=2\tilde{\boldsymbol{\Lambda}}^{T}\boldsymbol{B}_{q}\tilde{\boldsymbol{\Lambda}}\boldsymbol{V}_{q}+\lambda_{q}\boldsymbol{I}=0\\
			\frac{\partial\mathcal{L}}{\partial\lambda_{q}}=\text{Tr}\left(\boldsymbol{V}_{q}\right)-1=0,
		\end{cases}
	\end{equation}\normalsize
	which has the solution \par\noindent\small
	\begin{equation}
		\begin{cases}
			\boldsymbol{V}_{q}^{\star}=\frac{\left(\tilde{\boldsymbol{\Lambda}}^{T}\boldsymbol{B}_{q}\tilde{\boldsymbol{\Lambda}}\right)^{-1}}{\text{Tr}\left[\left(\tilde{\boldsymbol{\Lambda}}^{T}\boldsymbol{B}_{q}\tilde{\boldsymbol{\Lambda}}\right)^{-1}\right]}\succeq\boldsymbol{0}\\
			\lambda_{q}^{\star}=\frac{-2}{\text{Tr}\left[\left(\tilde{\boldsymbol{\Lambda}}^{T}\boldsymbol{B}_{q}\tilde{\boldsymbol{\Lambda}}\right)^{-1}\right]}.
		\end{cases}
	\end{equation}\normalsize
	Substituting the expression of $\boldsymbol{V}_{q}^{\star}$ into the objective function of problem \eqref{eq:subpro_z_maximin}, we have\par\noindent\small
	\begin{equation}
		\sum_{q=1}^{Q}\text{Tr}\left(\left(\boldsymbol{V}_{q}^{\star}\right)^{T}\tilde{\boldsymbol{\Lambda}}^{T}\boldsymbol{B}_{q}\tilde{\boldsymbol{\Lambda}}\boldsymbol{V}_{q}^{\star}\right)=\sum_{q=1}^{Q}\frac{1}{\text{Tr}\left(\boldsymbol{\Lambda}\boldsymbol{B}_{q}^{-1}\boldsymbol{\Lambda}^{T}\right)},
	\end{equation}\normalsize
	which is the objective function of problem \eqref{eq:RA opt problem}. This completes the proof.
	
\subsection{Proof of Lemma \ref{lemma2}}\label{Appendix-B}
Define $w\left(\boldsymbol{z},\boldsymbol{V}_{q}\right)=\text{Tr}\left(\boldsymbol{V}_{q}^{T}\tilde{\boldsymbol{\Lambda}}^{T}\boldsymbol{B}_{q}\tilde{\boldsymbol{\Lambda}}\boldsymbol{V}_{q}\right)$
and $\tilde{w}\left(\boldsymbol{z}\right)=\underset{\left\{ \boldsymbol{V}_{q}\right\} }{\min}\sum_{q=1}^{Q}w\left(\boldsymbol{z},\boldsymbol{V}_{q}\right)$,
which is the objective function of problem \eqref{eq:subpro_z_maximin}. The function $w\left(\boldsymbol{z},\left\{ \boldsymbol{V}_{q}\right\} \right)$ is differentiable in $\boldsymbol{z}$ and the feasible set $\mathcal{Z}$ is convex. Furthermore, $w\left(\boldsymbol{z},\boldsymbol{V}_{q}\right)$
is strongly convex in $\boldsymbol{V}_{q}$ and the set $\mathcal{V}_{q}=\left\{\boldsymbol{V}_{q}|\text{Tr}\left(\boldsymbol{V}_{q}\right)=1,\boldsymbol{V}_{q}\succeq\boldsymbol{0}\right\}$ is compact. Thus, by applying the Danskin's theorem \cite{danskin2012theory}, we have $\nabla\tilde{w}\left(\boldsymbol{z}\right)=\sum_{q=1}^{Q}\partial_{\boldsymbol{V}_{q}}w\left(\boldsymbol{z},\boldsymbol{V}_{q}\right)|_{\boldsymbol{V}_{q}=\boldsymbol{V}_{q}^{\star}}$, where $\boldsymbol{V}_{q}^{\star}=\arg\underset{\text{Tr}\left(\boldsymbol{V}_{q}\right)=1,\boldsymbol{V}_{q}\succeq\boldsymbol{0}}{\text{min}}\left\{ \text{Tr}\left(\boldsymbol{V}_{q}^{T}\tilde{\boldsymbol{\Lambda}}^{T}\boldsymbol{B}_{q}\tilde{\boldsymbol{\Lambda}}\boldsymbol{V}_{q}\right)\right\} $.
From this perspective, the update rules of Algorithm \ref{alg:Alg_subprob_z} can be merged into $\boldsymbol{z}^{n+1}=\mathcal{P}_{\mathcal{Z}}\left(\boldsymbol{z}^{n}-\eta\nabla\tilde{w}\left(\boldsymbol{z}^{n}\right)\right)$.
By applying \cite[Theorem 31]{jin2020local} with $\epsilon=0$ (due to the availability of the optimal $\boldsymbol{V}$), we arrive at the conclusion that the sequence $\left\{ \boldsymbol{z}^{n}\right\} $ will converge to the stationary point of $\tilde{w}\left(\boldsymbol{z}\right)$ for $\boldsymbol{z}\in\mathcal{Z}$.

\subsection{Proof of Theorem \ref{theorem3}}\label{Appendix-C}
At the $\ell$-th iteration, the objective value of problem \eqref{eq:RA opt problem} is $g\left(\boldsymbol{z}_{\ell},\boldsymbol{F}_{\ell}^{c}\right)$. We have 
%\[
$g\left(\boldsymbol{z}_{\ell},\boldsymbol{F}_{\ell}^{c}\right)\le g\left(\boldsymbol{z}_{\ell},\boldsymbol{F}_{\ell+1}^{c}\right)\le g\left(\boldsymbol{z}_{\ell+1},\boldsymbol{F}_{\ell+1}^{c}\right)$, 
%\]
where the first inequality holds because of Algorithm \ref{alg:Alg_subprob_F}, and the second inequality holds because Algorithm \ref{alg:Alg_subprob_z} converges to a stationary point. Thus, the sequence $\left\{ g\left(\boldsymbol{z}_{\ell},\boldsymbol{F}_{\ell}^{c}\right)\right\} $ is non-decreasing. Since the objective function $g\left(\boldsymbol{z},\boldsymbol{F}^{c}\right)$ is upper-bounded, the non-decreasing $\left\{ g\left(\boldsymbol{z}_{\ell},\boldsymbol{F}_{\ell}^{c}\right)\right\}$ will converge to a finite value.

\bibliographystyle{IEEEtran}
\bibliography{IEEEabrv,reference}

\vspace{-20pt}
\begin{IEEEbiography}[{\includegraphics[width=1in,height=1.25in,clip,keepaspectratio]{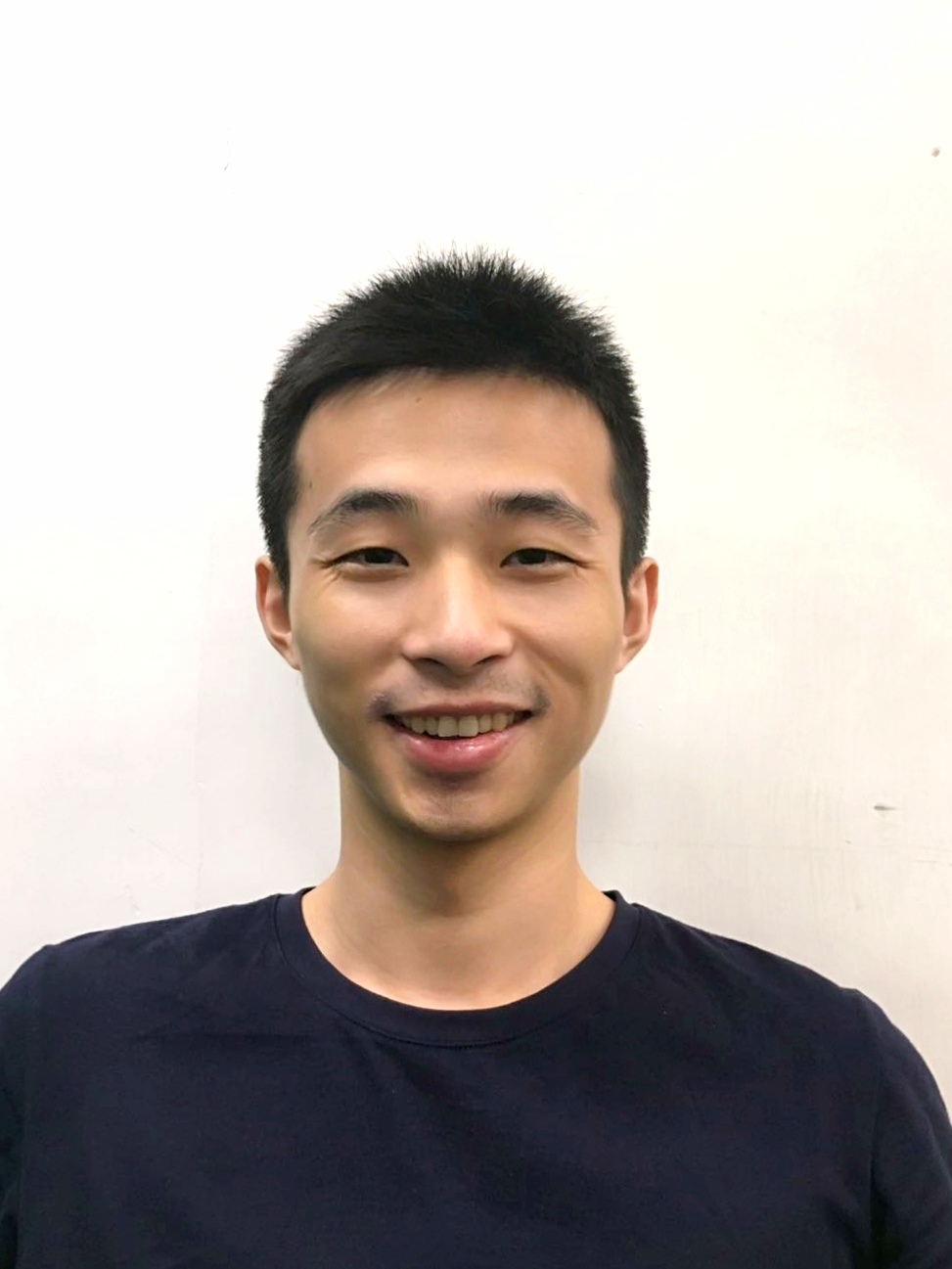}}]{Linlong Wu} received the B.E. degree in electronic information from Xi’an Jiaotong University (XJTU), Xi’an, China, in 2014, and the Ph.D. degree in electronic and computer engineering from Hong Kong University of Science and Technology (HKUST), Hong Kong, in 2018. He was with the wireless network group of Alibaba Cloud from November 2018 to October 2020 as a Research Engineer working on designing and building commercial RFID based localization systems. Since November, 2020, he has been with the Interdisciplinary Centre for Security, Reliability and Trust (SnT), University of Luxembourg and is currently a Research Associate in the Signal Processing Applications in Radar and Communications (SPARC) group. His research interests are signal processing, optimization and machine learning with applications in waveform design, integrated sensing and communication and IoT networks.\end{IEEEbiography}

%\begin{IEEEbiographynophoto}[{\includegraphics[width=1in,height=1.25in,clip,keepaspectratio]{Fig1.png}}]{Kumar Vijay Mishra} Biography text here
%\end{IEEEbiographynophoto}

\begin{IEEEbiography}[{\includegraphics[width=1in,height=1.25in,clip,keepaspectratio]{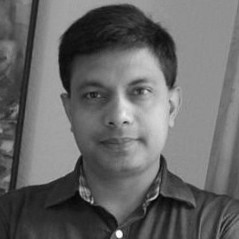}}]{Kumar Vijay Mishra} (S’08-M’15-SM’18) obtained a Ph.D. in electrical engineering and M.S. in mathematics from The University of Iowa in 2015, and M.S. in electrical engineering from Colorado State University in 2012, while working on NASA’s Global Precipitation Mission Ground Validation (GPM-GV) weather radars. He received his B. Tech. summa cum laude (Gold Medal, Honors) in electronics and communication engineering from the National Institute of Technology, Hamirpur (NITH), India in 2003. He is currently Senior Fellow at the United States Army Research Laboratory (ARL), Adelphi. He is the recipient of U. S. National Academies Harry Diamond Distinguished Fellowship (2018-2021), Royal Meteorological Society Quarterly Journal Editor's Prize (2017), Viterbi Postdoctoral Fellowship (2015, 2016), Lady Davis Postdoctoral Fellowship (2017), and DRDO LRDE Scientist of the Year Award (2006). Dr. Mishra is Vice-Chair (2021-present) of the newly constituted IEEE Synthetic Aperture Standards Committee of the IEEE Signal Processing Society. Since 2020, he has been Associate Editor of IEEE Transactions on Aerospace and Electronic Systems. He is Vice Chair (2021-2023) and Chair-designate (2023-2026) of International Union of Radio Science (URSI) Commission C. He is the lead/corresponding co-editor of three upcoming books on radar: \textit{Signal Processing for Joint Radar-Communications} (Wiley-IEEE Press), \textit{Next-Generation Cognitive Radar Systems} (IET Press), and \textit{Advances in Weather Radar Volumes 1, 2 and 3} (IET Press). His research interests include radar systems, signal processing, remote sensing, and electromagnetics.
\end{IEEEbiography}

\begin{IEEEbiography}[{\includegraphics[width=1in,height=1.25in,clip,keepaspectratio]{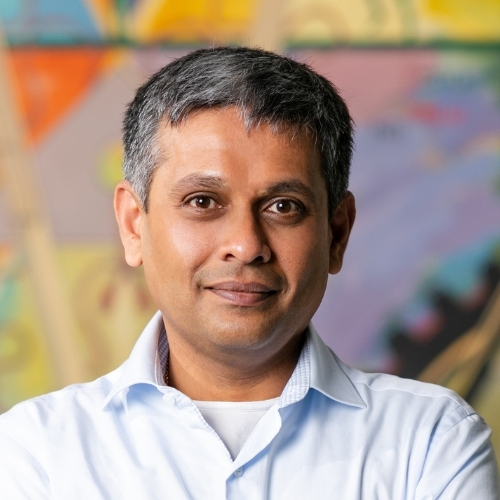}}]{M. R. Bhavani Shankar} received Masters and Ph.D. in Electrical Communication Engineering from Indian Institute of Science, Bangalore in 2000 and 2007 respectively. He was a Post Doc at the ACCESS Linnaeus Centre, Signal Processing Lab, Royal Institute of Technology (KTH), Sweden from 2007 to September 2009. He joined SnT in October 2009 as a Research Associate and is currently Assistant Professor leading the Signal Processing Applications in Radar and Communications (SPARC) group. He was with Beceem Communications, Bangalore from 2006 to 2007 as a Staff Design Engineer working on Physical Layer algorithms for WiMAX compliant chipsets. He was a visiting student at the Communication Theory Group, ETH Zurich, headed by Prof. Helmut B\"{o}lcskei during 2004. Prior to joining Ph. D, he worked on Audio Coding algorithms in Sasken Communications, Bangalore as a Design Engineer from 2000 to 2001. His research interests include Design and Optimization of MIMO Communication Systems, Automotive Radar and Array Processing, polynomial signal processing, Satellite communication systems, Resource Allocation, Game Theory and Fast Algorithms for Structured Matrices. He is currently on the Executive Committee of the IEEE Benelux joint chapter on communications and vehicular technology and serves as handling editor for Elsevier Signal Processing. He was a co-recipient of the 2014 Distinguished Contributions to Satellite Communications Award, from the Satellite and Space Communications Technical Committee of the IEEE Communications Society.\end{IEEEbiography}

\begin{IEEEbiography}[{\includegraphics[width=1in,height=1.25in,clip,keepaspectratio]{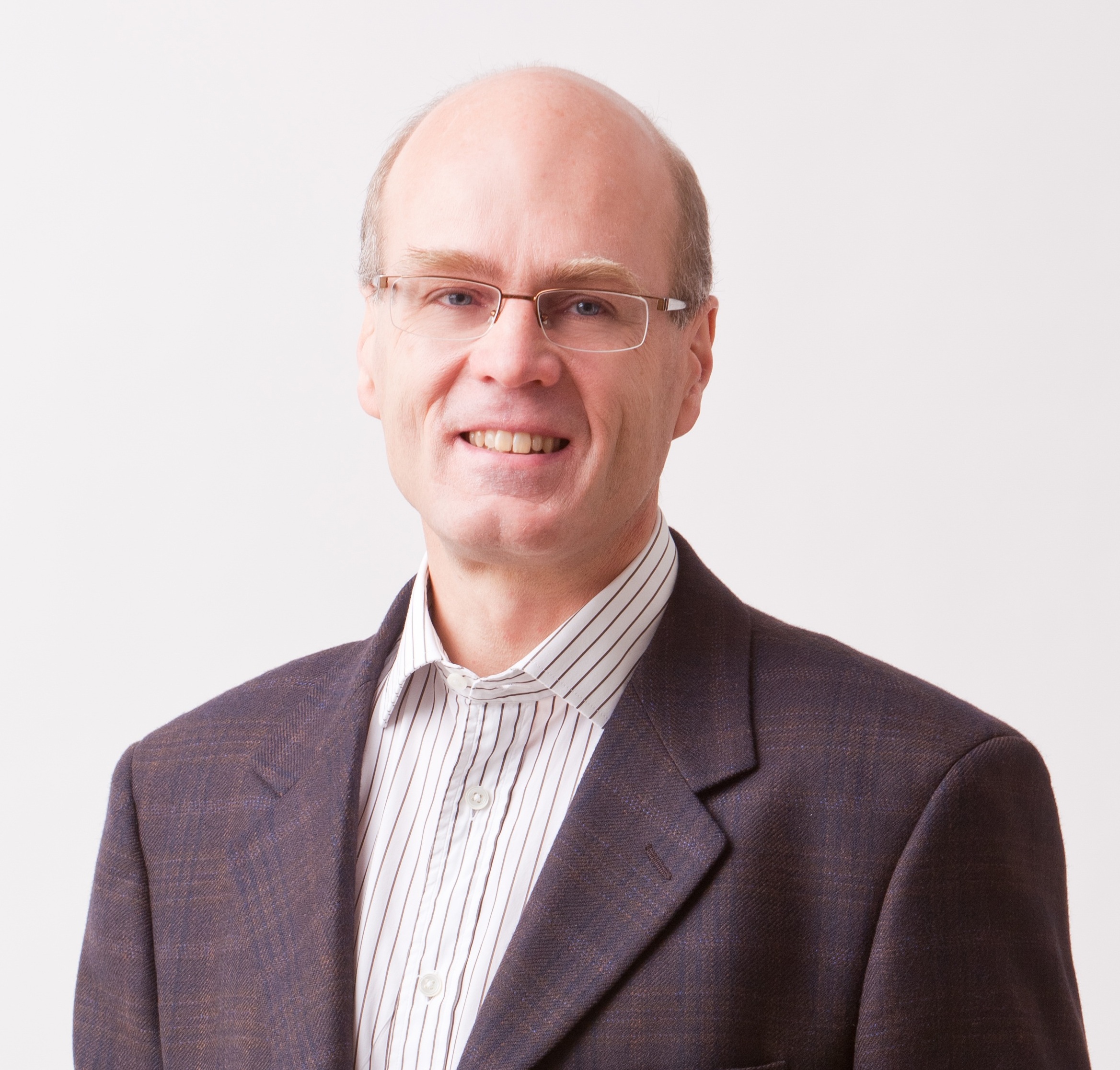}}]{Bj\"{o}rn Ottersten} (S’87–M’89–SM’99–F’04) received the M.S. degree in electrical engineering and applied physics from Linköping University, Linköping, Sweden, in 1986, and the Ph.D. degree in electrical engineering from Stanford University, Stanford, CA, USA, in 1990. He has held research positions with the Department of Electrical Engineering, Linköping University, the Information Systems Laboratory, Stanford University, the Katholieke Universiteit Leuven, Leuven, Belgium, and the University of Luxembourg, Luxembourg. From 1996 to 1997, he was the Director of Research with ArrayComm, Inc., a start-up in San Jose, CA, USA, based on his patented technology. In 1991, he was appointed Professor of signal processing with the Royal Institute of Technology (KTH), Stockholm, Sweden. Dr. Ottersten has been Head of the Department for Signals, Sensors, and Systems, KTH, and Dean of the School of Electrical Engineering, KTH. He is currently the Director for the Interdisciplinary Centre for Security, Reliability and Trust, University of Luxembourg. He is a recipient of the IEEE Signal Processing Society Technical Achievement Award, the EURASIP Group Technical Achievement Award, and the European Research Council advanced research grant twice. He has co-authored journal papers that received the IEEE Signal Processing Society Best Paper Award in 1993, 2001, 2006, 2013, and 2019, and 8 IEEE conference papers best paper awards. He has been a board member of IEEE Signal Processing Society, the Swedish Research Council and currently serves of the boards of EURASIP and the Swedish Foundation for Strategic Research. Dr. Ottersten has served as Editor in Chief of EURASIP Signal Processing, and acted on the editorial boards of IEEE Transactions on Signal Processing, IEEE Signal Processing Magazine, IEEE Open Journal for Signal Processing, EURASIP Journal of Advances in Signal Processing and Foundations and Trends in Signal Processing. He is a fellow of EURASIP.\end{IEEEbiography}

\end{document}